\documentclass[aps,groupedaddress,floatfix,nofootinbib,eqsecnum]{revtex4}
\usepackage{amssymb,amsmath,graphicx,multirow}

\begin{document}

\title{Dynamical Dark Matter:~ II. An Explicit Model}
\author{Keith R. Dienes$^{1,2,3}$\footnote{E-mail address:  {\tt dienes@physics.arizona.edu}},
        Brooks Thomas$^{4}$\footnote{E-mail address:  {\tt thomasbd@phys.hawaii.edu}}}
\affiliation{
     $^1$ Physics Division, National Science Foundation, Arlington, VA  22230  USA\\
     $^2$ Department of Physics, University of Maryland, College Park, MD  20742  USA\\
     $^3$ Department of Physics, University of Arizona, Tucson, AZ  85721  USA\\
     $^4$ Department of Physics, University of Hawaii, Honolulu, HI 96822  USA}

\begin{abstract}
In a recent paper~\cite{DynamicalDM1}, we introduced ``dynamical dark matter,'' a new 
framework for dark-matter physics, and outlined its underlying theoretical 
principles and phenomenological possibilities.  Unlike most traditional 
approaches to the dark-matter problem which hypothesize the existence of 
one or more stable dark-matter particles, our dynamical dark-matter 
framework is characterized by the fact that the requirement of stability 
is replaced by a delicate balancing between cosmological abundances and 
lifetimes across a vast ensemble of individual dark-matter components. 
This setup therefore collectively produces a time-varying cosmological 
dark-matter abundance, and the different dark-matter components can 
interact and decay throughout the current epoch.  While the goal of our 
previous paper was to introduce the broad theoretical aspects of this 
framework, the purpose of the current paper is to provide an explicit 
model of dynamical dark matter and demonstrate that this model satisfies 
all collider, astrophysical, and cosmological constraints.  The results of 
this paper therefore constitute an ``existence proof'' of the 
phenomenological viability of our overall dynamical dark-matter framework, 
and demonstrate that dynamical dark matter is indeed a viable alternative 
to the traditional paradigm of dark-matter physics.  Dynamical dark 
matter must therefore be considered alongside other approaches to the 
dark-matter problem, particularly in scenarios involving large extra 
dimensions or string theory in which there exist large numbers of 
particles which are neutral under Standard-Model symmetries.
\end{abstract}

\maketitle

\newcommand{\newc}{\newcommand}
\newc{\gsim}{\lower.7ex\hbox{$\;\stackrel{\textstyle>}{\sim}\;$}}
\newc{\lsim}{\lower.7ex\hbox{$\;\stackrel{\textstyle<}{\sim}\;$}}
\makeatletter
\newcommand{\biggg}{\bBigg@{3}}
\newcommand{\Biggg}{\bBigg@{4}}
\makeatother

\def\vac#1{{\bf \{{#1}\}}}

\def\beq{\begin{equation}}
\def\eeq{\end{equation}}
\def\beqn{\begin{eqnarray}}
\def\eeqn{\end{eqnarray}}
\def\calM{{\cal M}}
\def\calV{{\cal V}}
\def\calF{{\cal F}}
\def\half{{\textstyle{1\over 2}}}
\def\quarter{{\textstyle{1\over 4}}}
\def\ie{{\it i.e.}\/}
\def\eg{{\it e.g.}\/}
\def\etc{{\it etc}.\/}


\def\inbar{\,\vrule height1.5ex width.4pt depth0pt}
\def\IR{\relax{\rm I\kern-.18em R}}
 \font\cmss=cmss10 \font\cmsss=cmss10 at 7pt
\def\IQ{\relax{\rm I\kern-.18em Q}}
\def\IZ{\relax\ifmmode\mathchoice
 {\hbox{\cmss Z\kern-.4em Z}}{\hbox{\cmss Z\kern-.4em Z}}
 {\lower.9pt\hbox{\cmsss Z\kern-.4em Z}}
 {\lower1.2pt\hbox{\cmsss Z\kern-.4em Z}}\else{\cmss Z\kern-.4em Z}\fi}
\def\thbar{\bar{\theta}}
\def\fhatPQ{\hat{f}_{\mathrm{PQ}}}
\def\fPQ{f_{\mathrm{PQ}}}
\def\mPQ{m_{\mathrm{PQ}}}
\def\wtl{\widetilde{\lambda}}
\def\ta{\widetilde{a}}
\def\TBBN{T_{\mathrm{BBN}}}
\def\OmegaCDM{\Omega_{\mathrm{CDM}}}
\def\OmegaDM{\Omega_{\mathrm{CDM}}}
\def\Omegatot{\Omega_{\mathrm{tot}}}
\def\rhocrit{\rho_{\mathrm{crit}}}
\def\alMRE{a_{\lambda,\mathrm{M}}}
\def\aldotMRE{\dot{a}_{\lambda,\mathrm{M}}}
\def\tMRE{t_{\mathrm{MRE}}}
\def\TMRE{T_{\mathrm{MRE}}}
\def\tQCD{t_{\mathrm{QCD}}}
\def\tauMRE{\tau_{\mathrm{M}}}
\def\mPQdot{\dot{m}_{\mathrm{PQ}}}
\def\mPQddot{\ddot{m}_{\mathrm{PQ}}}
\def\mPQbar{\overline{m}_{\mathrm{PQ}}}
\def\mXdot{\dot{m}_X}
\def\mXddot{\ddot{m}_X}
\def\mXbar{\overline{m}_X}
\def\TRH{T_{\mathrm{RH}}}
\def\tRH{t_{\mathrm{RH}}}
\def\LambdaQCD{\Lambda_{\mathrm{QCD}}}
\def\fhatX{\hat{f}_X}
\def\tnow{t_{\mathrm{now}}}
\def\Omvac{\Omega_{\mathrm{vac}}^{(0)}}
\def\arcsinh{\mbox{arcsinh}}
\def\zRH{z_{\mathrm{RH}}}
\def\zMRE{z_{\mathrm{MRE}}}
\def\zinit{z_{\mathrm{init}}}
\def\tinit{t_{\mathrm{init}}}
\def\sinit{s_{\mathrm{init}}}
\def\sRH{s_{\mathrm{RH}}}
\def\sMRE{s_{\mathrm{MRE}}}
\def\snow{s_{\mathrm{now}}}
\def\BRgamma{\mathrm{BR}_{\lambda}^{(2\gamma)}}
\def\te{t_{\mathrm{early}}}
\def\tl{t_{\mathrm{late}}}
\def\Ehi{E_{\mathrm{high}}}
\def\Elo{E_{\mathrm{low}}}
\def\tBBN{t_{\mathrm{BBN}}}
\def\tosc{t_{\mathrm{osc}}}
\def\Tosc{T_{\mathrm{osc}}}
\def\Tnow{T_{\mathrm{now}}}
\def\Tmax{T_{\mathrm{max}}}
\def\fhatX{\hat{f}_X}
\def\LambdaG{\Lambda_G}
\def\mX{m_X}
\def\tX{t_X}
\def\tG{t_G}
\def\OmegaDM{\Omega_{\mathrm{CDM}}}
\def\ntrans{n_{\mathrm{trans}}}
\def\nosc{n_{\mathrm{osc}}}
\def\ninf{n_{\mathrm{inf}}}
\def\nG{n_{G}}
\def\ndec{n_{\mathrm{dec}}}
\def\ncut{n_{\mathrm{cut}}}
\def\nexpl{n_{\mathrm{expl}}}
\def\tLS{t_{\mathrm{LS}}}
\def\Ech{T_{\mathrm{ch}}}
\def\Omegavac{\Omega_{\mathrm{vac}}}
\def\etanow{\eta_\ast}
\def\lambdadec{\lambda_{\mathrm{dec}}}
\def\lambdatrans{\lambda_{\mathrm{trans}}}
\def\Omegatotnow{\Omega_{\mathrm{tot}}^\ast}


\input epsf



\tableofcontents


\section{Introduction}


The nature of what constitutes the non-baryonic dark matter in our universe 
remains one of the most fundamental mysteries in 
particle physics~\cite{DMReviews}.  
The most precise measurements of the relic abundance of this dark matter to date 
are those derived from WMAP data~\cite{WMAP}, which yield a value
\begin{equation}
  \Omega_{\mathrm{CDM}} h^2 ~=~ 0.1131 \pm 0.0034~,
  \label{eq:OmegaWMAP}
\end{equation}
where $h \approx 0.72$ is the Hubble constant.  Beyond this, we know very little about
the properties of this dominant constituent of the matter density in our universe,
save that its interactions with the fields of the Standard Model (SM) are extremely weak.  
One of the reasons why the nature of the dark matter remains so elusive is 
its apparent stability.
Observational constraints on the lifetime $\tau_\chi$ of any decaying 
dark-matter candidate $\chi$ are quite stringent.  Indeed, for
any particle with a relic abundance $\Omega_\chi \sim \OmegaDM$, 
current limits~\cite{DMDecayChenKamionkowski} 
from cosmic microwave background (CMB) measurements, \etc, require that
\begin{equation}
  \tau_\chi ~\gtrsim~  10^{26}\mathrm{~s}~.
  \label{eq:DecayingDMLifetimeLimit}
\end{equation}
For this reason, most models of the dark sector posit the existence of a 
single dark-matter particle (or, in the case of certain multi-component dark-matter 
scenarios~\cite{MultiCompDMBlock,Winslow}, a small number of such particles) which is 
either absolutely stable (with that stability usually conferred by some additional 
symmetry, such as R-parity in supersymmetric models, 
KK-parity~\cite{KKParity} in universal extra 
dimensions~\cite{Antoniadis,DDGLargeED,UED}, or T-parity~\cite{TParity} 
in little-Higgs theories~\cite{LittleHiggs}), 
or else sufficiently long-lived as to satisfy the bound in 
Eq.~(\ref{eq:DecayingDMLifetimeLimit}).  Indeed, the phenomenological consequences 
of dark-matter decays in models with unstable dark-matter candidates~\cite{DecayingDM} 
can be quite significant.  

Recently, an alternative framework for addressing the dark-matter question 
has been proposed~\cite{DynamicalDM1}.  In this so-called ``dynamical dark matter'' 
paradigm, the dark sector comprises not one or merely a few
particle species, but rather a vast ensemble of different fields 
$\phi_i$, each of which 
contributes only a fraction $\Omega_i$ of the total dark-matter relic 
abundance $\OmegaDM$.  None of these fields is presumed to be absolutely stable, and 
thus a non-zero decay width $\Gamma_i$ is associated with each field.
However, in this framework, the individual relic abundances of the $\phi_i$ fields 
are presumed to be generated in such a way that the most stable members of 
that ensemble are the most abundant.  By contrast, the abundances of the more unstable 
members are suppressed according to the size of their decay widths.  It is this balancing 
between $\Gamma_i$ and $\Omega_i$ which makes it possible for the phenomenological 
constraints relating to the effects of dark-matter decays to be satisfied. 

In Ref.~\cite{DynamicalDM1}, we focused on the model-independent aspects
of our dynamical dark-matter framework, discussing its broad theoretical
properties, without any detailed phenomenological analysis or 
comparison with data.  By contrast, in this work, as a ``proof of concept,'' we 
provide an explicit model of dynamical dark matter.   
In this model, the fields which collectively constitute the dynamical dark-matter 
ensemble are the KK excitations of a light axion-like field propagating in 
the bulk of a spacetime with one or more large, flat extra dimensions.  In this 
model, the fields of the SM, as well as the gauge fields 
associated with some additional, non-Abelian gauge group $G$ which confines at a scale
$\Lambda_G$, are taken to be 
localized on a four-dimensional subspace of that bulk.  The axion field is assumed 
to couple to the gauge fields of $G$ (and also potentially to one or more of the 
SM fields) via non-renormalizable operators suppressed by some effective, 
four-dimensional cutoff scale $\fhatX$.  
We shall demonstrate that within this setup, the resulting ensemble of axion KK
modes naturally satisfies all applicable observational constraints on dark-matter 
decays --- even if the stability of this ``dark tower'' is entirely unprotected. 

Another advantage of this particular dynamical dark-matter model is that it is
not only phenomenologically viable, but also theoretically well motivated.
In any theory in which the SM fields 
reside on a brane, the KK excitations of any bulk field are, from the point 
of view of the four-dimensional theory, massive particles neutral under the 
SM gauge group.  Thus, were it not for the lack of a stabilizing symmetry, 
any of these particles would be a natural candidate for dark matter.  However, our results 
demonstrate that a lack of stability is not an insurmountable impediment to
such fields serving as dark matter {\it collectively}, rather than individually.
This is an exciting prospect, for it provides a novel way of addressing the
dark-matter question in theories with extra dimensions.  
Furthermore, our model also demonstrates that realizing a viable  
dynamical dark-matter ensemble does not require an overly complicated dark sector, 
a large number of independent mass scales, or an excessive degree of fine-tuning.  
Indeed, the model presented here involves only three independent physical scales: 
the effective four-dimensional cutoff scale $\fhatX$, 
the confinement scale $\Lambda_G$, and the 
compactification scale $M_c$.  Together, these three scales determine the mass 
spectrum and decay properties of the entire ensemble.   
 
The outline of this paper is as follows.  
In Sect.~\ref{sec:AxionsInED}, we briefly review the formalism for discussing
axions and axion-like fields, beginning with the standard, four-dimensional
case and then moving on to the generalized, five-dimensional bulk-axion case.  In 
Sect.~\ref{sec:AxionDecayRates}, we calculate the decay widths of the KK 
modes of such a bulk axion-like field and investigate how these decay 
widths scale with the mass of
the mode.  In the process, we show that the decays of the lighter modes 
to SM fields experience a natural suppression, but that such decays
nevertheless dominate over decays to other, lighter bulk fields in the theory.
In Sect.~\ref{sec:Production}, we examine the various mechanisms through 
which a population of axion modes may be generated in the early universe, and demonstrate 
that the abundances generated for those modes by misalignment production are 
indeed balanced against their decay widths in precisely the manner required for 
dynamical dark matter.  In Sect.~\ref{sec:Abundances}, we examine the collective 
properties of the ensemble of axion KK modes.  We show that such an 
ensemble can collectively reproduce the observed dark-matter relic density 
given in Eq.~(\ref{eq:OmegaWMAP}), and that it possesses the appropriate equation of
state to be regarded as dark matter.  In Sect.~\ref{sec:Constraints}, we summarize
the experimental, astrophysical, and cosmological constraints on scenarios 
involving bulk axions in large, flat extra dimensions.  We demonstrate that 
these constraints can be satisfied in a model which also simultaneously 
yields the correct total relic abundance --- in other words, that our model 
truly constitutes a 
viable model of dynamical dark matter.  Finally, in the Conclusions, we 
summarize the results of the previous sections and discuss several further 
directions for future investigation.

As we have indicated, this paper is the second part of a two-part series that
began with Ref.~\cite{DynamicalDM1}.
Consequently, we shall assume that the reader is familiar with the ideas,
notation, and conventions established in Ref.~\cite{DynamicalDM1} in what follows.      


\section{Bulk Axions as Dynamical Dark Matter\label{sec:AxionsInED}}


As discussed in the Introduction, the model for dynamical dark
matter that we shall consider in this paper is a model in 
which the KK excitations of a bulk axion constitute the 
dark-matter ensemble.  In this section, therefore, we briefly review the 
formalism relevant for describing the dynamics of axions in four or 
more dimensions.  We begin with a brief summary of the relevant properties of the 
four-dimensional QCD axion (more detailed reviews of 
which may be found, \eg, in 
Refs.~\cite{KolbAndTurner,PecceiReview,KimReview2,HertzbergAxionCosmology}),
and then discuss how this formalism can be generalized to a broader class of axions 
and axion-like fields.  Finally, we summarize the formalism for embedding such 
fields in the bulk in theories with extra dimensions.      

\subsection{Axions in Four Dimensions}

The QCD axion emerges as a consequence of the Peccei-Quinn (PQ) 
mechanism~\cite{PecceiQuinn}, a mechanism which provides an elegant, dynamical 
solution to the strong CP problem.
The strong CP problem arises due to the non-trivial vacuum structure of QCD.  
Specifically, the QCD Lagrangian can in principle contain an additional term 
\begin{equation}
  \mathcal{L}_{\mathrm{QCD}}~\ni~ \overline{\Theta}\frac{g_s^2\xi}{32\pi^2}
    G^{\mu\nu a}\widetilde{G}_{\mu\nu}^a~,
\end{equation}
where $g_s$ is the $SU(3)$ coupling,
$\xi$ is a numerical factor of $\mathcal{O}(1)$, 
$G_{\mu\nu}^{a}$ is the field-strength tensor for the gluon field, and
$\widetilde{G}_{\mu\nu}^a =\frac{1}{2}\epsilon_{\mu\nu\rho\sigma}G^{\rho\sigma a}$ 
is its dual.  The parameter $\overline{\Theta}$ is given by  
$\overline{\Theta} \equiv \Theta + \mathrm{Argdet}M$, where $\Theta$ is the 
strong-interaction theta-angle and $M$ is the 
Cabbibo-Kobayashi-Maskawa (CKM) matrix.  In principle, 
$\overline{\Theta}$ can take any value.
However, experimental bounds on the electric dipole moment $d_n$ of the neutron 
serve to constrain $\overline{\Theta}$.  The most stringent limit is currently 
$|d_n|\leq 2.9 \times 10^{-26}$~e~cm~\cite{NeutronEDM}, which translates into
a bound
\begin{equation}
  \overline{\Theta} ~ <~ 0.7 \times 10^{-11}~.
\end{equation}   
While there is, in principle, no problem with $\overline{\Theta}$ taking so small
a value, there is no particular reason why it should be so small.  This fine-tuning
issue is what is commonly referred to as the strong CP problem.    

In the Peccei-Quinn solution to the strong CP problem, the 
effective $\overline{\Theta}$-parameter associated with the gluon field 
relaxes to zero dynamically as a consequence of the spontaneous breaking 
of an anomalous, global $U(1)$ symmetry, usually dubbed 
$U(1)_{\mathrm{PQ}}$, at some high scale $\fPQ$.     
The spontaneous breaking of this $U(1)_{\mathrm{PQ}}$ symmetry 
implies the presence of
a pseudo-Nambu-Goldstone boson: a real pseudoscalar commonly known as the QCD axion~\cite{WeinbergWilczekAxion}, which necessarily interacts
with the gluon field via the Lagrangian
\begin{equation}
  \mathcal{L} ~\ni~ \frac{1}{2}\partial_\mu a\partial^\mu a +
     \frac{g_s^2\xi}{32\pi^2 \fPQ} a G^{\mu\nu a}\widetilde{G}_{\mu\nu}^a~, 
  \label{eq:LagPQAxion}
\end{equation} 
where $a$ denotes the axion field.  Furthermore, this pseudoscalar may also 
have interactions with the other fields of the SM.  The presence of the
anomalous $U(1)_{\mathrm{PQ}}$ symmetry in the high-scale theory 
determines the effective Lagrangian for these interactions (at 
leading order) to be    
\begin{equation}
  \mathcal{L}_{\mathrm{int}} ~=~ 
     \frac{g_s^2\xi}{32\pi^2\fPQ} a G_{\mu\nu}^a\tilde{G}^{a\mu\nu}+
     \sum_i\frac{c_i}{\fPQ}
     (\partial_\mu a)\overline{\psi}_i\gamma^\mu\gamma^5\psi_i 
     +\frac{e^2 c_{\gamma}}{32\pi^2\fPQ} 
     a F_{\mu\nu}\tilde{F}^{\mu\nu}~,
  \label{eq:PQAxion4DInts}
\end{equation} 
where $\psi_i$ are the SM fermions and
$c_\gamma$ and $c_i$ are dimensionless coefficients.  These coefficients
depend on the charge assignments of the SM fields (and potentially of  
additional fields in the theory as well) under $U(1)_{\mathrm{PQ}}$, and
are therefore substantially more model-dependent than $\xi$.

At high temperatures, the axion field is effectively massless, as befits a
Nambu-Goldstone boson; however, it acquires a small, temperature-dependent mass 
$m_a(T)$ at lower scales due to QCD instanton effects.  A number of computations
of this mass have been performed, and while the results depend to some extent
on the assumptions and calculational techniques involved, $m_a(T)$ is often
assumed to have the rough form~\cite{QCDInstantonGross,Turner}
\begin{equation}
   m_a(T) ~\approx~
          \frac{g_s\xi}{4\sqrt{2}\pi} \frac{\LambdaQCD^2}{\fPQ}\times
      \begin{cases}
        \displaystyle\vspace{0.1cm} b \left(\frac{\LambdaQCD}{T}\right)^4 
        &\mathrm{for~} T\gtrsim \Lambda_{\mathrm{QCD}}\\ 
        \displaystyle 1  &\mathrm{for~}  T\lesssim \Lambda_{\mathrm{QCD}}~,
       \end{cases}
  \label{eq:InstantonMassThermal}
\end{equation}
where $\Lambda_{\mathrm{QCD}}\approx 250$~MeV is the QCD confinement scale, 
$b$ is a numerical coefficient of $\mathcal{O}(10^{-2})$, and $\xi$ is an 
$\mathcal{O}(1)$ numerical factor.  

The fact that the axion necessarily couples to the gluon field implies
that it will also have effective couplings to hadrons.  The most important
such couplings, phenomenologically speaking, are those of the axion to
pions and nucleons.  These couplings take the form~\cite{ChangChoiThermalNucleonRate}
\begin{eqnarray}
   \mathcal{L}_{\mathrm{had}} &=& \frac{C_{a\pi}}{f_\pi\fPQ}(\partial_\mu a)
       \left[(\partial^\mu \pi^+)\pi^-\pi^0 + (\partial^\mu \pi^-)\pi^+\pi^0
       -2(\partial^\mu \pi^0)\pi^+\pi^-\right] \nonumber\\ & & ~+
       \frac{C_{an}}{\fPQ}(\partial_\mu a)\overline{n}\gamma^\mu\gamma^5n +
         \frac{C_{ap}}{\fPQ}(\partial_\mu a)\overline{p}\gamma^\mu\gamma^5p + 
         \frac{iC_{a\pi N}}{f_\pi\fPQ}(\partial_\mu a)
         \left[\pi^+(\overline{p}\gamma^\mu n) - \pi^-(\overline{n}\gamma^\mu p)\right]~.
   \label{eq:HadronAxionCouplings4D}
\end{eqnarray}
The precise values for the effective nucleon-nucleon-axion couplings $C_{ap}$ and $C_{an}$,
the nucleon-pion-axion coupling $C_{a\pi N}$, and the axion-pion-pion coupling $C_{a \pi}$
depend on the $U(1)_{\mathrm{PQ}}$ charges of the quark fields.  For the case of a
so-called hadronic axion~\cite{KSVZ}, which does not couple directly to the SM quarks, the 
coefficients for the axion-nucleon-nucleon interaction are   
\begin{equation}
  C_{ap} ~=~ 0.24~ \frac{z}{(1+z)} + 0.15~ \frac{z-2}{(1+z)} + 0.02~,
  ~~~~~~~~
  C_{an} ~=~ 0.24~ \frac{z}{(1+z)} + 0.15~ \frac{1-2z}{(1+z)} + 0.02~,
\label{eq:CapAndCan}\\
\end{equation}
where $z=m_u/m_d\approx 0.56$ denotes the ratio of the up-quark to down-quark masses.
Similarly, the coefficients for the interactions involving pions are given by
\begin{equation}
  C_{a\pi N} ~=~ \frac{1-z}{2\sqrt{2}(1+z)}~,~~~~~~~
  ~~~~~~C_{a\pi} ~=~ \frac{1-z}{3(1+z)}~,
\label{eq:CaNAndCapi}
\end{equation}
where $m_\pi\approx 135.0$~MeV is the neutral pion mass, and 
$f_{\pi}\approx 93$~MeV is the pion decay constant. 
These hadronic couplings play an important role in constraining 
the parameter space of axion models.

The QCD axion is the prototypical example of a light pseudoscalar field 
whose mass arises solely due to non-perturbative effects associated with 
instanton dynamics, and whose interactions with the SM fields are highly suppressed. 
It is by no means the only example, however.  Indeed, 
a wide variety of additional particles possessing these same properties have appeared
in the literature in a number of beyond-the-Standard-Model (BSM) contexts, 
and are often generically referred to as 
axion-like particles (ALPs).  One particularly well-motivated example is the 
model-independent axion~\cite{WittenStringAxion} in string theory.  Since
axions of this more general sort are, by and large, no less viable as dark-matter
candidates than the QCD axion, it behooves us to extend our focus to encompass 
such fields as well. 

For the remainder of this work, then, we will use the term ``axion'' to refer
to any pseudoscalar field whose mass is generated by the instanton dynamics
associated with an arbitrary non-Abelian gauge group $G$.  This gauge group 
could be the SM $SU(3)$ color group, as it is for the QCD axion, but alternatively 
it could be some additional group which either resides in a hidden sector, 
or else confines at a very high scale.  As with the QCD axion,
any axion we consider will be assumed to be a 
pseudo-Nambu-Goldstone boson associated with the breaking of some global symmetry 
$U(1)_X$ at a scale $f_X$ by the vacuum expectation value (VEV) of some scalar field. 
The axion field is assumed to couple to the field strength 
$\mathcal{G}_{\mu\nu}^a$ associated with $G$ via a term of the form
\begin{equation}
  \mathcal{L}_{\mathrm{int}} ~\ni~ 
     \frac{g_G^2\xi}{32\pi f_X} a \mathcal{G}_{\mu\nu}^a
     \widetilde{\mathcal{G}}^{\mu\nu a}~,
\end{equation} 
where $a$ again denotes the axion field, $g_G$ is the 
coupling constant associated with $G$, $\widetilde{\mathcal{G}}^{\mu\nu a}$ is
the dual of $\mathcal{G}^{\mu\nu a}$, and $\xi$ is a model-dependent coefficient
which parameterizes the strength of the effective interaction between the 
axion and the gauge fields.  We will also assume that $G$ goes through a 
confining phase transition at some scale $\Lambda_G$, and that a potential 
analogous to that appearing in Eq.~(\ref{eq:InstantonPotential}) is
thereby generated for $a$.  In other words, this general axion couples to $G$   
in a manner completely analogous to that in which the QCD axion couples to the
SM $SU(3)$.  It therefore follows that all of the QCD-axion formalism 
outlined above continues to hold for axions in the broader sense of the word, 
provided one makes the substitutions 
$\LambdaQCD\rightarrow\Lambda_G$, $g_3\rightarrow g_G$, $\fPQ \rightarrow f_X$,
\etc, where appropriate.

There is, however, one crucial physical distinction between axions in general
and the QCD axion in specific: for general axions, the confinement 
scale $\Lambda_G$ is essentially a free parameter.  The properties 
of such an axion are therefore far less constrained than those of a QCD axion, 
simply because the axion mass is not uniquely determined by $f_X$ alone.  
Moreover, the vast majority of the experimental bounds on axions depend 
crucially on the charge assignments of the SM gauge fields under the global $U(1)_X$
symmetry.  For a generic axion, these charges need not have any relationship 
to the $U(1)_{\mathrm{PQ}}$ charge assignments for these fields.  An important 
consequence of this is that a generalized axion need not couple directly to the gluon 
field at leading order.  Moreover, other scenarios could be realized in 
such a framework that cannot arise for a QCD axion.  For example, one can imagine a 
purely ``photonic'' axion which couples to the photon field
at leading order, but not to the 
gluon field or to any of the SM fermions.  In what follows, we will focus on several
different concrete coupling scenarios.  One of these will be such a photonic axion; 
another will be a ``hadronic'' axion which couples to the gluon and photon fields, 
but not directly to any of the SM fermions.  However, we note that numerous other 
possibilities exist, and that the laboratory, astrophysical, and
cosmological constraints on any given model depend sensitively on
the couplings between the axion and the fields of the SM. 

It is also worth noting that certain details of any scenario of this sort 
will depend on the details of the instanton dynamics 
associated with the particular gauge group $G$ in question.  
The scaling behavior of $\mX(T)$ as a function of $T$, for example, may not
be identical to the scaling behavior quoted in Eq.~(\ref{eq:InstantonMassThermal}) 
for QCD instantons.  However, none of 
these details plays a crucial role in the dark-matter phenomenology of the 
our model.  We will therefore assume for the remainder
of this work that, except for the values of $\Lambda_G$, $g_G$, \etc, the 
standard axion results derived in the context of QCD-instanton dynamics apply 
to $G$-instanton dynamics as well.      

\subsection{Axions in Extra Dimensions}

Having summarized the formalism applicable to a four-dimensional axion, 
we now consider how the situation changes when the axion in question is    
allowed to propagate in the extra-dimensional bulk of a theory with more than
four dimensions.
As was originally pointed out in Ref.~\cite{DDGAxions}, the 
dynamics of such an axion is far richer than that of a purely four-dimensional 
axion, due both to the presence of an entire KK tower of axion 
excitations and to a non-trivial mixing between these excitations due to the
presence of brane mass terms, which explicitly violate KK mode-number conservation.  
Indeed, as we shall see, it is those KK excitations which 
will constitute the dark-matter ensemble in our model, 
and it is their mixing which gives this ensemble the appropriate 
properties to be a viable dynamical dark-matter candidate.  
Of course scenarios involving large extra dimensions 
have many other attractive features as well: they provide 
a geometric interpretation of the hierarchy between the weak scale 
and the Planck scale~\cite{ADD,ADDPhenoBounds,RS}, between the weak
scale and the grand-unification scale~\cite{DDGLargeED}, and between the
weak scale and the string scale~\cite{WeakStringHierarchy,DDGLargeED}.
Moreover, a higher-dimensional 
axion field can be accommodated quite naturally in such a brane/bulk 
framework.  Indeed,       
while only gravity is required to propagate in the bulk, the
propagation of SM-gauge-singlet fields there, including axions of 
all varieties, is, in a sense, almost expected. 

In what follows, we present the setup for a generic axion field in the 
bulk.  This parallels the setup for a QCD axion put forth in
Ref.~\cite{DDGAxions}.  For concreteness, we choose to
focus on the case in which the axion is allowed to propagate in a single, 
large extra dimension compactified on a $S_1/\IZ_2$ orbifold of radius $R$, 
while the fields of the SM
and the gauge fields associated with the additional symmetry group $G$  
are confined to the brane located at $x_5=0$.
However, we emphasize that the setup described here can easily be extended to 
scenarios in which the axion in question is allowed to propagate in multiple 
extra dimensions, or in which the background geometry is more 
complicated~\cite{AxionsInRS}.  

At scales below the weak scale but above the confinement scale $\Lambda_G$, 
the effective action for a bulk axion in five dimensions takes the form
\begin{equation}
   S_{\mathrm{eff}} ~=~ \int d^4x\int_0^{2\pi R} dx_5
     \left[\frac{1}{2}\partial_M a \partial^M a + 
     \delta(x_5)\,\big(\mathcal{L}_{\mathrm{brane}}+\mathcal{L}_{\mathrm{int}}\big)\right]~.
\end{equation}
Here, we have divided the brane-localized terms in the Lagrangian into two parts.  The first,
$\mathcal{L}_{\mathrm{brane}}$, contains the terms involving the brane fields alone --- both 
the fields of the SM and any additional fields, including the gauge fields associated
with the gauge group $G$.  The second, $\mathcal{L}_{\mathrm{int}}$, contains the 
interaction terms involving the brane-localized fields and the five-dimensional axion.  This 
second piece is given by
\begin{equation}
  \mathcal{L}_{\mathrm{int}} ~=~ 
     \frac{g_G^2\xi}{32\pi^2f_X^{3/2}} a \mathcal{G}_{\mu\nu}^a\tilde{\mathcal{G}}^{a\mu\nu}+
     \sum_i\frac{c_i}{f_X^{3/2}}
     (\partial_\mu a)\overline{\psi}_i\gamma^\mu\gamma^5\psi_i 
     +\frac{g_s^2 c_g}{32\pi^2f_X^{3/2}} a G_{\mu\nu}^a\tilde{G}^{a\mu\nu}
     +\frac{e^2 c_{\gamma}}{32\pi^2f_X^{3/2}} 
     a F_{\mu\nu}\tilde{F}^{\mu\nu}~,
  \label{eq:HighT5DAxionAction}
\end{equation} 
where $e$ and $g_s$ are the respective
couplings for $U(1)_{\mathrm{EM}}$ and $SU(3)$ color,
and $f_X$ is the fundamental five-dimensional scale associated with the 
breaking of $U(1)_X$ (the analogue of the Peccei-Quinn scale $\fPQ$ in 
Ref.~\cite{DDGAxions}). 
   
The first term in $\mathcal{L}_{\mathrm{int}}$ is the requisite coupling between 
the five-dimensional axion $a$ and the gauge fields of $G$.  The second term
represents the derivative couplings between the five-dimensional axion $a$ 
and the SM fermion fields $\psi_i$, with model-dependent coefficients 
$c_{i}$ that depend on the $U(1)_X$ charges of the $\psi_i$. 
The remaining two terms represent the interactions between the 
axion and the gluon and photon fields, the field-strength tensors for which
are here respectively denoted $G^{\mu\nu a}$ and $F^{\mu\nu}$, with 
(once again model-dependent) coefficients $c_{\gamma}$ and $c_g$.

The five-dimensional axion field can be 
represented as a tower of KK excitations via the decomposition
\begin{equation}
  a(x^\mu,x_5) ~=~ \frac{1}{\sqrt{2\pi R}}\sum_{n=0}^{\infty} 
     r_n a_n(x^\mu)\cos\left(\frac{nx_5}{R}\right)~,
  \label{eq:AxionModeDecomp}
\end{equation}    
where the factor
\begin{equation}
  r_n ~\equiv~ \begin{cases}
  1 &\mathrm{for~} n=0\\
  \sqrt{2}          &\mathrm{otherwise}
  \end{cases}
  \label{eq:rmDef}
\end{equation}
ensures that the kinetic term for each mode is canonically normalized.
Substituting this expression into Eq.~(\ref{eq:HighT5DAxionAction}) and 
integrating over $x_5$, we obtain
\begin{eqnarray}
  S_{\mathrm{eff}} &=& \int d^4x\Bigg[\sum_{n=0}^\infty
     \bigg(\frac{1}{2}\partial_\mu a_n \partial^\mu a_n 
     +\frac{g_G^2\xi}{32\pi^2\fhatX} 
     r_n a_n \mathcal{G}^a_{\mu\nu}\tilde{\mathcal{G}}^{a\mu\nu}
     +\sum_i\frac{c_i}{\fhatX}
     r_n(\partial_\mu a_n)\overline{\psi}_i\gamma^\mu\gamma^5\psi_i 
     \nonumber\\ & & ~~~~~~~~~~~     
     +~\frac{g_s^2c_g}{32\pi^2\fhatX} 
     r_n a_n G^a_{\mu\nu}\tilde{G}^{a\mu\nu}
     +\frac{e^2 c_{\gamma}}{32\pi^2\fhatX} 
     r_n a_n F_{\mu\nu}\tilde{F}^{\mu\nu}\bigg)
     -V(a)\Bigg]~,
  \label{eq:HighT4DAxionAction}
\end{eqnarray}  
where 
\begin{equation}
  V(a) ~=~ \sum_{n=0}^\infty\frac{1}{2}\frac{n^2}{R^2}a_n^2~,
\end{equation}
and where the quantity $\fhatX$, defined by the relation
\begin{equation}
  \fhatX^2 ~\equiv~ 2 \pi Rf_X^3~,
  \label{eq:fhatInTermsOff}
\end{equation}
represents the effective 
four-dimensional $U(1)_X$-breaking scale.  Note that each mode in the KK tower couples
to the SM fields with a strength inversely proportional to $\fhatX$.
Note also that at these scales, the axion mass-squared matrix 
\begin{equation}
  \mathcal{M}^2_{mn} ~\equiv~ \frac{\partial^2 V(a)}{\partial a_m\partial a_n}
\end{equation}  
is purely diagonal.

The effective action in Eq.~(\ref{eq:HighT4DAxionAction}) is valid at
high scales where $T \gg \Lambda_G$.  Around $T\sim \Lambda_G$,
however, instanton effects give rise to an additional contribution 
to the effective axion potential.  In the low-temperature regime,
the full potential takes the form
\begin{equation}
   V(a) ~=~ \sum_{n=0}^\infty\frac{1}{2}\frac{n^2}{R^2}a_n^2+
          \frac{g_G^2}{32\pi^2}\Lambda_G^4
          \left[1-\cos\left(\frac{\xi}{\fhatX}
          \sum_{n=0}^\infty r_n a_n + \overline{\Theta}_G\right)\right]~, 
  \label{eq:InstantonPotential}
\end{equation}     
where $\overline{\Theta}_G$ is the analogue of the QCD theta-parameter  
$\overline{\Theta}$.
Minimizing the potential yields the vacuum configuration 
$\langle a_0\rangle =\fhatX (-\overline{\Theta}_G+\pi\ell)/\xi$ for 
$\ell \in 2\mathbb{Z}$, with $\langle a_n\rangle = 0$ for all $n>0$.  
This additional potential term modifies the axion mass-squared matrix at scales
$T\lesssim \Lambda_G$ to
\begin{equation}
  \mathcal{M}^2_{mn} ~=~ M_c^2n^2\delta_{mn}
  +\frac{g_G^2\xi^2}{32\pi^2}\frac{\Lambda_G^4}{\fhatX^2}
  r_mr_n\cos\left(\frac{\xi}{\fhatX}\sum_{k=0}^\infty r_k a_k +\overline{\Theta}_G\right)~,
  \label{eq:MassMixMatmn}
\end{equation}
where $M_c \equiv 1/R$ is the compactification scale.
We see here that the terms originating from the instanton-induced potential in 
Eq.~(\ref{eq:InstantonPotential}) include off-diagonal contributions, which
result in mixings among the KK eigenstates. 
In the vicinity of the minimum of $V(a_n)$, the axion mass-squared matrix above 
takes the form~\cite{DDGAxions}
\begin{equation}
  \mathcal{M}^2 ~=~ \mX^2\left(\begin{array}{ccccc}
  1 & \sqrt{2} & \sqrt{2} & \sqrt{2} & \ldots \\
  \sqrt{2} & 2 + y^2 & 2 & 2 & \ldots \\
  \sqrt{2} & 2 & 2+4y^2 & 2 & \ldots \\
  \sqrt{2} & 2 & 2 & 2+9y^2 & \ldots \\
  \vdots & \vdots & \vdots & \vdots & \ddots 
  \end{array}\right)~,
  \label{eq:AxionMassMatrixExplicit}
\end{equation}   
where 
\begin{equation}
  y~\equiv~ \frac{M_c}{\mX}~~~~~~~ \mathrm{and}~~~~~~~
  \mX^2 ~\equiv~ \frac{g_G^2\xi^2}{32\pi^2}\frac{\Lambda_G^4}{\fhatX^2}~.
  \label{eq:DefsOfyandmPQ}
\end{equation}
The eigenvalues $\lambda^2$ of this mass-squared matrix are the set of
solutions to the transcendental equation
\begin{equation}
  \frac{\pi\lambda\mX}{y}\cot\left(\frac{\pi\lambda}{\mX y}\right) ~=~ \lambda^2~.
  \label{eq:TranscendentalEqForLambdas}
\end{equation}
The normalized mass eigenstate $a_\lambda$ corresponding to each $\lambda$
may be written as a sum of the KK eigenstates $a_n$:
\begin{equation}
  a_\lambda ~=~ \sum_{n=0}^\infty U_{\lambda n} a_n
    ~\equiv~\sum_{n=0}^\infty\left(
    \frac{r_n\widetilde{\lambda}^2}{\widetilde{\lambda}^2-n^2y^2}\right)A_\lambda a_n~,
  \label{eq:DefOfalambda}
\end{equation}
where $\widetilde{\lambda}\equiv\lambda/\mX$, and where
\begin{equation}
  A_\lambda ~\equiv~ \frac{\sqrt{2}}{\wtl}\left[1+ \wtl^2 + \pi^2/y^2\right]^{-1/2}~.
  \label{eq:DefOfCapitalAlambda}
\end{equation}
The quantity $A_\lambda$ can be shown to obey the sum rules~\cite{DDGAxions}
\begin{equation}
  \sum_\lambda A_\lambda^2 ~=~ 1~, ~~~~~~
  \sum_\lambda \wtl^2A_\lambda^2 ~=~ 1~,
  \label{eq:AlambdaSqdID}
\end{equation}  
which follow directly from the unitarity of $U_{\lambda n}$.  
Upon rewriting Eq.~(\ref{eq:HighT4DAxionAction}) in this mass eigenbasis, we obtain
the axion effective action at temperatures $T\lesssim\Lambda_G$, 
which, up to $\mathcal{O}(a_\lambda^6/\fhatX^6)$, is given by 
\begin{eqnarray}
  S_{\mathrm{eff}} &=& \int d^4x
     \Bigg[\sum_{\lambda}\bigg(\frac{1}{2}\partial_\mu a_\lambda 
     \partial^\mu a_\lambda -\frac{1}{2}\wtl^2\mX^2 a_\lambda^2
     +\frac{e^2 c_{\gamma}\wtl^2A_\lambda}{32\pi^2\fhatX} 
     a_\lambda F_{\mu\nu}\tilde{F}^{\mu\nu}
     +\frac{g_s^2 c_g\wtl^2A_\lambda}{32\pi^2\fhatX} 
     a_\lambda G_{\mu\nu}^a\tilde{G}^{\mu\nu a}  
     \nonumber\\ & & ~~~~~ 
     + \sum_i\frac{c_i\wtl^2A_\lambda}{\fhatX}
     (\partial_\mu a_\lambda)\overline{\psi}_i\gamma^\mu\gamma^5\psi_i\bigg) 
     +\frac{g_G^2\xi^4\Lambda_G^4}{768\pi^2\fhatX^4}
     \sum_{\lambda_i,\lambda_j,\lambda_k,\lambda_\ell}
     \hspace{-0.4cm}\wtl_i^2\wtl_j^2\wtl_k^2\wtl_\ell^2
     A_{\lambda_i}A_{\lambda_j}A_{\lambda_k}A_{\lambda_\ell}
     a_{\lambda_i}a_{\lambda_j}a_{\lambda_k}a_{\lambda_\ell}
     \Bigg]~.
  \label{eq:ActionInMassEigenbasis}
\end{eqnarray}
The quartic axion self-interaction terms shown above originate from the 
instanton-induced potential in Eq.~(\ref{eq:InstantonPotential}). 
Other, higher-order terms not shown may be safely neglected when 
$T \ll \fhatX$. 

If a non-trivial coupling exists between the bulk axion and the gluon field, 
effective interactions will also arise between 
the $a_\lambda$ and the hadron fields at temperatures below $\LambdaQCD$.  
(In the case of the QCD axion, of course, such couplings are mandatory.)  
The Lagrangian which describes these interactions
is just the five-dimensional analogue of Eq.~(\ref{eq:HadronAxionCouplings4D}):    
\begin{eqnarray}
   \mathcal{L}_{\mathrm{had}} &=& \wtl^2A_\lambda
       \frac{C_{a\pi}}{f_\pi\fhatX}(\partial_\mu a_\lambda)
       \Big[(\partial^\mu \pi^+)\pi^-\pi^0 + (\partial^\mu \pi^-)\pi^+\pi^0
       -2(\partial^\mu \pi^0)\pi^+\pi^-\Big] ~+~
       \wtl^2A_\lambda\frac{C_{an}}{\fhatX}(\partial_\mu 
         a_\lambda)\overline{n}\gamma^\mu\gamma^5n \nonumber\\ & & ~+~
         \wtl^2A_\lambda\frac{C_{ap}}{\fhatX}(\partial_\mu 
         a_\lambda)\overline{p}\gamma^\mu\gamma^5p ~+~ 
         i\wtl^2A_\lambda\frac{C_{a\pi N}}{f_\pi\fhatX}(\partial_\mu a_\lambda)
         \Big[\pi^+\overline{p}\gamma^\mu n - \pi^-\overline{n}\gamma^\mu p\Big]~,
   \label{eq:HadronAxionCouplings5D}
\end{eqnarray} 
where the coefficients $C_{a\pi}$, $C_{an}$, \etc, depend on the details of the
theory, and may differ from those given in Eqs.~(\ref{eq:CapAndCan}) 
and~(\ref{eq:CaNAndCapi}).
 
As discussed in Ref.~\cite{DynamicalDM1}, the mass spectrum of the model
reduces to a reasonably simple form in certain limiting cases which
depend on the value of the ratio $y$ defined in Eq.~(\ref{eq:DefsOfyandmPQ}).  
The first of these is weakly-mixed regime, in which $y\gg 1$.  In this regime, the 
$a_\lambda$ are all very nearly equivalent to the KK modes $a_n$, with
masses $\lambda\approx n M_c$, where $n$ is an integer.  The extent to
which any given value of $\lambda$ differs from $M_c$ is set by the size of
the off-diagonal terms in Eq.~(\ref{eq:AxionMassMatrixExplicit}), and in particular, 
the lightest mass eigenstate $a_{\lambda_0}$ has a mass 
$\lambda_0 \approx M_c/y = \mX$. 
In short, the KK tower essentially comprises a single light mode plus a tower 
of massive KK excitations of that mode.  In the extreme limit, in which 
$M_c \rightarrow \infty$, the theory reduces to an effectively four-dimensional 
theory with a single light axion whose mass is precisely equal to $\mX$, as expected.       

In the opposite limit, in which $y\ll 1$, the 
situation is markedly different~\cite{DDGAxions}.
The lighter mass eigenstates in the theory have masses 
$\lambda\approx (n+1/2)M_c \ll\mX$
(where $n=\{0,1,2,\ldots\}$ is an integer), while the heavier mass eigenstates have 
masses $\lambda\approx nM_c$.  The transition region between the two regimes occurs
at around
\begin{equation}
   \lambda ~\sim~ \lambdatrans ~\equiv~ \frac{\pi\mX^2}{M_c}~,
   \label{eq:LambdaRegimesTransPt}
\end{equation}
which corresponds to a value $n = \pi/y^2$.  In this regime, the states with masses
below this threshold are highly mixed, owing to the large, off-diagonal terms in
Eq.~(\ref{eq:AxionMassMatrixExplicit}) proportional to $\mX$.  We dub this the
``strongly-mixed'' regime.  In this latter regime, as we shall soon see, the full 
KK tower plays a far larger role in the dark-matter phenomenology 
of a given model than in weakly-mixed scenarios, in which the dark-matter 
phenomenology is, more or less, the phenomenology of the zero mode.  Indeed, 
the bulk-axion scenarios which give rise to dynamical dark-matter ensembles in
which the full tower contributes significantly tend to be
those in which $y$ is small.
   
It is evident from Eqs.~(\ref{eq:ActionInMassEigenbasis}) 
and~(\ref{eq:HadronAxionCouplings5D}) that the parameter combination
$\wtl^2 A_\lambda/\fhatX$ plays a critical role in bulk-axion dynamics. 
Indeed, it is this combination which determines the strength of the 
interaction between a given mass eigenstate $a_\lambda$ and any 
of the SM fields. 
It turns out to be phenomenologically quite significant 
that not all $a_\lambda$ couple to the fields of the SM with the same 
strength.  We discuss the impact of such a coupling
structure on the decay properties of the tower states in 
Sect.~\ref{sec:AxionDecayRates}, and summarize its impact on
phenomenological constraints in Sect.~\ref{sec:Constraints}
(a more detailed analysis of which can be found in 
Ref.~\cite{DynamicalDM3}).  In addition, a plot of how
$A_\lambda$ and $\wtl^2A_\lambda$ depend on $\lambda$ is provided in 
Ref.~\cite{DynamicalDM1}.  It is worth remarking that the
non-universality of the $a_\lambda$ couplings is yet another 
direct consequence of the non-trivial mixing between axion KK modes 
implied by Eq.~(\ref{eq:AxionMassMatrixExplicit}).  This effect
does not arise for bulk fields in the absence of such mixing: the 
couplings of the KK excitations of the graviton to the SM fields in
theories of this sort, for example, are identical for all modes.

In summary, our model for dynamical dark matter consists of an axion 
propagating in the bulk of an extra dimension of radius $R$, with the SM
living on a brane.  From the perspective of an observer on the brane, our
dynamical dark-matter ensemble consists of the KK modes of this bulk axion 
field.  As we have discussed above, our model involves three important 
dimensionful parameters: $\fhatX$, $M_c$, and $\Lambda_G$.  We have 
also shown that the physics of this model depends crucially on one 
particular dimensionless combination of these parameters, namely $y$, which
governs the extent to which the individual KK modes mix when forming the 
constituents of our dynamical dark-matter ensemble.


\section{Characterizing the Constituents:~ Decay Widths\label{sec:AxionDecayRates}}


Now that we have reviewed the setup underlying our model for 
dynamical dark matter, we may begin to assess its phenomenological 
ramifications.  As discussed in Ref.~\cite{DynamicalDM1}, the essence of the 
dynamical dark-matter framework lies in the balance between the decay 
widths and relic abundances
of the fields which contribute to $\OmegaDM$.  Therefore, our principal aim must be
to evaluate the decay widths $\Gamma_\lambda$ and relic abundances $\Omega_\lambda$ of 
the fields $a_\lambda$ which our dark-matter ensemble comprises, and examine how these
two quantities scale with $\lambda$.  In this section, we focus on decays:
we calculate the partial-width contribution associated with each of the potentially 
relevant decay channels for a generic $a_\lambda$ and assess how the total width 
$\Gamma_\lambda$ depends on the dimensionful parameters of the model, namely
$\fhatX$, $M_c$, and $\Lambda_G$.  In the subsequent section, we focus on abundances.
     
\subsection{Decays to Standard-Model States}

We begin our discussion of axion decays by computing the decay widths 
of the $a_\lambda$ directly to SM states on the brane.
The first step is to derive Feynman rules for the relevant
interactions, which can be obtained directly from 
the interaction terms given in Eq.~(\ref{eq:ActionInMassEigenbasis}) 
and~(\ref{eq:HadronAxionCouplings5D}).
For those $a_\lambda$ with $\lambda$ below a few GeV, 
the relevant vertices are:
\begin{eqnarray*}
\raisebox{0.1cm}{
\resizebox{1.65in}{!}{\includegraphics{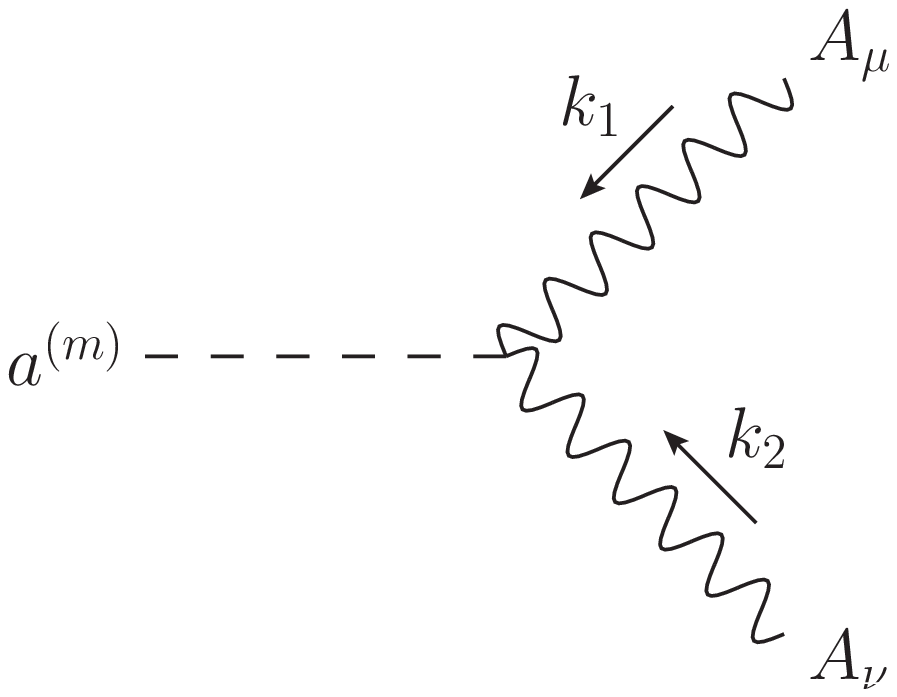}}}&
   \raisebox{1.7cm}{
   $ \hspace{-1.0cm} 
   \displaystyle ~=~
   -\frac{ie^2c_\gamma}{8\pi^2\fhatX}\wtl^2A_\lambda\epsilon_{\mu\nu\rho\sigma}
   k_1^\rho k_2^\sigma$}\\ 
\resizebox{1.65in}{!}{\includegraphics{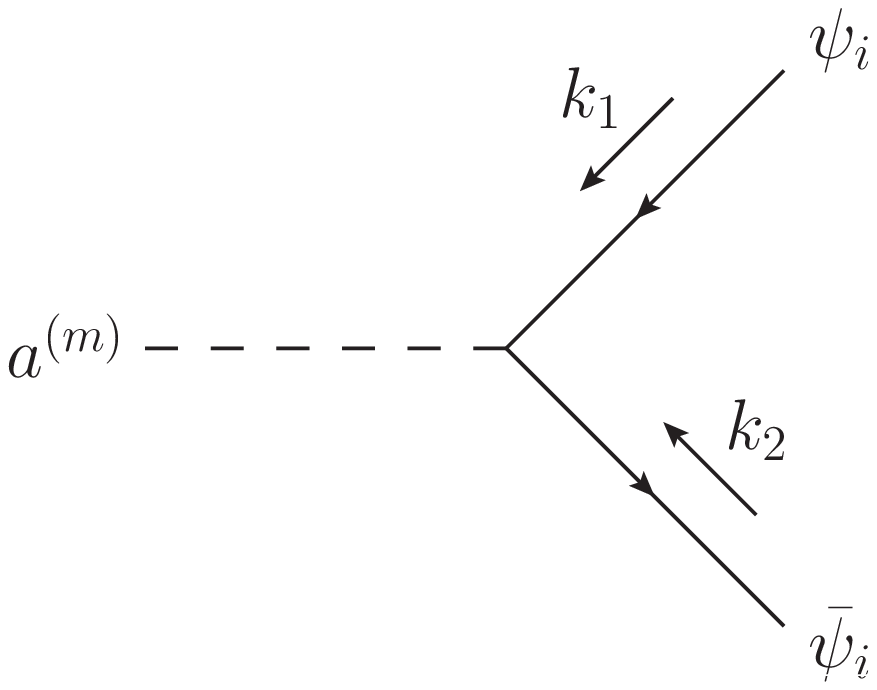}}&
   \raisebox{1.7cm}[3.25cm]{
   $ \hspace{-2.0cm}
   \displaystyle ~=~
   \frac{c_i\widetilde{\lambda}^2A_\lambda}{\fhatX}
   (\displaystyle{\not}k_1 + \displaystyle{\not}k_2)\gamma^5$}\\  
\resizebox{1.65in}{!}{\includegraphics{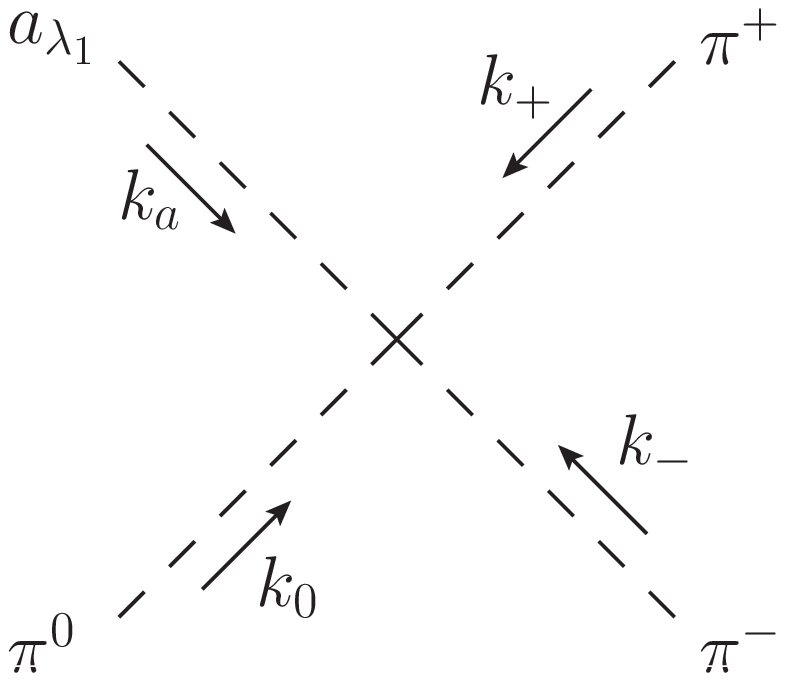}}& 
   \raisebox{1.7cm}{
   $ \hspace{0.5cm} 
   \displaystyle ~=~
   -\frac{C_{a\pi}}{f_\pi f_X}\wtl^2 A_\lambda k_a\cdot(k_- + k_+ - 2k_0)$~.}
\end{eqnarray*}
Here, and throughout the rest of this work, the symbol 
$M_P$ represents the effective, four-dimensional {\it reduced} Planck mass.
Since the coupling of each $a_\lambda$ to the SM fields is suppressed by
$\fhatX$, the contribution to the total decay width $\Gamma_\lambda$ of 
each $a_\lambda$ which comes from decays
to SM fields will be suppressed by $\fhatX^{-2}$.  At low
temperatures, the most relevant decay processes (depending, of course, 
on the precise values of $c_i$ and $c_\gamma$ in any given model) will 
be $a_\lambda\rightarrow \gamma\gamma$, $a_\lambda \rightarrow e^+e^-$, and
$a_\lambda \rightarrow \nu_i\overline{\nu}_i$, where $i=\{1,2,3\}$ labels the three
light neutrino mass eigenstates.  

The one decay channel which is kinematically accessible for all axion modes,
regardless of their mass (provided $c_\gamma$ is non-vanishing), is 
$a\rightarrow \gamma\gamma$.  The partial decay width of an axion mass eigenstate 
$a_\lambda$ into a pair of photons is
\begin{equation}
  \Gamma(a_\lambda\rightarrow\gamma\gamma) ~=~
    \frac{c_\gamma^2\alpha^2\lambda^3}{256\pi^3\fhatX^2}
    (\wtl^2A_\lambda)^2
    ~=~ G_\gamma(\wtl^2A_\lambda)^2 \frac{\lambda^3}{\fhatX^2}~, 
  \label{eq:GammaDecayToPhotons}
\end{equation}
where $\alpha\equiv e^2/4\pi$, and where we have defined the quantity 
$G_\gamma \equiv c_\gamma^2\alpha^2/256\pi^3$.  Note that 
$\Gamma(a_\lambda\rightarrow\gamma\gamma)$ includes an overall 
factor $(\wtl^2 A_\lambda)^2$.  This factor is a direct 
consequence of mixing in the axion mass matrix, and it appears
universally in all partial-width expressions for axion decays to
states on the brane.  For those modes with masses 
$\lambda \gtrsim \pi \mX^2/M_c$, this factor is $\mathcal{O}(1)$; 
however, for those modes with $\lambda \lesssim \pi \mX^2/M_c$, this
factor can be much smaller.  We therefore see that
the mixing factor $(\wtl^2 A_\lambda)^2$ suppresses the decay widths of
the lighter $a_\lambda$, while leaving the widths of the heavier $a_\lambda$
unsuppressed. 
This decay-width suppression for the light modes plays a crucial role 
in bulk-brane models of dynamical dark matter, as discussed in 
Ref.~\cite{DynamicalDM1}: since these light modes also turn out to have
the largest relic abundances, their decays are generally the most 
dangerous from a phenomenological perspective.

As $\lambda$ increases,
additional decay channels open up in which a given $a_\lambda$ decays to a light 
fermion-antifermion pair, provided that direct couplings exist 
between the axion and the fermionic species $\psi_i$ in question.  The partial width 
for any decay of this sort is given by
\begin{equation}
  \Gamma(a_\lambda\rightarrow \psi_i\overline{\psi}_i) ~=~
    \frac{c_i^2\lambda m_\psi^2}{2\pi\fhatX^2}(\wtl^2A_\lambda)^2
    \left(1-\frac{4m_\psi^2}{\lambda^2}\right)^{1/2}~,
  \label{eq:GammaDecayToFermions}
\end{equation}
where $m_\psi$ is the mass of the fermion in question.  Note that if
the five-dimensional axion field couples to the SM neutrinos, the 
$\Gamma_\lambda$ will have a non-trivial dependence on the neutrino mass spectrum.
While the precise masses of the three neutrino species are as yet unknown, measurements
of the solar and atmospheric squared-mass splittings place lower limits on two
of the three $m_{\nu_i}$.  The current best-fit values for these splittings 
are~\cite{PDG}
\begin{eqnarray}
  \Delta m^2_{\odot} &=& 7.59^{+0.19}_{-0.21} \times 10^{-5}\mbox{ eV}^2~,
   \nonumber\\
  |\Delta m^2_A|     &=& 2.43^{+0.13}_{-0.13} \times 10^{-3}\mbox{ eV}^2~.
  \label{eq:NeutSplitAtmo}
\end{eqnarray}
In what follows we shall assume a normal hierarchy, and we will take the mass of 
the lightest neutrino to be vanishingly small.  In this case, the masses of the
heavier two neutrinos are $m_{\nu_2} \simeq 8.7\times 10^{-3}$~eV and 
$m_{\nu_3} \simeq 4.9\times 10^{-2}$~eV, which are comparable to the lower 
bound~\cite{KapnerModGrav}
\begin{equation}
  M_c ~\gtrsim~ 3.9\times 10^{-3}~\mbox{eV}
  \label{eq:MinimumMc}
\end{equation} 
on the compactification scale $M_c$
from modified-gravity experiments.  Therefore, if $M_c$ lies only
slightly above this bound, the masses $\lambda$ of certain light $a_\lambda$ will be
comparable to $m_{\nu_{2,3}}$.  The partial width for the decays of those 
$a_\lambda$ to neutrinos can therefore in principle be quite large compared 
to their partial widths for decays to photons, as can be seen by comparing 
Eqs.~(\ref{eq:GammaDecayToPhotons}) and~(\ref{eq:GammaDecayToFermions}).  Note that
if $c_{\nu_i} = 0$ for all $i$, the decay width loses all sensitivity to the 
neutrino mass spectrum.  This is indeed the case for the photonic and hadronic axions 
which will serve as our primary examples in what follows. 

As $\lambda$ increases still further, decays of the $a_\lambda$ to hadrons become
kinematically accessible --- provided, of course, that either $c_g \neq 0$, or 
else that $c_{q_i} \neq 0$ for some quark species $q_i$.  The lowest 
such threshold is that for decays of the form
$a\rightarrow \pi^+\pi^-\pi^0$, which are kinematically allowed whenever 
$\lambda > 2m_{\pi^\pm} + {m_\pi^0}$.  
The relevant interaction vertex is that
appearing in the top line of Eq.~(\ref{eq:HadronAxionCouplings5D}), and the 
corresponding contribution to the decay width of $a_\lambda$
from this three-body decay takes the form 
\begin{equation}
  \Gamma(a_\lambda\rightarrow \pi^+\pi^-\pi^0) ~=~ \frac{C_{a\pi}^2(\wtl^2A_\lambda)^2}
     {1024\pi^3\lambda^3f_\pi^2\fhatX^2} ~\mathcal{I}(\lambda)~, 
     \label{eq:Gamma3piFull}
\end{equation}
where $\mathcal{I}(\lambda)$ denotes the phase-space integral 
\begin{equation}
   \mathcal{I}(\lambda) ~=~
     \int_{4m_{\pi^\pm}^2}^{(\lambda-m_{\pi_0})^2} dm_{12}^2~
     \big(\lambda^2+m_{12}^2 - m_{\pi_0}^2\big)^2
     \left(1-\frac{4m_{\pi^\pm}^2}{m_{12}^2}\right)^{1/2} 
     \Big[m_{12}^4+2m_{12}^2(\lambda^2-3m_{\pi^0}^2)+(\lambda^2-m_{\pi^0}^2)^2
     \Big]^{1/2}~,
    \label{eq:Gamma3piIntegralDef}
\end{equation}  
with integration variable $m_{12}^2 \equiv (k_+ + k_-)^2$.
For $\lambda \gg 2m_{\pi^\pm} + m_{\pi^0}$, this expression takes the 
asymptotic form
\begin{equation}
   \Gamma(a_\lambda\rightarrow \pi^+\pi^-\pi^0) ~\approx~ 
     (2.07\times 10^{-2})
     \, C_{a\pi}^2       
     \left(\frac{\lambda^5}{f_\pi^2\fhatX^2}\right)(\wtl^2A_\lambda)^2~.
\end{equation}
In practice, this asymptotic expression is a good approximation for 
$\Gamma(a_\lambda\rightarrow \pi^+\pi^-\pi^0)$ as long as $\lambda$ roughly 
exceeds a few GeV.

For even larger values of $\lambda$, decays to nucleons become kinematically accessible.
In the present treatment, however, any $a_\lambda$
with masses this large will not play a significant role in the phenomenology of such 
``dark-tower'' scenarios, nor will they have a significant impact on the observational and 
experimental constraints on such scenarios.  Indeed, as was observed in
Ref.~\cite{TroddenKKGravitons} for the related case of KK-graviton decays, such
modes are innocuous precisely {\it because} of the large contributions to their
decay widths from hadronic decays.  
The axion case under consideration here differs qualitatively from this KK-graviton case 
only in that the decay width $\Gamma_\lambda$ of each $a_\lambda$ contains an
additional factor $(\wtl^2 A_\lambda)^2$.  However,
as we shall see in Sect.~\ref{sec:Constraints},
the quantity $\pi \mX^2/M_c$ is never much larger than a few GeV in realistic 
axion models of dynamical dark matter.  Consequently, this factor
will be $\mathcal{O}(1)$ for modes with $\lambda$ roughly exceeding a few GeV ---
those modes for which decays to nucleons are kinematically allowed --- and the 
partial widths for those decays will therefore be unsuppressed.    
We therefore refrain from explicitly 
calculating the partial-width contribution from such decays, and simply 
acknowledge that any $a_\lambda$ with $\lambda$ roughly exceeding a few GeV will decay 
quite early --- \ie, before the big-bang-nucleosynthesis (BBN) epoch.  
  
Thus far, we have obtained partial-width expressions for all of the relevant decay 
channels for those $a_\lambda$ with $\lambda$ less than roughly a few GeV directly 
into final states involving SM fields alone.  The total width for the decays of 
$a_\lambda$ into this set of final states is obtained by 
combining the expressions in Eqs.~(\ref{eq:GammaDecayToPhotons}), 
(\ref{eq:GammaDecayToFermions}), and~(\ref{eq:Gamma3piFull}):
\begin{equation}
  \Gamma_\lambda ~=~ 
     \frac{\lambda^3}{8\pi\fhatX^2}(\wtl^2 A_\lambda)^2\Bigg[
     \frac{\alpha^2c_\gamma^2}{32\pi^2} + \sum_{i}\Theta(\lambda - 2m_{\psi_i})
     \frac{4c_i^2m_{\psi_i}^2}{\lambda^2} 
     \left(1-\frac{4m_{\psi_i}^2}{\lambda^2}\right)^{1/2}
     +\Theta(\lambda - 2m_{\pi^\pm}-m_{\pi^0})
     \frac{C_{a\pi}^2\mathcal{I}(\lambda)}{128\pi^2f_\pi^2\lambda^6}    
     \Bigg]~,
  \label{eq:GammaTotFormula} 
\end{equation}
where the Heaviside functions $\Theta(\lambda - 2m_\psi)$ and 
$\Theta(\lambda - 2m_{\pi^\pm}-m_{\pi^0})$ enforce that only 
kinematically accessible decays contribute in the sum.

\subsection{Intra-Ensemble Decays}

Up to this point, we have only been considering the decay-width contributions 
from decays of a given $a_\lambda$ directly to the fields of the SM.
We have yet to address the issue of {\it intra-ensemble} decays --- \ie, decays in 
which a given state in the dynamical dark-matter ensemble decays to a final state
containing one or more other, lighter states in that ensemble.  Indeed, as discussed
in Ref.~\cite{DynamicalDM1}, such decays can have a significant impact on the
phenomenology of a given dynamical dark-matter model:  
not only will they alter the individual relic abundances $\Omega_i$ of the particles $\phi_i$ 
in a given ensemble, but they will also alter the phase-space distributions 
$f_i(\vec{p}_i,t)$ of those particles, potentially generating a sizable population 
of $\phi_i$ with relativistic three-momenta $\vec{p}_i$.  For these reasons, it is crucial
to assess whether such decays occur at a substantial rate in the bulk-axion model presented 
here, or whether the net rate for these decays is negligible, in which case   
the quantity $\Gamma_\lambda$ given in Eq.~(\ref{eq:GammaTotFormula}) 
truly represents the total decay width of a given $a_\lambda$ with $\lambda$ less than a
few GeV.

In the model under discussion here, the dark sector properly comprises 
KK axions, KK gravitons, and graviscalars.  A complete description of the dynamics of the 
ensemble would therefore involve solving the coupled system of Boltzmann 
equations which describes the evolution of the respective phase-space distributions 
$f_\lambda(\vec{p}_\lambda,t)$, $f_n(\vec{p}_n,t)$, and $f_s(\vec{p}_s,t)$ 
for the various axion modes $a_\lambda$, KK gravitons $G_{\mu\nu}^{(n)}$, and
graviscalars $\varphi_s$, as discussed in the Appendix of Ref.~\cite{DynamicalDM1}.
In the present work, our aim will not be to solve the Boltzmann equations in 
complete generality,
but rather to demonstrate that the effects of intra-ensemble decays on the abundances and 
phase-space distributions of the $\phi_i$ are sufficiently small that they may be safely
neglected for any otherwise phenomenologically reasonable choice of model parameters.
Some of these effects --- for example, 
the depletion of $\Omega_\lambda$ for a given $a_\lambda$ due to intra-ensemble decays --- 
depend only on the net contribution $\Gamma_\lambda^{(\mathrm{IE})}$ to the width of a 
given $a_\lambda$, obtained by summing over the partial widths from all such decays.   
On the other hand, certain other effects, 
such as the increase in the abundances $\Omega_\lambda$ and alteration of the 
phase-space distributions $f_\lambda(\vec{p}_\lambda,t)$ of the lighter $a_\lambda$ due 
to the decays of the heavier $a_\lambda$, depend on these partial widths in a different manner. 
Determining the precise magnitude of these effects therefore requires a more thorough 
analysis of the Boltzmann equations.  Nevertheless, 
$\Gamma_\lambda^{(\mathrm{IE})}$ can still be useful as a rough rubric for assessing
whether or not the effects in question are likely to be significant.       
In this work, then, we simply 
demonstrate that $\Gamma_\lambda^{(\mathrm{IE})}$ for any given 
$a_\lambda$ is sufficiently small in comparison with the result in 
Eq.(\ref{eq:GammaTotFormula}) in otherwise phenomenologically reasonable regions of 
model parameter space that its effect on total decay widths may safely be neglected.
This result provides a good initial indication that the additional effects mentioned 
above are also unimportant.  A more rigorous justification for neglecting these effects
will appear in Ref.~\cite{DynamicalDM3}.

We begin our discussion of intra-ensemble decays by discussing the
partial-width contribution arising from the
decays of a given $a_\lambda$ into multiple, lighter axion modes.
The leading contribution to
this partial width comes from the quartic interaction term appearing in
Eq.~(\ref{eq:ActionInMassEigenbasis}), which gives rise to decays of the
form $a_{\lambda}\rightarrow a_{\lambda_1}a_{\lambda_2}a_{\lambda_3}$,
where $a_{\lambda_i}$, with $i=\{1,2,3\}$, are lighter axion states whose
masses satisfy the constraint $\lambda \geq \lambda_1+\lambda_2+\lambda_3$.
The Feynman rule for the corresponding four-point interaction vertex is    
\begin{eqnarray*}
\resizebox{1.65in}{!}{\includegraphics{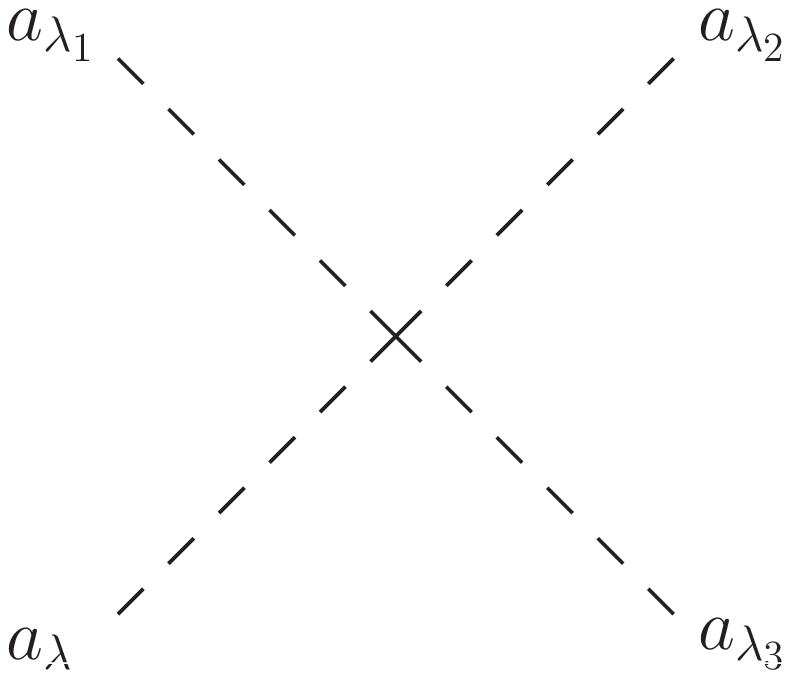}}&
   \raisebox{1.75cm}{
   $\displaystyle ~=~
   -\frac{ig_G^2\xi^4}{32\pi^2}\left(\frac{\LambdaG}{\fhatX}\right)^4
    (\wtl^2 A_\lambda) \prod_{i=1}^{3}(\wtl^2_iA_{\lambda_i}$)~,}
\end{eqnarray*}
from which the contribution to $\Gamma_\lambda$ from this three-body decay is
found to be
\begin{equation}
  \Gamma(a_\lambda \rightarrow a_{\lambda_1}a_{\lambda_2}a_{\lambda_3}) ~=~
  \frac{4\pi}{\lambda^3}\left(\frac{g_G\xi^2\LambdaG^2}{32\pi^2\fhatX^2}\right)^4
  (\wtl^2 A_\lambda)^2\left(\prod_{i=1}^{3}\wtl^2_iA_{\lambda_i}\right)^2
  \int^{(m_{12}^2)_{\mathrm{max}}}_{(m_{12}^2)_{\mathrm{min}}}
  \int^{(m_{23}^2)_{\mathrm{max}}}_{(m_{23}^2)_{\mathrm{min}}}
  dm^2_{12}\,dm^2_{23}~,
  \label{eq:Gamma3aIntermediate}
\end{equation}
where now $m_{12}^2 = (p_1 + p_2)^2$ and $m_{23}^2 = (p_2 + p_3)^2$, with 
$p_i$ representing the four-momentum of the final-state $a_{\lambda_i}$.  The 
limits of integration for the $dm_{12}^2$ integral are 
$(m_{12})^2_{\mathrm{max}} = (\lambda - \lambda_3)^2$ and 
$(m_{12})^2_{\mathrm{min}} = (\lambda_1 + \lambda_2)^2$, and 
since we are primarily interested in the parametric dependence of the partial width
on $\lambda$, $M_c$, \etc, it will be sufficient for our purposes to
construct an upper bound on 
$\Gamma(a_\lambda \rightarrow a_{\lambda_1}a_{\lambda_2}a_{\lambda_3})$
by setting 
$(m_{23})^2_{\mathrm{max}} \rightarrow (\lambda - \lambda_1)^2$ and 
$(m_{23})^2_{\mathrm{min}} \rightarrow (\lambda_2 + \lambda_3)^2$. 
Doing so, we obtain the result
\begin{eqnarray}
  \Gamma(a_\lambda \rightarrow a_{\lambda_1}a_{\lambda_2}a_{\lambda_3}) & \leq &
    \frac{4\pi}{\lambda^3}
    \left(\frac{g_G\xi^2\LambdaG^2}{32\pi^2\fhatX^2}\right)^4
    (\wtl^2 A_\lambda)^2\left(\prod_{i=1}^{3}\wtl^2_iA_{\lambda_i}\right)^2
    \nonumber\\ & & ~~~\times~
    \Big(\lambda_1 + \lambda_2 + \lambda_3 - \lambda \Big)^2
    \Big(\lambda^2-\lambda_1^2 +\lambda_2^2-\lambda_3^2 + 
       2\lambda\lambda_2 + 2\lambda_1\lambda_3\Big)~.
    \label{eq:Gammaato3aPerSpecies}
\end{eqnarray}
Note that the asymmetry of this expression under permutations of the $\lambda_i$ 
is due to the asymmetric limits of integration we have adopted in order to 
construct this bound. 
In order to obtain the total contribution $\Gamma_\lambda(a\rightarrow 3a)$ 
to the partial width of a given $a_\lambda$ from decays of this form, 
we need to sum over the contributions
from all kinematically allowed decays of the form 
$a_{\lambda}\rightarrow a_{\lambda_1}a_{\lambda_2}a_{\lambda_3}$.  Next, we 
approximate the sums over the different allowed final-state axions with 
integrals over $d\lambda_1$, $d\lambda_2$, and $d\lambda_3$.
Furthermore, 
Eq.~(\ref{eq:DefOfCapitalAlambda}) implies that $\wtl^2A_\lambda<\sqrt{2}$ for all $\lambda$.  
Therefore, in order to obtain an upper bound on $\Gamma_\lambda(a\rightarrow 3a)$, 
we make the replacements $\wtl^2_iA_{\lambda_i}\rightarrow\sqrt{2}$ 
and $\wtl^2A_{\lambda}\rightarrow \sqrt{2}$.  Doing so, we obtain our final
result: 
\begin{equation}
   \Gamma_\lambda(a\rightarrow 3a) ~\leq~ \frac{g_G^4\xi^8}{45(4\pi)^7}
     \frac{\lambda^4}{M_c^3}\left(\frac{\LambdaG}{\fhatX}\right)^8~.
   \label{eq:Gammaato3a} 
\end{equation} 

Given this result, we are now prepared to address
the question of whether axion decays to other bulk axions can ever contribute 
significantly to the total width $\Gamma_\lambda$ of any $a_\lambda$.  For 
example, the bound in Eq.~(\ref{eq:Gammaato3a}) implies that the ratio of
$\Gamma_\lambda(a\rightarrow 3a)$ to the decay rate 
$\Gamma_\lambda(a\rightarrow \gamma\gamma)$ given in Eq.~(\ref{eq:GammaDecayToPhotons})
is bounded from above by 
\begin{equation}
  \frac{\Gamma_\lambda(a\rightarrow 3a)}{\Gamma_\lambda(a\rightarrow \gamma\gamma)}
    ~\leq~ \frac{4g_G^4\xi^8}{45(4\pi)^4\alpha^2c_\gamma^2}
    \left(\frac{\lambda \LambdaG^8}{M_c^3\fhatX^6}\right)
     ~\approx~ (6.69 \times 10^{-2}) \times (g_G\xi^2)^4 
      \left(\frac{\lambda \LambdaG^8}{M_c^3\fhatX^6}\right)~.
  \label{eq:Gamma3aIntermediateStep} 
\end{equation} 
As we shall see in Sect.~\ref{sec:Constraints}, for $\mathcal{O}(1)$ values of 
$\xi$ and $g_G$, the phenomenologically preferred ranges for $\fhatX$ and 
$\Lambda_G$ turn out to be  $\fhatX \sim 10^{14} - 10^{15}$~GeV and 
$\Lambda_G \sim 10^{3} - 10^{5}$~GeV,
while $M_c$ is bounded from below by Eq.~(\ref{eq:MinimumMc}).  Within this
parameter-space regime, we find that the ratio in Eq.~(\ref{eq:Gamma3aIntermediateStep})
will be vanishingly small, as desired, unless $\lambda \gtrsim 10^{10}$~GeV.~ 
Since this is far larger than the cutoff scale $f_X$ in this
same regime, we conclude that decays of the form 
$a_\lambda\rightarrow a_{\lambda_1}a_{\lambda_2}a_{\lambda_3}$ will not play a
significant role in the phenomenology of realistic bulk axion models of dynamical 
dark matter.  Thus such decays can be safely neglected in computing the total 
decay width of a given $a_\lambda$.

Not only can the $a_\lambda$ decay to final states comprising lighter axion modes 
alone, but they can also decay into final states which include other bulk 
states.  In the minimal bulk-axion theory under discussion here, these
include KK graviscalars and KK gravitons.  (Note that the vector degrees of 
freedom $h_{\mu 5}^{(n)}$ with $n>0$ in the gravity multiplet do not couple to the 
$a_\lambda$ in the linearized-gravity limit in the unitary gauge, and the zero-mode
$h_{\mu 5}^{(0)}$ vanishes due to the orbifold projection.)  Therefore, we must
also assess whether decay channels involving these KK gravitons and KK graviscalars 
can provide an appreciable contribution
to $\Gamma_\lambda^{(\mathrm{IE})}$.  In the five-dimensional theory 
under discussion here, in the unitary gauge, the only physical
graviscalar present is a single radion mode, which we assume here to be sufficiently 
massive (\eg, as the result of some stabilization mechanism) as
not to be relevant for $a_\lambda$ decays.  As for decays involving KK gravitons in
the final state, a rough upper bound on their contribution to $\Gamma_\lambda$ 
for the case of a single, flat extra dimension will be given in 
Ref.~\cite{DynamicalDM3} within the framework of linearized gravity.  As we will
see, the leading contribution comes from two-body decays of the form 
$a_\lambda\rightarrow G_{\mu\nu}^{(n)}a_{\lambda'}$, where $G_{\mu\nu}^{(n)}$ denotes
the KK graviton with KK mode number $n$.  The total contribution from decays of this 
sort, summed over all kinematically accessible combinations of $n$ and $\lambda'$, 
is found to be approximately~\cite{DynamicalDM3}   
\begin{eqnarray}
  \Gamma(a_\lambda\rightarrow Ga) &\lesssim&  
      \frac{8\mX^4(\wtl^2A_\lambda)^2}{9\pi\lambda^3 M_c^2 M_P^2} 
      \int_0^\lambda d\lambda' (\wtl'^2A_{\lambda'})^2(\lambda+\lambda')
      \Bigg[(\lambda^2 +\lambda'^2) 
      E(x_\lambda)
      -2\lambda\lambda'
      K(x_\lambda)
      \Bigg]~,
  \label{eq:GammaaatoGaSum}
\end{eqnarray}
where $K(x_\lambda)$ and $E(x_\lambda)$ respectively 
denote the complete elliptic integrals of the first and second 
kind, with $x_\lambda \equiv (\lambda-\lambda')^2/(\lambda+\lambda')^2$.
    
In order to compare $\Gamma(a_\lambda\rightarrow Ga)$ to the
rate for $a_\lambda$ decays to SM fields, it is necessary to 
integrate Eq.~(\ref{eq:GammaaatoGaSum}) numerically, as a function of 
$\fhatX$, $\Lambda_G$, and $M_c$.  The results of such an analysis are 
given in Ref.~\cite{DynamicalDM3}.  Here, however, 
to illustrate our point, we simply 
choose a set of benchmark values typical of a
phenomenologically consistent scenario for which  
$\Gamma(a_\lambda\rightarrow Ga)/\Gamma(a_\lambda\rightarrow \gamma\gamma)$ 
is roughly maximal.  Specifically, we take
$\Lambda_G=1$~TeV and $M_c=10^{-11}$~GeV, with $g_G = \xi = 1$.
We then find that $\Gamma(a_\lambda\rightarrow Ga)$ always remains
several orders of magnitude smaller than 
$\Gamma(a_\lambda\rightarrow \gamma\gamma)$ for all values of 
$\fhatX \lesssim 10^{15}$~GeV.~ 
We therefore conclude that the decays of $a_\lambda$ to KK gravitons
will not have a significant impact on the total widths of the $a_\lambda$. 
 
Taken together, these results strongly suggest that intra-ensemble decays do not 
play a significant role in the phenomenology of bulk-axion dynamical dark matter, 
and can therefore be neglected.  Therefore,
from this point forward we will ignore intra-ensemble decays and 
identify $\Gamma_\lambda$, as given in Eq.~(\ref{eq:GammaTotFormula}), with the total
width of any given $a_\lambda$.    

\begin{figure}[ht!]
\centerline{
  \epsfxsize 3.25 truein \epsfbox {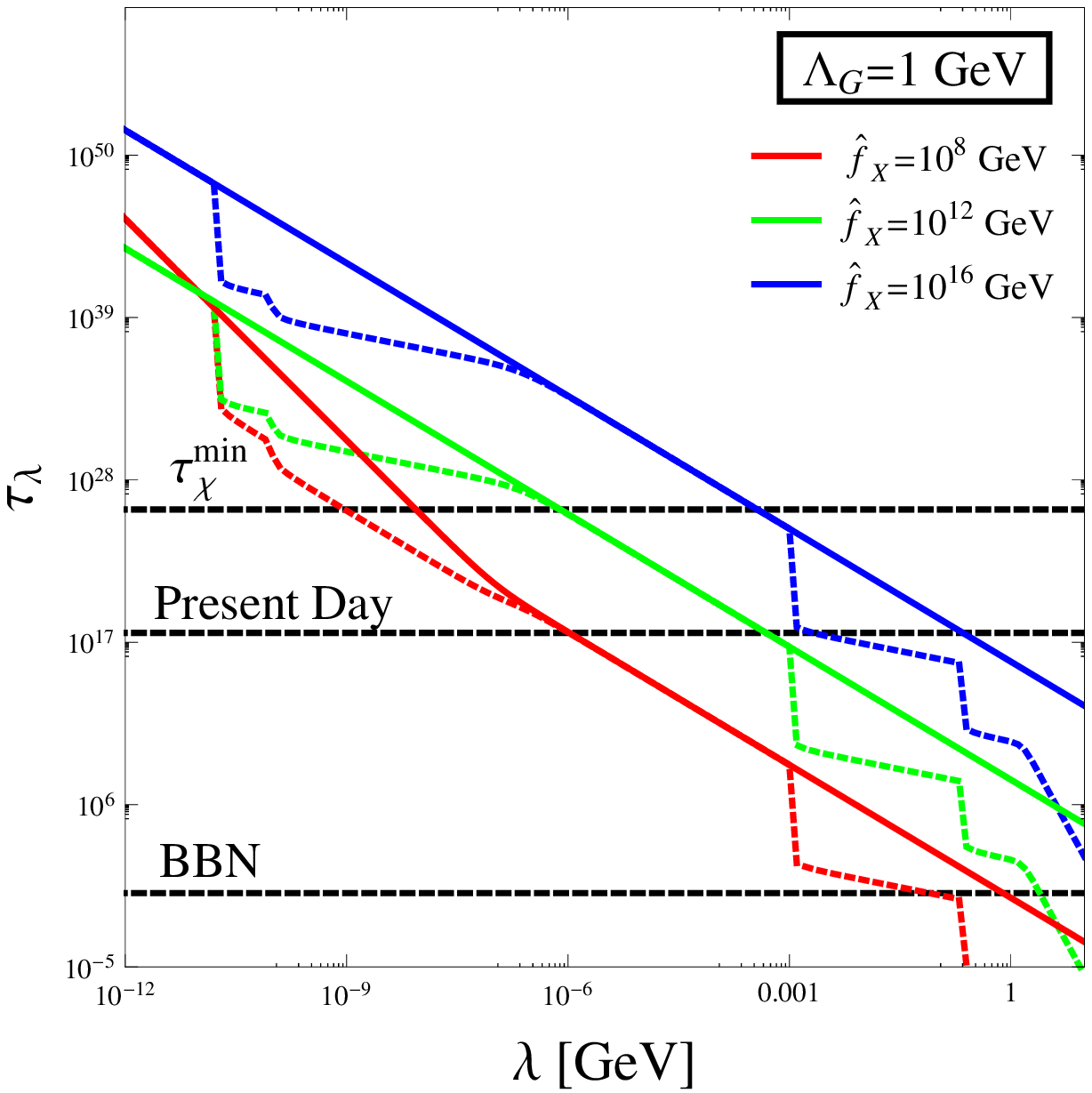}~~~~
  \epsfxsize 3.25 truein \epsfbox {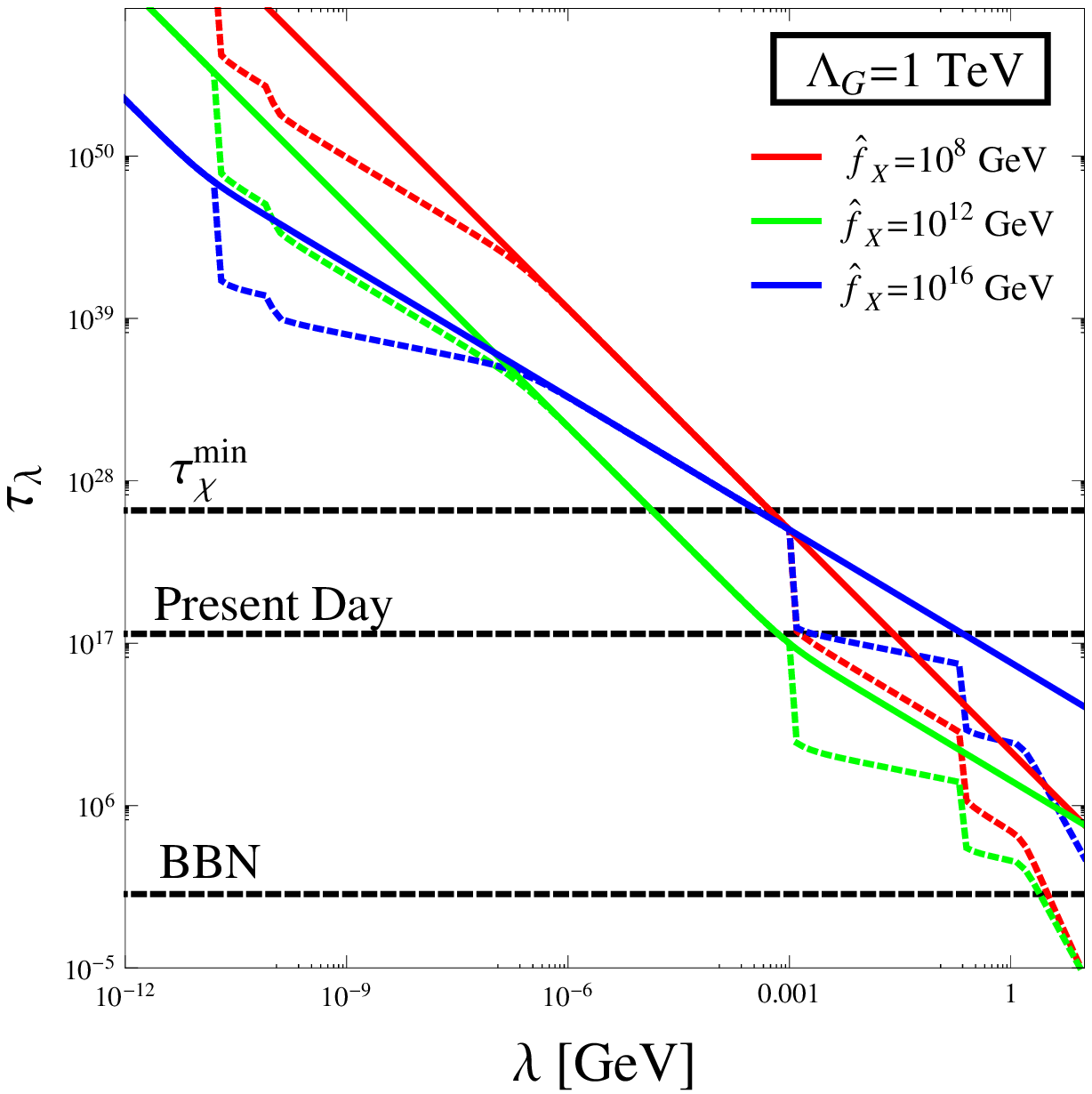} }
\caption{The lifetime $\tau_\lambda$ of the axion mass 
eigenstate $a_\lambda$, shown as a function of its mass
$\lambda$ for $\Lambda_G = 1$~GeV (left panel) 
and an axion with $\Lambda_G = 1$~TeV (right panel).  In both panels,
we have set $\xi= g_G = 1$ and have chosen $M_c = 10^{-11}$~GeV --- a value 
just above the lower bound imposed by modified-gravity experiments.    
The three solid curves in each panel shown correspond to different choices 
of $\fhatX$ for a ``photonic'' axion with the coupling-coefficient assignments 
$c_\gamma = 1$, $c_g = 0$, and $c_i = 0$ for all fermions 
$i = \{e,\nu_e,\nu_\mu,\nu_\tau\}$.  The solid red curve corresponds 
to $\fhatX = 10^{8}$~GeV, the solid green curve 
corresponds to $\fhatX = 10^{12}$~GeV, and the solid blue curve corresponds to 
$\fhatX = 10^{16}$~GeV.~  The dashed curves correspond to the same choices of
$\fhatX$ for an axion with the coupling assignments 
$c_\gamma = c_g= 1$ and $c_i = 1$ for all $i$.  
The kinks in the curves reflect the opening up
of new decay channels as $\lambda$ is increased past a series of kinematic 
thresholds associated with decays to neutrino pairs,  
electron pairs, muon pairs, and $\pi^+\pi^-\pi^0$.  The horizontal lines
indicate the time scales associated with the onset of BBN ($\tBBN\sim 1$~s),  
the present age of the universe ($\tnow\sim 4.3\times 10^{17}$~s), 
and the usual lower limit given in Eq.~(\protect\ref{eq:DecayingDMLifetimeLimit}) 
on the lifetime $\tau_\chi$ of a single-particle dark-matter candidate.
\label{fig:DecayVsLambdaPlot}}
\end{figure}

\subsection{Axion Lifetimes Across the Ensemble}

In Fig.~\ref{fig:DecayVsLambdaPlot}, we show how the 
lifetime $\tau_\lambda \equiv 1/\Gamma_\lambda$ 
of an axion mass eigenstate $a_\lambda$ behaves as a function of $\lambda$.  
The left panel shows the
results for an axion with $\Lambda_G = 1$~GeV, while the right panel shows the results for 
an axion with $\Lambda_G = 1$~TeV.~  In each case, we have taken $g_G = \xi = 1 $ and
set $M_c=10^{-11}$~GeV.~  In each of the two panels, the three solid curves correspond to 
three different choices of $\fhatX$ for a photonic axion with $c_\gamma = 1$.
The solid red curve corresponds to $\fhatX = 10^{8}$~GeV, the solid green curve 
corresponds to $\fhatX = 10^{12}$~GeV, and the solid blue curve corresponds to 
$\fhatX = 10^{16}$~GeV.~   The dashed curves correspond to the same choices of
$\fhatX$ for an axion with $c_\gamma = c_g = 1$ and $c_i = 1$ 
for $i = \{e,\nu_e,\nu_\mu,\nu_\tau\}$.
The series of kinks which are evident in each dashed curve correspond to the thresholds at
$m_{\nu_2}$, $m_{\nu_3}$, $m_e$, and $m_\mu$ above which new decay channels for 
$a_\lambda$ open up.  The sharp drop in $\tau_\lambda$ at around $\lambda \sim 400$~MeV
is the result of the $a\rightarrow \pi^+\pi^-\pi^0$ decay channel opening up.

One significant property of the decay rates of the $a_\lambda$ in bulk-axion 
scenarios can be readily appreciated upon comparing the curves appearing
in the two panels of Fig.~\ref{fig:DecayVsLambdaPlot}: the total width $\Gamma_\lambda$ 
is independent of $\mX$ (and therefore independent of $\Lambda_G$) in the limit in 
which $\lambda \gg \pi \mX^2/M_c$.  This can also be seen from 
Eq.~(\ref{eq:GammaTotFormula}).  It therefore follows that  
the corresponding curves appearing in the two panels of this figure should 
coincide for values of $\lambda$ above the threshold at which 
this condition is met for both of the selected values of $\Lambda_G$.  Indeed,
we see that this is in fact the case.  The $\fhatX = 10^{16}$~GeV curves
coincide for nearly the entirety of the range of $\lambda$ shown, since 
$\pi\mX^2/M_c$ is approximately $9.95\times 10^{-12}$~GeV for 
$\Lambda_G=1$~TeV and is far smaller for $\Lambda_G =1$~GeV.  The 
$\fhatX = 10^{12}$~GeV curves, on the other hand, begin to coincide only for 
$\lambda \sim 10^{-3}$~GeV, which is just above the threshold 
$\pi\mX^2/M_c \approx 9.95\times 10^{-4}$~GeV for $\Lambda_G = 1$~TeV.~

There is, however, an even more important message to be gleaned from 
Fig.~\ref{fig:DecayVsLambdaPlot}.
Note that in each panel, we have included for reference a set of
horizontal, dashed lines indicating the time scales associated with the beginning 
of BBN ($\tBBN\sim 1$~s), the present age of the universe  
($\tnow\sim 4.3\times 10^{17}$~s), and the usual limit on the lifetime $\tau_\chi$ of
a single decaying dark-matter candidate $\chi$ given in 
Eq.~(\ref{eq:DecayingDMLifetimeLimit}).  {\it These benchmark times are absolutely 
critical for the survival of our dynamical dark-matter model.}  Any $a_\lambda$ with
a lifetime that falls between $\tBBN$ and $\tau_\chi$ has the potential to disrupt 
BBN predictions for the abundances of light elements, distort the CMB 
to an unacceptable degree~\cite{ADDPhenoBounds,CosmoConstraintsLargeED}, 
produce too large a flux of X-ray or gamma-ray photons, \etc~  
For this reason, the success of our dynamical dark-matter model rests upon the
assumption that such $a_\lambda$ have sufficiently small relic abundances 
$\Omega_\lambda$ that these decays are harmless.  {\it It is in this manner that lifetimes
must be balanced against abundances across our dark-axion towers.}

\begin{figure}[ht!]
\begin{center}
  \epsfxsize 4.5 truein \epsfbox {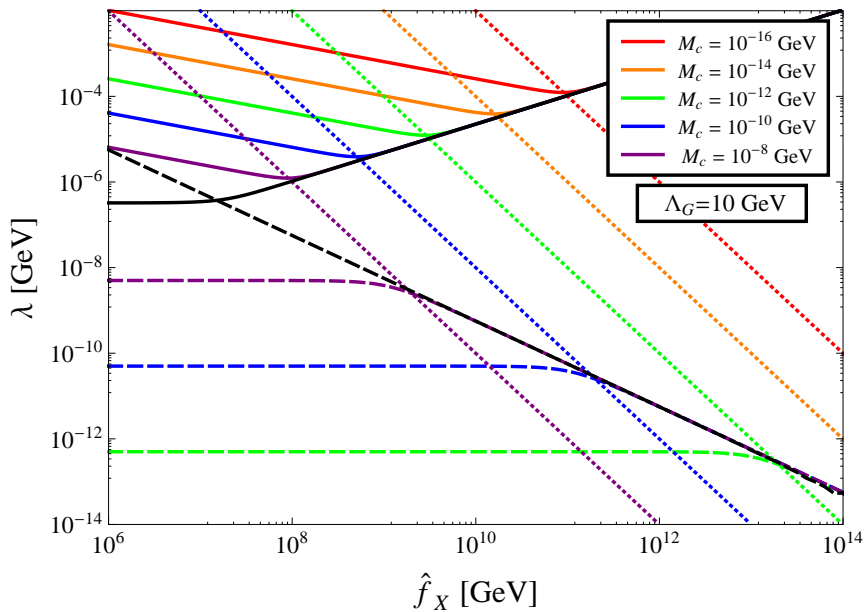} 
\end{center}
\caption{Curves showing a variety of critical values of $\lambda$ as
  functions of $\fhatX$ for $\Lambda_G = 10$~GeV and several different
  choices of $M_c$ ranging from $10^{-16}$~GeV to $10^{-8}$~GeV.  Each dashed 
  curve indicates the mass $\lambda_{0}$ of the 
  of the lightest axion mass eigenstate $a_{\lambda_0}$ for a given choice of 
  $M_c$.  Each solid curve
  indicates the mass $\lambda_{\mathrm{dec}}$ of the heaviest $a_\lambda$ 
  for that choice of $M_c$ which has not decayed by present time, assuming
  a photonic axion with $c_\gamma = 1$.  Each 
  dotted curve marks the transition point $\lambdatrans$ between the 
  small-$\lambda$ and large-$\lambda$ regimes for the same choice of $M_c$, 
  as defined in in Eq.~(\protect\ref{eq:LambdaRegimesTransPt}).  The black curves
  indicate the asymptotic behavior of $\lambda_{0}$ and $\lambdadec$ for 
  large $M_c$.   
\label{fig:LambdaMaxAndLambdaMin}}
\end{figure} 

The results in Fig.~\ref{fig:DecayVsLambdaPlot} also highlight another 
important aspect of our dynamical dark-matter ensemble, which is that 
at a given time $t$, only a fraction of the $a_\lambda$ --- those which
have not already decayed --- can contribute significantly to $\OmegaDM$.
Therefore, since $\Gamma_\lambda$ increases monotonically as a function of 
$\lambda$, there exists a maximum value $\lambdadec$ for which $a_\lambda$ 
may be considered stable for particular time scale $t$ (in the sense that 
$\Gamma_\lambda t < 1$) and which is potentially capable of contributing 
significantly to $\OmegaDM$. 
For a photonic axion, for example, the approximate form of $\lambdadec$ can 
readily be obtained in both the $y \ll 1$ and $y \gg$ regimes by inverting
Eq.~(\ref{eq:GammaDecayToPhotons}):
\begin{equation}
  \lambdadec ~\approx~
    \begin{cases} \vspace{0.15cm}
      \displaystyle\left(\frac{g_G^4\xi^4}
      {2(32\pi)^2G_\gamma\tnow}\right)^{1/5}
      \left(\frac{\Lambda_G^{8/5}}{\fhatX^{2/5}M_c^{2/5}}\right)~~~~
       & y\ll 1 \\
      \displaystyle\left(\frac{1}{2G_\gamma\tnow}\right)^{1/3}\fhatX^{2/3}
       & y \gg 1~.
    \end{cases}
   \label{eq:LambdaDecRegimes}
\end{equation}  
 
In Fig.~\ref{fig:LambdaMaxAndLambdaMin}, we illustrate the behavior of  
$\lambdadec$ (solid curves) for a photonic axion a function 
of $\fhatX$ for several choices of $M_c$.  
In each case, we have fixed $\Lambda_G = 10$~GeV and set 
$\xi = g_G = c_\gamma = 1$.  For reference, we have also included the 
corresponding curves for two other critical values of $\lambda$ in
any given axion KK tower for each $M_c$.  These are the mass 
$\lambda_{0}$ (dashed curves) of the lightest axion mass 
eigenstate $a_{\lambda_0}$ and the mass $\lambdatrans$ (dotted curves) 
defined in Eq.~(\ref{eq:LambdaRegimesTransPt}) which delineates the 
transition point between the small-$\lambda$ and large-$\lambda$ regimes.
We have also included a pair of black curves indicating the asymptotic 
behavior of $\lambda_{0}$ and $\lambdadec$ for large $M_c$.  Note that
in this limit, mixing is negligible, and $\lambda_{0}\approx\mX$ for all
values of $\fhatX$.       

The behavior of $\lambda_0$, $\lambdadec$, and $\lambdatrans$ depends
primarily on the value of the mixing parameter $y$. 
When $\fhatX$ is sufficiently large that $y\gg 1$ for a given value of
$M_c$ (\ie, in the lower right portion of Fig.~\ref{fig:LambdaMaxAndLambdaMin}), 
we find that $\lambda_0 \approx \mX$, as expected.  At the same time, since
mixing is negligible in this regime, increasing $\fhatX$ results in a
uniform suppression of the decay widths of all $a_\lambda$, and hence    
$\lambdadec$ increases with increasing $\fhatX$ (\ie, in the upper right portion
of the figure).  However, as $\fhatX$ 
decreases past the point at which the $\lambda_0$ and $\lambdatrans$ curves
intersect, we pass from the $y\gg 1$ to the $y \ll 1$ regime.  In this latter
regime, mixing is significant and $\lambda_0 \approx M_c/2$.  At first, $\lambdadec$ 
still decreases with decreasing $\fhatX$ in this regime, as those modes with 
$\lambda \lesssim \pi \mX^2/M_c$ continue to destabilize down the tower.
However, as $\fhatX$ decreases still further, to the point 
at which the $\lambdadec$ and $\lambdatrans$ curves intersect,
the $a_\lambda$ in the $\lambda \lesssim \pi \mX^2/M_c$ regime begin to 
destabilize as well.  The dependence of $\Gamma_\lambda$ on $\fhatX$
is qualitatively different for such modes, as indicated in 
Eq.~(\ref{eq:LambdaDecRegimes}), and consequently $\lambdadec$ 
actually begins to {\it increase} with decreasing $\fhatX$.  These 
observations will turn out to be critical in interpreting the 
results to be derived in Sect.~\ref{sec:Abundances}.


\section{Characterizing the Constituents:~ Relic Abundances\label{sec:Production}}


In the previous section, we focused on one of the two crucial properties of the 
particles that constitute a dynamical dark-matter ensemble: their stability.  
In particular, 
we examined the decay widths of the various $a_\lambda$ and investigated how these 
widths scale with $\lambda$.  In this section, we focus on the other property: their
abundances.  We begin by establishing a consistent cosmological context in which to
situate our bulk-axion theory.  We then proceed to 
evaluate the various mechanisms through which a population of $a_\lambda$ can be produced
in the early universe, and explain why 
misalignment production is favored from the perspective of dynamical dark matter.  This
thereby justifies the emphasis placed on this mechanism in Ref.~\cite{DynamicalDM1}.  We 
then derive explicit formulae for $\Omega_\lambda$, and demonstrate that the proper
balance between $\Gamma_\lambda$ and $\Omega_\lambda$ is indeed realized in the
context of misalignment production.  In the following section, we will then use
these results for $\Gamma_\lambda$ and $\Omega_\lambda$ to characterize the 
aggregate properties of the entire ensemble, such as its equation of state, its 
total relic abundance, and the way in which that abundance is partitioned among
its constituents. 
 
\subsection{Standard and Low-Temperature Reheating (LTR) Cosmologies}

Before embarking on a discussion of axion production in the early universe, 
however, we must first specify the cosmology in which that production occurs.
This is particularly relevant in the context of theories involving large extra 
dimensions, since the properties of the early universe in such theories can 
differ dramatically from those which characterize the standard cosmology.  For 
example, the presence of substantial energy density in the bulk can alter the 
expansion rate of the universe in a significant way~\cite{BinetruyHubbleInBraneworld}, 
and late decays of KK excitations of the graviton (or other bulk fields) can 
disrupt BBN, produce visible distortions in the diffuse photon spectrum, \etc~
For this reason, such scenarios must obey stringent 
constraints~\cite{ADDPhenoBounds} on the so-called ``normalcy temperature'' 
$T_\ast$: the temperature below which the universe is effectively four-dimensional, 
in the sense that the bulk is essentially empty of energy density and the radii of 
the extra dimensions can be regarded as fixed.  These bounds come from a diverse 
array of considerations and leave a very 
narrow window of $4\mathrm{~MeV}\lesssim T_\ast \lesssim 30$~MeV for this normalcy
temperature.  The most attractive solution for arranging such a value of $T_\ast$ is
to posit a very late period of cosmic inflation, precipitated by a brane-localized
inflaton~\cite{ADDPhenoBounds}, with a reheating temperature $\TRH \simeq T_\ast$.  

If $\TRH$ is indeed to be identified with $T_\ast$, then the 
universe must be described not by the standard cosmology, but by an 
alternative framework commonly dubbed the low-temperature reheating (LTR) 
cosmology~\cite{LTRCosmo}.  In the LTR framework, inflation occurs very
late, and the energy density of the universe remains dominated by coherent 
oscillations of the inflaton field $\phi$.  Such oscillations behave like massive matter down 
to very low temperatures --- potentially as low as a few MeV.~  Consequently,
the universe undergoes an additional epoch of matter domination (MD) at early times,
which ends only once decays of $\phi$ into SM fields in the radiation bath 
have depleted the energy density $\rho_\phi$ to the point that
$\rho_\phi = \rho_{\mathrm{rad}}$.  This condition defines the reheating
temperature $\TRH$, which is determined solely by the decay width 
$\Gamma_\phi$ of the inflaton:    
\begin{equation}
  \TRH ~\equiv~ \left[\frac{90}{\pi^2g_\ast(\TRH)}\right]^{1/4} \sqrt{\Gamma_\phi M_P}~,
  \label{eq:DefOfTRH}
\end{equation}
where $g_\ast(T)$ denotes the effective number of massless, interacting degrees of
freedom at temperature $T$.  At temperatures $T\lesssim \TRH$ the universe is 
effectively radiation dominated (RD), and maps onto the standard cosmology.  
This modification implies that the relationship between time and temperature in
the LTR cosmology is quite different from that obtained in the standard cosmology
at early times.  In particular, one finds that in any universe which underwent
a period of cosmic inflation immediately prior to the RD era, the relationship
between time and temperature is given by   
\begin{equation}
  t ~=~ 
  \begin{cases}
     \vspace{0.25cm}\displaystyle
     \sqrt{\frac{45}{2\pi^2}}
     \frac{g^{1/2}_{\ast}(\TRH)}{g_{\ast}(T)}\frac{\TRH^2M_P}{T^4}
       & ~~~ T_{\mathrm{max}} ~\gtrsim~T~ \gtrsim~ \TRH \\
    \vspace{0.25cm}\displaystyle
    \sqrt{\frac{45}{2\pi^2}}
       g^{-1/2}_{\ast}(T)\frac{M_P}{T^2}
       & ~~~\TRH ~\gtrsim ~T ~\gtrsim~ T_{\mathrm{MRE}} \\
    \displaystyle
    \sqrt{\frac{45}{2\pi^2}}
       g^{-1/8}_{\ast}(\TMRE)g_{\ast}^{-3/8}(T)
       \frac{M_P}{T_{\mathrm{MRE}}^{1/2}T^{3/2}}
       & ~~~ T ~\lesssim~ T_{\mathrm{MRE}}~,
  \end{cases}
  \label{eq:tTempRelLTR}
\end{equation}
where $\TMRE \sim \mathcal{O}(\mathrm{eV})$ is the temperature associated with
the usual matter/radiation transition, at which the energy density of the  
universe once again becomes dominated by the contributions from dark and 
baryonic matter. (For our purposes, it will be sufficient to approximate 
the present, $\Lambda$CDM universe as matter-dominated.)  
In the standard cosmology, $\TRH$ is high enough that
the universe will be radiation-dominated at all time scales relevant to   
axion dynamics, all the way down to the time scale $\tMRE$ 
associated with this transition.  By contrast, in a LTR cosmology with 
$\TRH\sim \mathcal{O}(\mathrm{MeV})$, much of the relevant dynamics will
occur while the universe is still dominated by coherent oscillations of the
inflaton field.  Such a cosmological modification can have profound effects on 
axion dynamics, even in the case of a four-dimensional 
axion~\cite{LTRAxionsKamionkowski,LTRAxions}. 

Since our model necessarily involves
large extra dimensions, constraints on $T_\ast$ would seem to require that
the universe be described by the LTR cosmology, rather than the standard 
cosmology, at early times.  Indeed, for this reason, we shall adopt such 
an LTR cosmology in what follows.  This will actually turn out to be advantageous, 
as the relic abundance of a light scalar generated via misalignment production
can differ substantially between the two cosmologies~\cite{LTRAxions}.     
To better facilitate comparison between the results obtained in these two 
cosmologies, we therefore find it instructive to present results for each 
in parallel.  The reader should keep in mind, however, that the results obtained 
for an LTR cosmology with $4\mathrm{~MeV}\lesssim T_\ast \lesssim 30$~MeV 
should be taken, in some sense, as the ``true'' ones, given the constraints 
on $T_\ast$.

In this paper, we shall be interested in values of the confinement 
scale $\Lambda_G$ which range between roughly $10$~MeV and $100$~TeV.
When operating in the standard cosmology, we shall assume that
$ \Lambda_G \ll \TRH$, so that confinement takes place {\it within the RD era}.  
Conversely, when operating within the LTR cosmology, we will assume that
$\Lambda_G > \TRH$, so that confinement takes place {\it before or during reheating}, 
when the universe is dominated by coherent oscillations of the inflaton
field.  Note, however, that for smaller confinement scales $\Lambda_G < \TRH$,
the results for the LTR cosmology are identical to those for the standard cosmology.   

\subsection{Axion Production Mechanisms}

Having now set the cosmological tableaux, let us begin our discussion of axion 
production in the early universe.  A number of production mechanisms can, in 
principle, contribute significantly to the axion relic density.  In generic models,
three such mechanisms typically tend to provide the dominant contribution to the
relic density of any individual axion field.   
One of these mechanisms is thermal production via the interactions of the $a_\lambda$ 
with the SM fields in the radiation bath.  The other two are 
non-thermal in nature and can generate a population of cold axions. 
These are misalignment production and production from the decays of topological 
defects (in particular, cosmic strings) associated with the breaking of the 
global $U(1)_X$ symmetry. 

We begin with a discussion of thermal production.
A number of processes can contribute appreciably to thermal axion production in 
the early universe, depending of course on the magnitudes of the couplings 
between the axion in question couples to the SM particles.  
Among hadronic processes, $q g \rightarrow q a_\lambda$, 
$q \bar{q} \rightarrow g a_\lambda$, $g g \rightarrow g a_\lambda$, \etc,
dominate for $T \gtrsim \LambdaQCD$, while pion-axion conversion off nuclei 
(including all processes of the form $N \pi\rightarrow N' a_\lambda$, where 
$N,N' = \{n,p\}$) and the purely pionic process $\pi \pi \rightarrow \pi a_\lambda$ 
dominate at lower temperatures.
Since we are interested in values of $M_c$ which are far below $\TRH$, 
it follows that $T\gg \lambda$ for a large number of the $a_\lambda$ for at least
some of the post-inflationary era.  These $a_\lambda$ can therefore be considered 
effectively massless at such temperatures.  In this massless limit, the rate for  
each of the axion-production processes enumerated above 
(except for the inverse-decay process, which is generally sub-leading) takes
the rough, parametric form
\begin{equation}
  \Gamma ~\propto~ \frac{T^3}{\fhatX^2}(\wtl^2 A_\lambda)^2~.
  \label{eq:ThermalRateProportionality}
\end{equation}
In other words, since kinematic distinctions between states for which $T\gg \lambda$ 
are unimportant, the dependence of their production rates on $\lambda$ occurs 
primarily through the coupling factor $\wtl^4 A_\lambda^2$.  This factor too is   
effectively independent of $\lambda$ when $\lambda\gtrsim\pi m_X^2/M_c$; hence, at 
a given temperature $T$, any $a_\lambda$ with a mass $\lambda$ in the range
$\pi m_X^2/M_c\lesssim \lambda \lesssim T$ will be produced at essentially the
same rate.  Moreover, production rates actually {\it increase} with increasing 
$\lambda$ for those modes with masses $\lambda \lesssim \pi m_X^2/M_c$.  
This means that at a given temperature $T$, the heavier $a_\lambda$ in this
mass range will actually be produced from the thermal bath at an equal or 
higher rate than the lighter $a_\lambda$ --- at least until $\lambda$ becomes 
comparable with $T$, and their production rates are Boltzmann suppressed.  
Since $\Gamma_\lambda$ also increases with $\lambda$, the 
less stable states will be thermally produced at an equal or higher rate 
than the more stable ones.  Such a relationship is  
clearly undesirable in models of dynamical dark matter. 
Moreover, the majority of light axions produced through interactions with the 
thermal bath would be relativistic at the time of production, and therefore
not cold.  

From these considerations, we conclude that if an ensemble of axion KK modes is to 
constitute the majority of the dark-matter relic density, thermal production must 
contribute only a negligible fraction of the total relic abundance of each 
$a_\lambda$, with the remainder of that abundance generated through
non-thermal means.  This requirement places a non-trivial constraint on scenarios of 
this sort, a detailed analysis of which appears in Ref.~\cite{DynamicalDM3}.  We 
will defer the discussion of how this constraint restricts the parameter
space of our model until Sect.~\ref{sec:Constraints}.  For the moment, we simply
note that this constraint exists, and proceed to discuss non-thermal 
mechanisms for axion production.  Note, however, that in traditional 
models of KK dark-matter (either single-component~\cite{KKParity} or 
multi-component~\cite{Winslow}), in which the dark-matter candidates are stable, 
thermal freeze-out can be a viable production method for generating relic abundances.    

One method in which a non-thermal population of axions may be generated in the
early universe is production via the decay of cosmic strings associated with 
the broken global $U(1)_X$ symmetry.  However, this mechanism can 
contribute significantly to axion production only if $H_I \gtrsim f_X$, 
where $H_I$ is the value of the Hubble parameter
during inflation, so that those cosmic strings are not inflated away.
Since the value of $H_I$ is relatively unconstrained, and since
astrophysical and cosmological constraints will turn out to require 
$\fhatX$ to be quite large, in the rest of this paper, we shall assume that 
$H_I \ll f_X$.  We will therefore not consider axion production from $U(1)_X$ 
string decay.  However, it should be noted that in other scenarios (or in other 
regions of parameter space), axions produced via cosmic-string decay could have 
important phenomenological consequences in dynamical dark-matter models, 
and this production mechanism therefore deserves further study.
 
Finally, we turn to non-thermal axion production via the misalignment mechanism.
As we shall see, this turns out to be the most promising axion-production 
mechanism from the perspective of dynamical dark matter.  The basis of this mechanism 
is that at temperatures $T\gg \Lambda_G$, the instanton-induced 
contribution to the axion potential in Eq.~(\ref{eq:InstantonPotential}) 
effectively vanishes.  This implies that the 
only contribution to the axion mass matrix at such high temperatures are the diagonal
contributions from the KK masses: no mixing occurs, and consequently the mass 
eigenstates are merely the KK eigenstates $a_n$.  While the potential for each 
$a_n$ with $n \neq 0$ is therefore non-vanishing, due to the presence of the
KK masses, and is minimized at $a_n = 0$, 
the potential for the zero mode 
$a_0$ vanishes.  In the absence of a potential for $a_0$, there is no preferred 
vacuum expectation value $\langle a_0\rangle$ which minimizes $V(a_0)$; indeed, any
$\langle a_0 \rangle \lesssim \fhatX$ is as good as any other.  
This means that when the $U(1)_X$ symmetry is broken, 
the value of $\langle a_0\rangle$ within a given 
domain is essentially arbitrary.  Thus, immediately after this
phase transition occurs, one would expect to find 
a set of domains, each with a different homogeneous background 
value for the axion field, which would generically be expected
to be $\mathcal{O}(\fhatX)$, but could in principle be smaller.  Our 
ignorance of this initial value of $\langle a_0\rangle$ is commonly
parameterized by a ``misalignment angle'' $\theta$, so that the 
initial conditions at the time at which the $U(1)_X$ symmetry is broken
can be written as~\cite{DDGAxions}
\begin{equation}
 \langle a_0\rangle = \theta\fhatX~, ~~~~~~~~~~~~~
 \langle a_n\rangle = 0 \mathrm{~~~~for~~~} n\neq 0~. 
 \label{eq:a0initcondits}
\end{equation}
Indeed, as was noted in Ref.~\cite{DynamicalDM1}, this initial condition 
follows from $U(1)_X$ invariance, which manifests itself here in the 
form of a five-dimensional shift symmetry under
which $a\rightarrow a+c$, where $c$ is a constant. 

At lower temperatures, however, the situation changes,
as instanton effects generate a brane mass $\mX(T)$ for the bulk axion.  
Here, we write $\mX(T)$ rather than $\mX$ in order to emphasize that this 
instanton-induced mass term is temperature dependent, and reserve the symbol
$\mX$ (without the argument) to refer to the constant, late-time 
(\ie, low-temperature) value of $\mX(T)$.  
Assuming that the instantons associated with the group $G$ 
behave analogously to QCD instantons,  
Eq.~(\ref{eq:InstantonMassThermal}) implies that $\mX(T)$ scales roughly 
like $(\Lambda_G/T)^4$.  Thus, when $T\sim \Lambda_G$, the off-diagonal   
terms in the mass matrix $\mathcal{M}^2_{mn}(T)$ become appreciable, and 
the $a_n$ are no longer mass eigenstates.
In this regime, the equations for $a_n$ form a coupled system~\cite{DDGAxions},
with the evolution of each such field governed by an equation of the form   
\begin{equation}
   \ddot{a}_n + 3H\dot{a}_n + 
     \sum_\lambda\Gamma_\lambda(T)U_{\lambda n}(T)\dot{a}_n +
     \sum_{m=0}^\infty\mathcal{M}^2_{nm}(T) a_m  ~=~ 0~,
   \label{eq:TheDoubleDotEqnMassEigenbasis}
\end{equation}
where $U_{\lambda n}(T)$ denotes the unitary matrix in Eq.~(\ref{eq:DefOfalambda}), 
a dot denotes a derivative with respect to the time $t$, and $H$ is the Hubble parameter.
Note that since the mass eigenvalues 
$\lambda(T)$, decay widths $\Gamma_\lambda(T)$, and even the rotation 
matrix $U_{\lambda n}(T)$ itself all depend on $\mX$, these quantities all
implicitly depend on temperature, and hence on $t$.

During any period in which $\Gamma_\lambda(T)$ and $\lambda(T)$ vary appreciably 
in time, it is in general not possible to write down an exact, closed-form solution
to the coupled system in Eq.~(\ref{eq:TheDoubleDotEqnMassEigenbasis}).
Fortunately, however, $\mX(T)$ can be regarded as effectively constant during most
of the history of the universe.  The only exception occurs at temperatures around 
$T\sim \Lambda_G$, where $\mX(T)$ rapidly rises from a negligible initial value
to the asymptotic value it attains at $T\ll \Lambda_G$.  At all other times,
$\mX(T)$ is well approximated either by zero or by $\mX$, and the 
system of equations therefore decouples in the mass-eigenstate basis.  In this basis, the 
time-evolution of each field $a_\lambda$ is governed by an equation of the form 
\begin{equation}
   \ddot{a}_\lambda + \frac{\kappa}{t}\dot{a}_\lambda +
     \Gamma_\lambda\dot{a}_\lambda + \lambda^2 a_\lambda  ~=~ 0~.
   \label{eq:TheDoubleDotEqnWithGammaLambda}
\end{equation} 
In arriving at this expression, we have used the fact that within an RD or MD era, 
$H$ is approximately given by the relation $3H \approx \kappa/t$, where
\begin{equation}
  \kappa ~\equiv~ 
     \begin{cases}
     3/2 &\mathrm{in~RD}\\
     2 & \mathrm{in~MD}~. 
     \end{cases}
  \label{eq:DefOfnForH}
\end{equation} 
Exact, closed-form solutions for $a_\lambda$ and $\dot{a}_\lambda$, given an 
evolution equation of this form, {\it do} exist, and we present 
these solutions in the Appendix.
It may be observed, however, that Eq.~(\ref{eq:TheDoubleDotEqnWithGammaLambda}) is simply the
equation of motion for a damped harmonic oscillator with a time-dependent damping
term.  As we shall see at the end of this section, 
it turns out that $\Gamma_\lambda \ll 3H$ at the time 
when $\lambda \approx 3H/2$ for all $a_\lambda$.  It then follows that the solutions 
for each $a_\lambda$ can be divided into two regimes,
depending on the relationship between $\lambda$ and $H$ at any given time $t$.  When
$\lambda\lesssim 3H/2$, $a_\lambda$ does not oscillate, and therefore its energy density 
scales approximately like vacuum energy.  By contrast, when $\lambda\gtrsim 3H/2$, 
$a_\lambda$ oscillates coherently around the minimum of its potential, with oscillations 
damped by a ``friction'' term with coefficient $(3H + \Gamma_\lambda)$.

Despite the fact that $\mX(T)$ is effectively constant both well before and well after
the time scale $t_G$ at which $T = \Lambda_G$, the non-trivial dynamics of the $a_\lambda$
at $t\sim t_G$ certainly can play a crucial role in establishing the initial relic 
abundances for these fields.  Nevertheless, while such a time-dependence 
can indeed have a significant
quantitative impact on the results of relic-density calculations in certain cases,
as has been shown to be the case with a standard, four-dimensional axion~\cite{Turner}, 
a great deal of information can be obtained by working in the ``rapid-turn-on'' 
approximation, in which we approximate
\begin{equation}
  \mX(t) ~=~ \mX \Theta(t - \tG)~.
  \label{eq:Heaviside}
\end{equation}
In this approximation, the procedure for calculating the background value
of each $a_\lambda$ at any time $t$ is clear~\cite{DynamicalDM1}.  
At early times, when $t < t_G$, $\mX(T) = 0$, and hence the $a_n$ remain 
fixed at the initial values given in Eq.~(\ref{eq:a0initcondits}).  At $t=t_G$, 
$\mX(T)$ immediately assumes its constant, non-zero, late-time value $\mX$, and each of 
the mass eigenstates $a_\lambda$ acquires a background value proportional to 
its overlap with $a_0$: 
\begin{equation}
  \langle a_\lambda\rangle = \theta\fhatX A_\lambda~,~~~
  \langle \dot{a}_\lambda \rangle = 0~~~~~~~ ~~{\rm at}~~t=t_G~.
  \label{eq:alambdaInitCondits}
\end{equation}
Even though all of the $a_\lambda$ acquire background values at
$t=t_G$, only those fields for which $\lambda \gtrsim 3H(t_G)/2$
begin oscillating immediately at the time of this phase transition.  As 
discussed above, all other, lighter $a_\lambda$ will begin oscillating
later, once the $\lambda\gtrsim 3H/2$ threshold is crossed.  The time 
$t_\lambda$ at which a given $a_\lambda$ begins to oscillate
is therefore given by
\begin{equation}
  t_\lambda ~\equiv~ \max\left\{\frac{\kappa_\lambda}{2\lambda},\tG\right\}~,
  \label{eq:tlambdaInBothRegimes}
\end{equation}
where $\kappa_\lambda$ is the value of $\kappa$ corresponding to the epoch during 
which this oscillation begins.  At times $t_G < t \lesssim t_\lambda$, a given 
$a_\lambda$ continues to behave like vacuum energy rather than like matter, and  
thus properly contributes not to $\OmegaDM$, but to the dark-energy abundance.
(Note that this definition of $t_\lambda$ is slightly different from the one given
in Ref.~\cite{DynamicalDM1}, where $t_\lambda$ was simply defined as 
$\kappa_\lambda/\lambda$, regardless of its relationship to $t_G$.) 

For any given $a_\lambda$, however, the relevant quantity for dark-matter phenomenology 
is not the value of $a_\lambda$ itself, but its energy density $\rho_\lambda$,
which is related to $a_\lambda$ and $\dot{a}_\lambda$ by the relation
\begin{equation}
  \rho_\lambda ~=~ \frac{1}{2}\big[\dot{a}_\lambda^2+\lambda^2a_\lambda^2\big]~.
  \label{eq:RhoLambdaDef}
\end{equation} 
At early times, when $t < t_G$, only $a_0$ has a non-zero background value in the
rapid-turn-on approximation, and since this field is massless, its energy density 
vanishes.  At $t = t_G$, however, Eq.~(\ref{eq:alambdaInitCondits}) implies that
each field acquires an initial energy density
\begin{equation}
  \rho_\lambda(t_G) ~=~ \frac{1}{2}\theta^2\fhatX^2 \lambda^2 A_\lambda^2~.
  \label{eq:RhoInitCondits}
\end{equation}
Since $a_\lambda$ remains effectively constant until $t = t_\lambda$ for any field
for which $t_\lambda > t_G$, we also see that 
$\rho_\lambda(t_\lambda) = \rho_\lambda(t_G)$.  This implies that the energy density
stored in such a field behaves like vacuum energy until $t = t_\lambda$, at which point
the field begins to oscillate coherently around the minimum of its potential.  The 
energy density stored in such oscillations, as is well known, scales like massive matter.      
At late times $t \gg t_\lambda$, when the time scale associated with these oscillations 
becomes rapid compared to the time scale over which the amplitude of 
$a_\lambda$ changes appreciably, the virial approximation implies that 
$\rho_\lambda \approx \langle \dot{a}_\lambda^2 \rangle$, where 
$\langle \dot{a}_\lambda^2 \rangle$ denotes the average of $\dot{a}_\lambda^2$ over one
cycle of oscillation.  In this regime, one finds that 
\begin{equation}
  \rho_\lambda(t) ~=~ \rho_\lambda(t_G)\left(\frac{t_\lambda}{t}\right)^{\kappa_\lambda}
      e^{-\Gamma_\lambda(t-t_G)}
  \label{eq:RhoOftEqnWithR}
\end{equation}  
for each $\rho_\lambda$ during the epoch
in which oscillation began, with $\rho_\lambda(t_G)$ given in
Eq.~(\ref{eq:RhoInitCondits}).  Computing $\rho_\lambda$ during subsequent
epochs is simply a matter of applying Eq.~(\ref{eq:RhoOftEqnWithR}) 
iteratively with the appropriate boundary conditions for 
$\rho_\lambda$ at the transition points at which $\kappa$ changes.

All that remains, then, in order to specify the energy density
$\rho_\lambda$ associated with a given $a_\lambda$ for any particular 
choice of model parameters is to determine the time scales $t_G$ and 
$t_\lambda$ as a function of those parameters.  Indeed, the results 
for $\rho_\lambda$ clearly depend sensitively both on the epoch 
during which abundances are established, and the epoch during which oscillation
begins.  In principle, $t_G$ could fall within the RD era, during the 
usual MD era, or during reheating; likewise, $t_\lambda$ could occur
at any time at or after $t_G$.  However, as will be shown in 
Sect.~\ref{sec:Abundances},  $\Lambda_G \gg 10$~MeV is required in order for
our ensemble of $a_\lambda$ to yield a relic abundance on the order of 
$\OmegaDM$.  For such values of $\Lambda_G$, $t_G \gg \tMRE$.  Moreover,
for $\mathcal{O}(1)$ values of $g_G$ and $\xi$, $t_\lambda \ll \tMRE$ as well.     
Therefore, when discussing the standard cosmology, we will focus on the case in
which $t_G \leq t_\lambda < \tMRE$ for all $a_\lambda$.
Furthermore, when operating within the context of the LTR cosmology, 
we will implicitly assume that $t_G \leq t_\lambda < \tRH$ for all $a_\lambda$, 
so that all fields begin oscillating while the energy density of the universe is still 
dominated by coherent oscillations of the inflaton field.  This is justified by the
fact that $\tRH \sim 10^{-1} - 10^{-4}$~s for reheating temperatures 
$\TRH$ within the phenomenologically allowed window 
$4\mathrm{~MeV}\lesssim T_\ast \lesssim 30$~MeV.  In conjunction with the 
experimental bound on $M_c$ given in Eq.~(\ref{eq:MinimumMc}), this implies 
that $t_\lambda \lesssim \tRH$ for all $a_\lambda$ in any given tower, unless 
$\Lambda_G \lesssim 10$~MeV.  In summary, we shall therefore assume that 
\begin{eqnarray}
   \mathrm{standard~cosmology:} & & \tRH < t_G \leq t_\lambda < \tMRE \nonumber \\
   \mathrm{LTR~cosmology:} & & t_P < t_G \leq t_\lambda < \tRH
   \label{eq:tGtlambdaRegimes}
\end{eqnarray}
in what follows, where $t_P$ is the Planck time.  Note that in the 
standard cosmology, $\tRH$ is assumed to be so early that all $t_G$
of interest will easily satisfy the lower bound.  Also note that in
the LTR cosmology, modes for which $t_\lambda$ occurs during or prior to
inflation will inflate away and therefore carry zero abundance at present
time. 

\subsection{Axion Relic Abundances}
 
We now provide explicit expressions for $\rho_\lambda$ in both 
the standard and LTR cosmologies.  
We begin by considering the case of the standard
cosmology, in which $\kappa=3/2$ at all relevant time scales prior to 
matter-radiation equality, and $\kappa = 2$ after the transition to 
matter-domination at $\tMRE$.  In this cosmological framework, given
the regimes for $t_G$ and $t_\lambda$ specified in Eq.~(\ref{eq:tGtlambdaRegimes}), 
it therefore follows that
\begin{equation}
   \rho_\lambda^{\mathrm{Std}}(t)~=~ 
      \frac{1}{2}\theta^2\fhatX^2 \lambda^2 A_\lambda^2
      e^{-\Gamma_\lambda(t-\tG)}\times\begin{cases}
      \displaystyle\vspace{0.25cm}
      \left(\frac{t_\lambda}{t}\right)^{3/2}~~ & t_\lambda~\lesssim~ t~\lesssim~\tMRE \\
      \displaystyle
      \left(\frac{t_\lambda^{3/2}\,\tMRE^{1/2}}{t^2}\right)~~ & t~\gtrsim~\tMRE~.
      \end{cases}
   \label{eq:RhoLambdaInStdCosmo}
\end{equation}  
By contrast, in the LTR cosmology,
the energy of the universe remains dominated by coherent oscillations of the 
inflaton field from the end of inflation until a very late time 
$\tRH\sim 1$~s.  Thus, if the axion fields begin oscillating at a time 
$t_\lambda < \tRH$, as specified in Eq.~(\ref{eq:tGtlambdaRegimes}), 
we initially have $\kappa=2$, followed by a transition at $\tRH$ 
to the usual RD era, in which $\kappa=3/2$.  This signifies that in the LTR 
cosmology, we have 
\begin{equation}
   \rho_\lambda^{\mathrm{LTR}}(t)~=~ 
      \frac{1}{2}\theta^2\fhatX^2 \lambda^2 A_\lambda^2
      e^{-\Gamma_\lambda(t-\tG)}\times\begin{cases}
      \displaystyle \vspace{0.25cm}
      \left(\frac{t_\lambda}{t}\right)^2~~ & t_\lambda~\lesssim~ t~\lesssim~\tRH \\
      \displaystyle \vspace{0.25cm} 
      \left(\frac{t_\lambda^2}{\tRH^{1/2}\,t^{3/2}}\right)~~ & \tRH~\lesssim~t~\lesssim~\tMRE\\
      \displaystyle
      \left(\frac{t_\lambda^2\,\tMRE^{1/2}}{t^2\,\tRH^{1/2}}\right)~~ &  t~\gtrsim~\tMRE~.
      \end{cases}
   \label{eq:RhoLambdaInLTRCosmo}
\end{equation}
In other words, Eq.~(\ref{eq:RhoLambdaInLTRCosmo}) replaces 
Eq.~(\ref{eq:RhoLambdaInStdCosmo}) in the context of the LTR cosmology, in which
the usual RD era is preceded by an initial period of matter domination.  
It is worth emphasizing here that the value of $t_\lambda$ for a given mass 
eigenvalue $\lambda$ will, in general, differ between the two cosmologies, due to
the differing relationship between $H$ and $t$ at times $t\lesssim\tRH$.

Comparing Eqs.~(\ref{eq:RhoLambdaInStdCosmo}) and~(\ref{eq:RhoLambdaInLTRCosmo}),
we see that the cosmological context in which the axion fields evolve can have
a potentially dramatic effect on the late-time results for the various 
$\rho_\lambda$.  However, for any given mode, 
Eq.~(\ref{eq:tlambdaInBothRegimes}) implies that the magnitude $\mathcal{E}_{\mathrm{LTR}}$ 
of that suppression depends on the relationship between $\lambda$ and $\tG$.  Comparing
Eqs.~(\ref{eq:RhoLambdaInStdCosmo}) and~(\ref{eq:RhoLambdaInLTRCosmo}), we find that 
the axion energy densities are 
suppressed in the LTR cosmology, relative to the standard cosmology, by a factor 
\begin{equation}
   \mathcal{E}_{\mathrm{LTR}} ~\equiv~ \frac{\rho_{\mathrm{LTR}}}{\rho_{\mathrm{Std}}} 
      ~\approx~ \begin{cases} 
        \displaystyle \vspace{0.25cm}
        \frac{g_\ast^{5/4}(\TRH)}{g_\ast^{5/4}(\Lambda_G)}
        \left(\frac{\TRH}{\Lambda_G}\right)^5
        &\lambda ~\geq~ 1/\tG\\
        \displaystyle 4\left(\frac{2}{3}\right)^{3/2}\left(\frac{2\pi^2}{45}\right)^{1/4}
        g_\ast^{1/4}(\TRH)\frac{\TRH}{M_P^{1/2}\lambda^{1/2}}~~~
        & \lambda ~<~ 1/\tG~.
        \end{cases}
  \label{eq:LTRSuppressionFactor}
\end{equation}
These results imply that the energy-density contributions from those modes 
which begin oscillating at $\tG$ are suppressed in the LTR cosmology, 
relative to their value in the standard cosmology, to a greater degree 
than the contributions from those modes which begin oscillating later.  
These results are analogous to those obtained
in Ref.~\cite{LTRAxions} for a standard, four-dimensional axion, where
the suppression factor is referred to as 
$V^{\mathrm{LTR}}/V^{\mathrm{Std}}$.  
    
It is also possible (and indeed when $\Lambda_G$ is large, more or less inevitable) 
that in the LTR cosmology, a great many of the heavier $a_\lambda$ will begin 
to oscillate either during or prior to the end of
inflation.  Since the scale factor $R$ grows exponentially during this epoch, the 
energy density in stored any such mode, which scales like $\rho_\lambda \propto R^{-3}$,
will effectively be diluted into irrelevance by this rapid expansion. 
As long as $\Lambda_G \lesssim \Tmax$, \ie, as long as confinement occurs only after
inflation is over, no energy density is stored in these modes during inflation.
As a result, the energy density is given by Eq.~(\ref{eq:RhoLambdaInLTRCosmo}) 
as usual.  However, if confinement occurs before or during inflation, $\rho_\lambda$ 
for any mode for which $\lambda > 3H_I/2$ will be exponentially damped by Hubble
dilution, and it is therefore reasonable to take 
$\rho_\lambda = 0$ for any such mode.  

From the results in Eqs.~(\ref{eq:RhoLambdaInStdCosmo}) and
(\ref{eq:RhoLambdaInLTRCosmo}), it is straightforward to obtain the
relic abundance $\Omega_\lambda \equiv \rho_\lambda/\rhocrit$ for each
$a_\lambda$.  Let us begin by addressing those modes for which 
$t_\lambda = \tG$.  Since the critical density for a flat universe 
is given by $\rhocrit = 3 H^2 M_P^2$, we find that in 
the rapid-turn-on approximation, the contribution to the dark-matter 
relic abundance from each such mode at a given time $t$ in the 
standard cosmology is 
\begin{equation}
  \Omega_\lambda^{\mathrm{Std}} ~ = ~
    3\left(\frac{\theta \fhatX \mX}{M_P}\right)^2     
      \tG^{3/2}
      \left[1+\frac{\lambda^2}{\mX^2}+
      \frac{\pi^2\mX^2}{M_c^2}\right]^{-1}
      e^{-\Gamma_\lambda(t-\tG)}\times
      \begin{cases}
        \displaystyle\vspace{0.25cm}
        \frac{4}{9}t^{1/2}~~ & t~\lesssim~\tMRE\\
        \displaystyle\frac{1}{4}\tMRE^{1/2}~~ & t~\gtrsim~\tMRE~.
      \end{cases}
  \label{eq:OmegaLambdaOftEqntG}
\end{equation} 
By contrast, in the LTR cosmology, the corresponding result is 
\begin{equation}
  \Omega_\lambda^{\mathrm{LTR}} ~=~
    3\left(\frac{\theta \fhatX \mX}{M_P}\right)^2  
      \tG^2  
      \left[1+\frac{\lambda^2}{\mX^2}+
      \frac{\pi^2\mX^2}{M_c^2}\right]^{-1}
      e^{-\Gamma_\lambda(t-\tG)}\times
      \begin{cases}
      \displaystyle\frac{1}{4} \vspace{0.25cm}~~
      & 1/\lambda ~\lesssim~ t~ \lesssim~\tRH \\
      \displaystyle\frac{4}{9}\left(\frac{t}{\tRH}\right)^{1/2} \vspace{0.25cm}~~
      & \tRH ~\lesssim~ t ~\lesssim~ \tMRE \\ 
      \displaystyle\frac{1}{4}\left(\frac{\tMRE}{\tRH}\right)^{1/2}~~
      & t ~\gtrsim~ \tMRE~.    
    \end{cases}
  \label{eq:OmegaLambdaOftEqnLTRtG}
\end{equation} 
For the rest of the $a_\lambda$ (\ie, those for which $t_\lambda \geq \tG$), 
the corresponding results in the context of the standard cosmology are  
\begin{equation}
  \Omega_\lambda^{\mathrm{Std}} ~ = ~
    3 \left(\frac{3}{4}\right)^{3/2}
      \left(\frac{\theta \fhatX \mX}{M_P}\right)^2
      \lambda^{-3/2}     
      \left[1+\frac{\lambda^2}{\mX^2}+
      \frac{\pi^2\mX^2}{M_c^2}\right]^{-1}
      e^{-\Gamma_\lambda(t-\tG)}\times
      \begin{cases}\vspace{0.25cm}
       \displaystyle\frac{4}{9}t^{1/2} ~~& t~\lesssim~\tMRE\\
       \displaystyle\frac{1}{4}\tMRE^{1/2}~~ & t~\gtrsim~\tMRE~,
      \end{cases}
  \label{eq:OmegaLambdaOftEqntlambda}
\end{equation} 
whereas in the context of the LTR cosmology, we instead have
\begin{equation}
  \Omega_\lambda^{\mathrm{LTR}} ~=~
    3\left(\frac{\theta \fhatX \mX}{M_P}\right)^2 
      \lambda^{-2}
      \left[1+\frac{\lambda^2}{\mX^2}+
      \frac{\pi^2\mX^2}{M_c^2}\right]^{-1}
      e^{-\Gamma_\lambda(t-\tG)}\times
      \begin{cases}
     \displaystyle\frac{1}{4} \vspace{0.25cm}~~
      & 1/\lambda ~\lesssim ~t ~\lesssim~ \tRH \\
      \displaystyle\frac{4}{9}\left(\frac{t}{\tRH}\right)^{1/2} \vspace{0.25cm}~~
      & \tRH ~\lesssim~ t ~\lesssim~ \tMRE \\ 
      \displaystyle\frac{1}{4}\left(\frac{\tMRE}{\tRH}\right)^{1/2}~~
      & t~\gtrsim ~\tMRE~.    
    \end{cases}
  \label{eq:OmegaLambdaOftEqnLTRtlambda}
\end{equation} 
Note that this is valid only for those $a_\lambda$ for which 
$\lambda \lesssim 3H_I/2$.  For those modes with $\lambda \gtrsim 3H_I/2$, 
as discussed above, effectively $\Omega_\lambda = 0$
due to Hubble dilution during inflation.
During periods of matter domination, we see that, 
$\rho_\lambda$ and $\rhocrit$ scale identically with time, and consequently 
$\Omega_\lambda$ remains constant in each of these expressions.
During periods of radiation domination, on the other hand, 
$\rho_\lambda$ falls faster with time than $\rhocrit$, and
$\Omega_\lambda$ grow like $t^{1/2}$.

Note that for $t=\tnow$, Eqs.~(\ref{eq:OmegaLambdaOftEqntG}) 
through~(\ref{eq:OmegaLambdaOftEqnLTRtlambda}) take the forms 
specified in Eqs.~(54) through~(60) in Ref.~\cite{DynamicalDM1}, 
where the brane mass $m$ appearing in these equations is identified
with $\mX$ for a bulk axion.  In other words, the distinction
between the $\Omega_\lambda$ expressions for the standard and LTR 
cosmologies given here is tantamount to identifying the era in which 
the initial abundances are established in Ref.~\cite{DynamicalDM1}.

Since we will be primarily interested in scenarios in which the dark-matter
relic abundance $\OmegaDM$ receives contributions from a large number of
different axion mass eigenstates, it is of critical importance to determine 
precisely how $\Omega_\lambda$ scales with $\lambda$, as this describes the relative
contributions to the total relic abundance $\OmegaDM$ from
the various states in the tower.  Fortunately, in the
rapid-turn-on approximation, the dependence of $\Omega_\lambda$ on $\lambda$ 
stems from only three factors, as discussed in Ref.~\cite{DynamicalDM1}.  The first,
of course, is the choice of cosmology.  
The second is how $\lambda$ compares to the mass scales $\mX$ and $\pi \mX^2/M_c$,
which determines the overlap between $a_\lambda$ and the KK zero mode $a_0$, 
and therefore the initial displacement of $a_\lambda$ at $t=t_\lambda$.   
For those modes for which $\lambda \gtrsim \max\{\mX,\pi\mX^2/M_c\}$, the
$\lambda^2/\mX^2$ term dominates in the factor in brackets appearing in each of 
Eqs.~(\ref{eq:RhoLambdaInStdCosmo}) and~(\ref{eq:RhoLambdaInLTRCosmo}), and 
$\Omega_\lambda$ acquires a factor of $\lambda^{-2}$.  By
contrast, for those modes for which $\lambda \lesssim \max\{\mX,\pi\mX^2/M_c\}$,
the constant terms dominate, and $\Omega_\lambda$ acquires no such dependence.

The third factor which determines how $\Omega_\lambda$ scales with $\lambda$
in the rapid-turn-on approximation is whether or not the axion mode in
question begins oscillating at $\tG$, or at some later time.  Indeed, in 
Ref.~\cite{DynamicalDM1}, these cases were referred to respectively 
as the ``instantaneous'' and ``staggered'' turn-on regimes.  Comparing the
expressions valid for $t_\lambda = \tG$ in 
Eqs.~(\ref{eq:OmegaLambdaOftEqntG}) and~(\ref{eq:OmegaLambdaOftEqnLTRtG}) with
those valid for $t_\lambda > \tG$ in 
Eqs.~(\ref{eq:OmegaLambdaOftEqntlambda}) and~(\ref{eq:OmegaLambdaOftEqnLTRtlambda}), 
we see that the $\Omega_\lambda$ for those $a_\lambda$ which begin oscillating after 
the confining phase transition takes place acquire an additional dependence on
$\lambda$.  The precise form of this dependence depends on the cosmological
context within which the model is embedded: in the standard cosmology, it
is $\lambda^{-3/2}$; in the LTR cosmology, it is $\lambda^{-2}$.  
Physically, this factor stems from the fact that prior to the time
it begins to oscillate coherently, the energy density in any given $a_\lambda$
remains constant, whereas after oscillation 
begins, it scales like massive matter.
Therefore, the later a given mode begins to oscillate, the longer the energy
density stored in that mode will remain unaffected by cosmic expansion, and
therefore the larger the present-day value of $\Omega_\lambda$ will be.   

\begin{figure}[p]
\begin{center}
\epsfxsize 2.75 truein \epsfbox {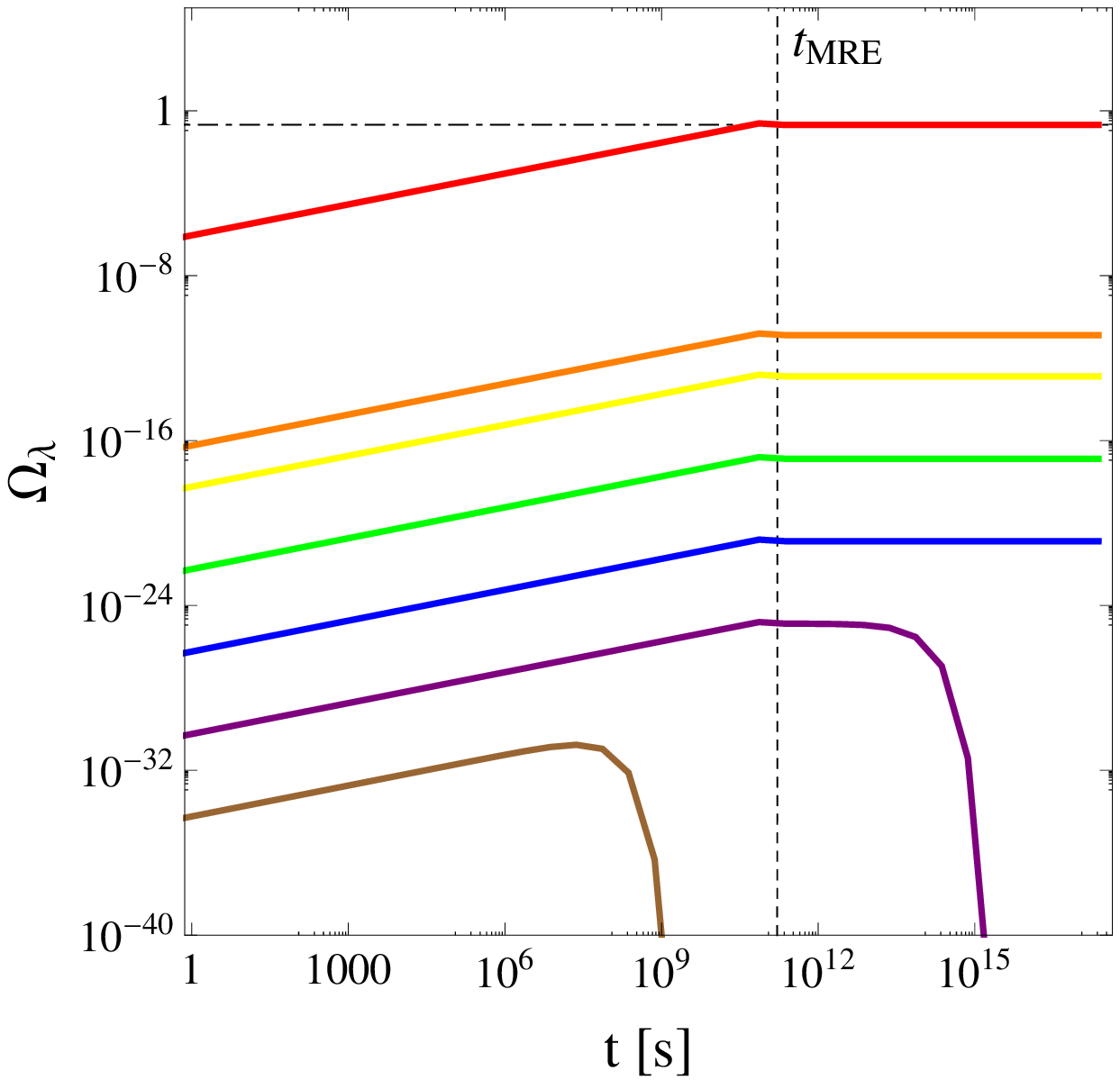}~~ 
\epsfxsize 2.75 truein \epsfbox {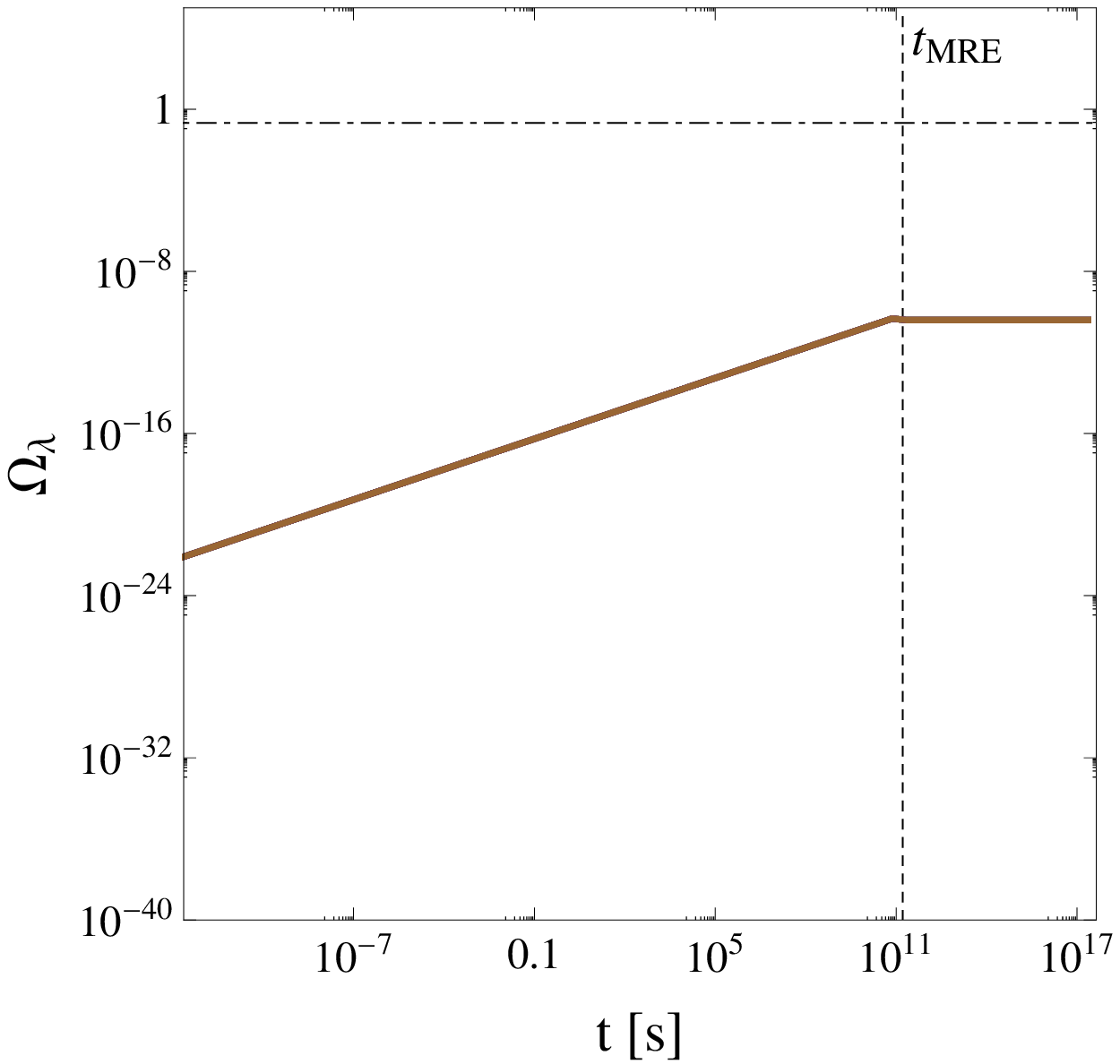}~~ 
\raisebox{1.5cm}{\epsfxsize 1.0 truein \epsfbox {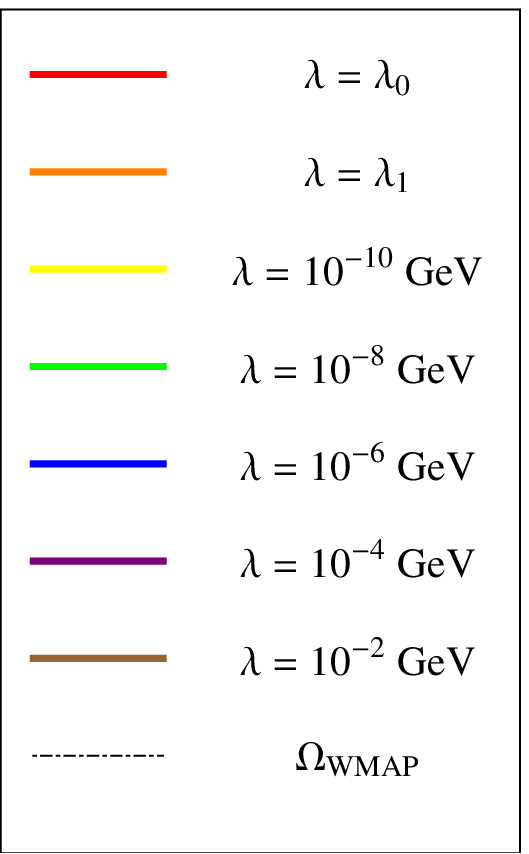}} 
\end{center}
\caption{The individual relic abundances $\Omega_\lambda$ associated with
  a variety of different $a_\lambda$ in the standard cosmology, shown as a function 
  of time $t$, for two scenarios with different values of the parameters 
  $\fhatX$, $\Lambda_G$, and $\theta$.  The left 
  panel corresponds to a choice of $\fhatX \approx 10^{9}$, 
  $\Lambda_G = 1$~MeV, and $\theta \approx 0.04$, while the 
  right panel corresponds to a choice of 
  $\fhatX \approx 2\times 10^{7}$~GeV, $\Lambda_G = 1$~TeV, and $\theta = 1$.  
  In both cases, we have taken $\xi = g_G = 1$, with
  $M_c = 10^{-11}$~GeV.~  The range of $t$ displayed in each panel spans
  from the corresponding confinement time scale $t_G$ for the hidden-sector gauge
  group $G$, and a vertical, dashed line indicating the time scale associated with
  matter-radiation equality has also been included for reference.  The horizontal,
  dash-dotted line indicates the total observed dark-matter relic abundance, 
  as measured by the WMAP satellite.  
  It should be emphasized that in both of these scenarios,
  the {\it total} present dark-matter relic abundance contribution from all 
  of the $a_\lambda$ in the tower reproduces this observed value 
  of $\OmegaDM$ to within the limits quoted in Eq.~(\protect\ref{eq:OmegaWMAP}). 
\label{fig:OmegaPanelsStd}}
\begin{center}
\epsfxsize 2.75 truein \epsfbox {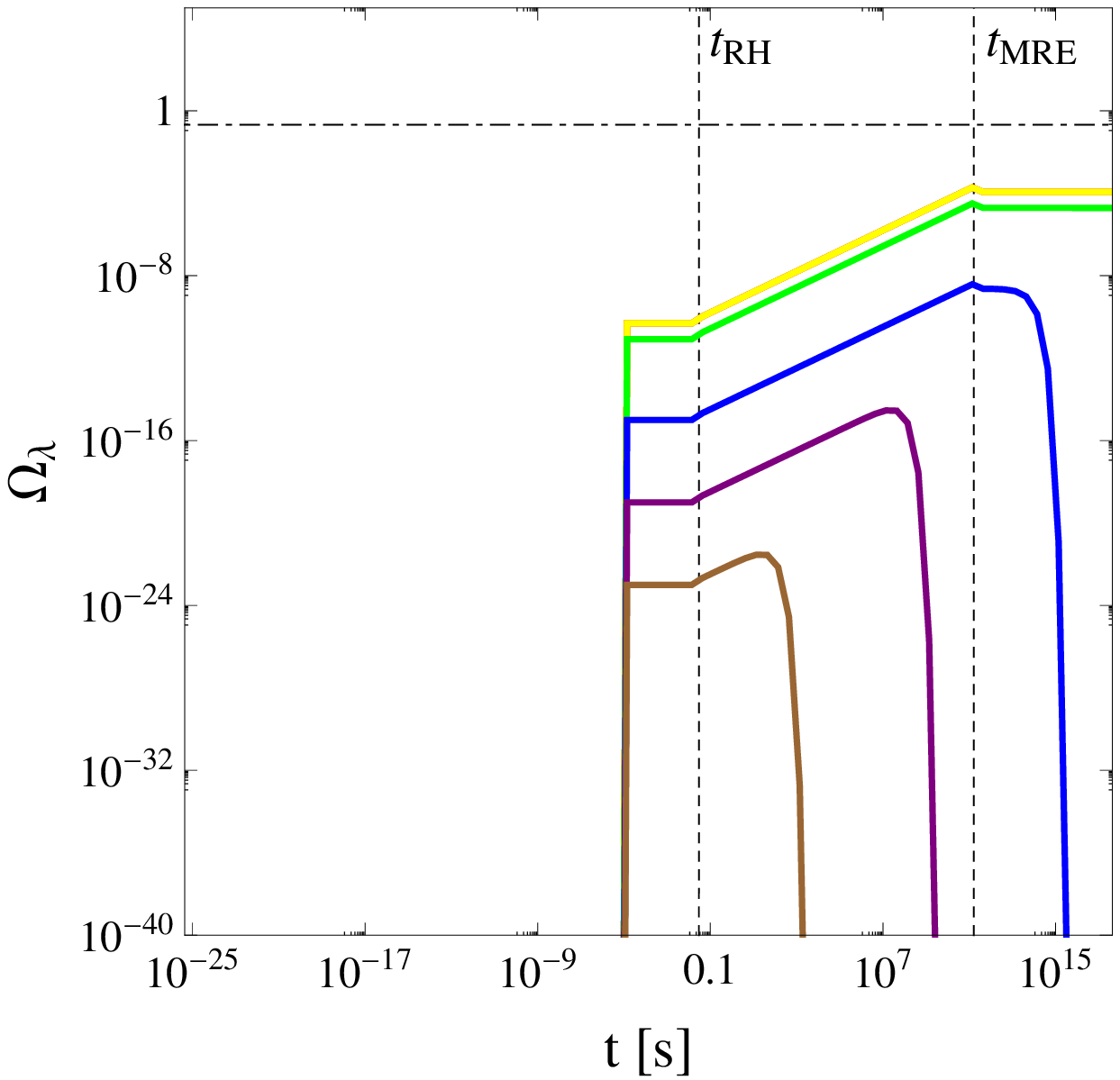}~~ 
\epsfxsize 2.75 truein \epsfbox {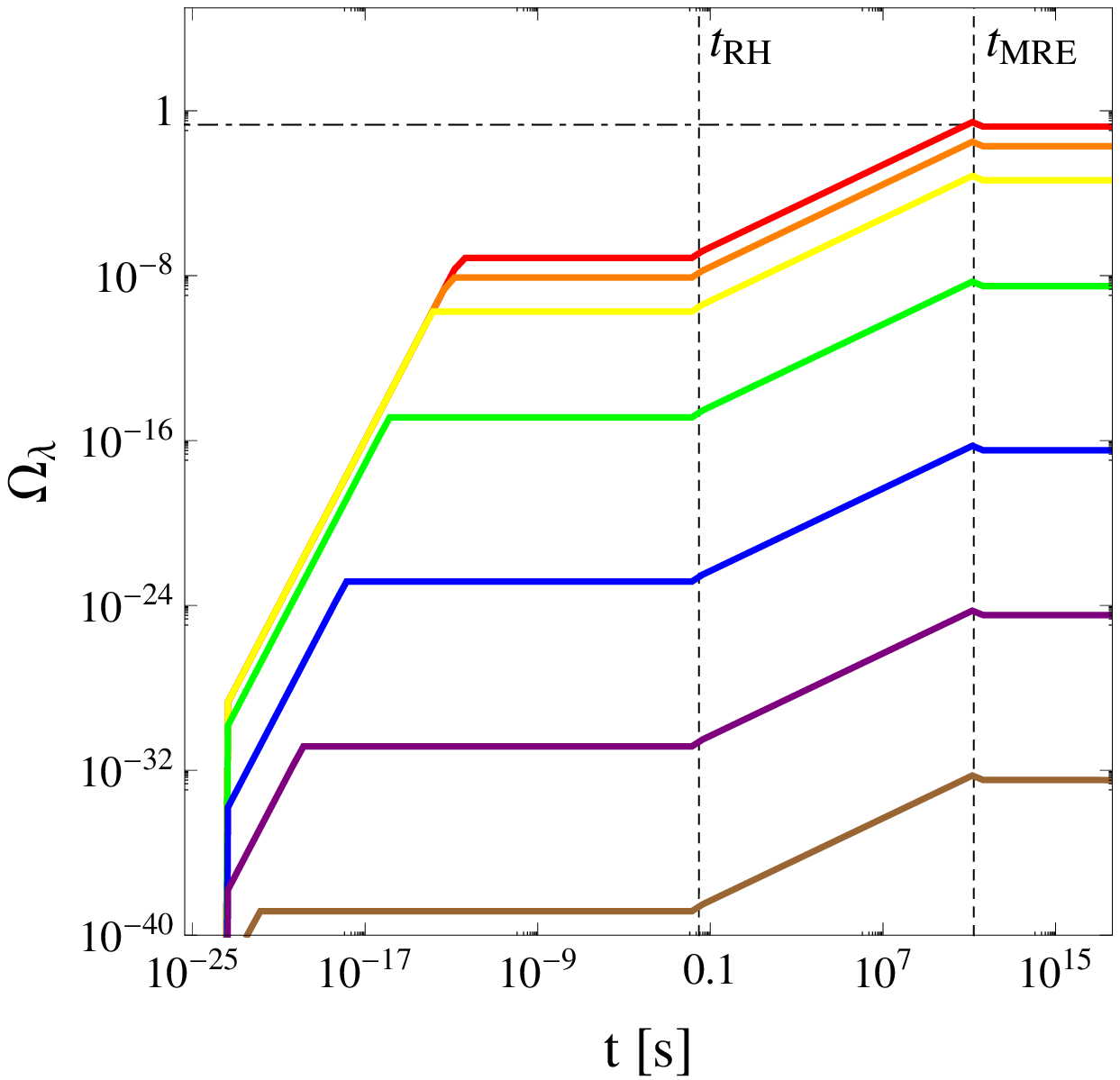}~~ 
\raisebox{1.5cm}{\epsfxsize 1.0 truein \epsfbox {KeyBox.eps}}
\end{center}
\caption{The individual relic abundances $\Omega_\lambda$ associated with
  a variety of different $a_\lambda$ in the LTR cosmology, shown as a function 
  of time $t$, for two scenarios with different values of $\fhatX$, $M_c$,
  and $\Lambda_G$.  The left panel corresponds to a choice of $\fhatX = 10^{6}$, 
  $M_c = 4\times 10^{-12}$~GeV,
  and $\Lambda_G \approx 37$~MeV, while the right panel corresponds to a choice of 
  $\fhatX \approx 6\times 10^{14}$~GeV, $M_c = 10^{-11}$~GeV and
  $\Lambda_G = 1$~TeV.~  In both cases, we have taken $\xi = g_G = \theta = 1$, with
  $\TRH = 5$~MeV and $H_I = 10$~GeV.~  
  The range of $t$ displayed in each panel spans
  from the end of cosmic inflation to present day, and vertical (dashed) lines 
  indicating the time scales associated with the end of reheating and with 
  matter-radiation equality have also been included for reference.  The horizontal,
  dash-dotted line indicates the total observed dark-matter relic abundance, 
  as measured by the WMAP satellite.  
  It should be emphasized that in both of these scenarios,
  the {\it total} present-day dark-matter relic abundance contribution from all 
  of the $a_\lambda$ in the tower reproduces this observed value 
  of $\OmegaDM$ to within the limits quoted in Eq.~(\protect\ref{eq:OmegaWMAP}).  
\label{fig:OmegaPanelsLTR}}
\end{figure}

In order to illustrate the implications of these effects,
in each of the panels of Figs.~\ref{fig:OmegaPanelsStd} and~\ref{fig:OmegaPanelsLTR} 
we track the evolution of $\Omega_\lambda$ for a representative 
sample of $a_\lambda$ within a given theory from $t_G$ to present time.
The curves shown in Fig.~\ref{fig:OmegaPanelsStd} reflect typical
results which arise in the context of the standard cosmology.
The left panel corresponds to a scenario
with a small confinement scale $\Lambda_G = 1$~MeV, a moderate value
$\fhatX = 10^9$~GeV for the effective four-dimensional $U(1)_X$-breaking scale, and
a small misalignment angle $\theta \approx 0.04$.  The
right panel corresponds to the opposite case: a scenario in which
$\Lambda_G = 1$~TeV, with $\fhatX \approx 2\times 10^{7}$~GeV and $\theta = 1$.  In
each scenario, we have taken $\xi = g_G = 1$, and set the compactification scale 
to be $M_c = 10^{-11}$~GeV.~  
The curves shown in each panel (from top to bottom) 
correspond to the lightest two values of $\lambda$,
here referred to as $\lambda_0$ and $\lambda_1$, 
corresponding to that particular choice of $\fhatX$ and $M_c$
($\lambda_0 \approx 6\times 10^{-17}$~GeV 
and $\lambda_1 \approx 10^{-11}$~GeV for the left panel;
$\lambda_0 \approx 5\times 10^{-12}$~GeV and 
$\lambda_1 \approx 2\times 10^{-11}$~GeV for the right panel), 
along with several additional values of $\lambda$, including
$\lambda = \{10^{-10},10^{-8},10^{-6},10^{-4},10^{-2}\}$~GeV.~  
It should be noted that in the right panel, all curves shown lie 
essentially on top of one another, and are thus not individually apparent.
It is worth remarking
that $y\gg 1$ in the scenario depicted in the left panel, while $y\ll 1$ 
in the scenario depicted in the right panel.  Also shown are a horizontal, 
dot-dashed line indicating the value for $\OmegaDM$ observed by the WMAP 
experiment~\cite{WMAP}, as quoted in Eq.~(\ref{eq:OmegaWMAP}), 
and a pair of vertical lines 
indicating the positions of $\tRH$ and $\tMRE$.  
 
Both of these scenarios yield a total dark-matter relic abundance
which is consistent with the WMAP results given in Eq.~(\ref{eq:OmegaWMAP}).
However, it should be noted that the parameter assignments used for these figures 
have been chosen exclusively for purposes of illustration.  In Sect.~\ref{sec:Constraints},
we will discuss the situation using the parameters which are consistent with all 
observational and phenomenological constraints.  

Note that the sets of curves shown in the two panels of 
Fig.~\ref{fig:OmegaPanelsStd} differ quite significantly.  For example,
in the left panel, the effect of the non-zero decay widths of the heavier 
$a_\lambda$ is readily apparent.  Indeed, the curves corresponding to masses
$\lambda \geq 10^{-4}$~GeV rapidly drop to zero at a time scale $t \sim \tau_\lambda$.
By contrast, in the right plot, which corresponds to a scenario with $y \ll 1$, 
the decay rate of each $a_\lambda$ is suppressed by a factor of $\wtl^4 A_\lambda^2$, 
which can be quite small in such a scenario.  Consequently, all of the $a_\lambda$ for 
which $\Omega_\lambda$ curves are shown in the plot are stable on cosmological time
scales.

The abundance curves shown in Fig.~\ref{fig:OmegaPanelsLTR}, on the other hand, 
display typical results obtained in the LTR cosmology.  More specifically, the
results shown here correspond to the parameter assignments 
$\TRH = 5$~MeV and $H_I = 10$~GeV.
As in Fig.~\ref{fig:OmegaPanelsStd}, the two panels shown in this figure correspond 
to two different choices for $\fhatX$, $M_c$, and $\Lambda_G$ which both yield a 
total present-day relic abundance consistent with WMAP data.  The curves shown 
in the left panel correspond to a scenario with $\fhatX = 10^{6}$~GeV, 
$M_c = 4\times 10^{-12}$~GeV, and $\Lambda_G \approx 37$~MeV.~  This scenario 
exemplifies the case in which $\fhatX$ and $\Lambda_G$ are both comparatively small.  
By contrast, the curves displayed in the right panel correspond to a
scenario with far larger values for these parameters: $\fhatX= 6\times 10^{14}$~GeV
and $\Lambda_G = 1$~TeV, with $M_c = 10^{11}$~GeV.~  In both cases, we have taken 
$\xi = g_G = \theta = 1$.  Once again, the curves shown in each plot correspond to 
$\lambda = \{ 10^{-10},10^{-8},10^{-6},10^{-4},10^{-2}\}$~GeV, as well as 
$\lambda_0$ and $\lambda_1$.  For the left plot, 
$\lambda_0 = 2\times 10^{-12}$~GeV and $\lambda_1 = 6\times 10^{-12}$~GeV; 
for the right plot, $\lambda_0 = 5\times 10^{-12}$~GeV and 
$\lambda_1 = 2\times 10^{-11}$~GeV.~  Again, in the left panel, the 
$\lambda_0$ and $\lambda_1$ curves are not apparent because they lie directly 
beneath the $\lambda=10^{-10}$~GeV curve.

The contrasting features between the two panels in Fig.~\ref{fig:OmegaPanelsLTR} are
predominately due to the differences between their $\fhatX$ and $\Lambda_G$ values.
First, as stated above, $\fhatX$ is quite small in the scenario displayed in the 
left panel, and consequently, the couplings between the $a_\lambda$ and
the fields of the SM are quite large.  This implies that the decay rates 
associated with the heavier $a_\lambda$ will be sizable in this scenario, and that
a large number of these heavier modes will decay before present time.  Indeed, the
precipitous drop in each of the curves corresponding to a mass eigenvalue in the range 
$\lambda \leq 10^{-6}$~GeV in this plot is a consequence of the decay of these modes to
SM fields.  By contrast, in the scenario displayed in the right plot, $\fhatX$ is
large enough that all of the $a_\lambda$ for which $\Omega_\lambda$ curves are shown
are stable on cosmological time scales, and no such effect is apparent. 
Second, $\Lambda_G$ is also quite small in the scenario displayed in the 
left panel, and the confinement time scale $t_G\approx 10^{-5}$~s
is consequently quite late.  As a result, $t_\lambda = t_G$ for all $a_\lambda$.
This implies not only that all of the modes begin
oscillating at the same time, but moreover, that the $\Omega_\lambda$ only 
become non-zero quite late.  By contrast, in the right panel, 
$\Lambda_G$ is large and $t_G$ is correspondingly quite early.  This results
in a situation in which the $t_\lambda$ for the lighter modes are staggered 
in time.  As discussed in Ref.~\cite{DynamicalDM1}, the 
primary consequence of this staggering is that the $\Omega_\lambda$ curves 
for these lighter modes depend more sensitively on $\lambda$. 

The most important implication of 
Eqs.~(\ref{eq:OmegaLambdaOftEqntG}) through~(\ref{eq:OmegaLambdaOftEqnLTRtlambda}),
however, is that $\Omega_\lambda$ decreases with 
increasing $\lambda$ regardless of the details of the cosmological framework. 
On the other hand, we saw in Sect.~\ref{sec:AxionDecayRates} 
that $\Gamma_\lambda$ increases monotonically with $\lambda$.  This  
observation is indeed encouraging, in that it suggests that $\Gamma_\lambda$ and 
$\Omega_\lambda$ possess the appropriate, reciprocal relationship needed 
for an ensemble of $a_\lambda$ to serve as dynamical dark matter.  In the following 
section, we will quantify more precisely the relationship between 
$\Gamma_\lambda$ and $\Omega_\lambda$ and demonstrate that this is indeed the case.   
The results of the present section therefore attest that misalignment production, 
in stark contrast to thermal production, is an ideal mechanism for the 
generation of axion relic 
abundances in dynamical dark-matter models.
Moreover, as we shall soon demonstrate, this mechanism dominates in the regime 
of model-parameter space in which an ensemble of $a_\lambda$ tends to be 
phenomenologically viable, in the sense that it correctly reproduces 
the observed dark-matter relic abundance, while at the same time satisfying all
relevant constraints from experiment, astrophysical observation, and cosmology. 

Before we proceed to analyze the collective properties of such ensembles, 
however, two brief comments are in order.   
The first of these concerns the validity of the rapid-turn-on approximation.
As we have stated above, Eq.~(\ref{eq:TheDoubleDotEqnWithGammaLambda}) is 
strictly valid only when $\lambda$ and 
$\Gamma_\lambda$ are essentially independent of temperature. 
However, there turn out to be certain situations in which the time-dependence of
$\lambda$ and $\Gamma_\lambda$ at temperatures $T\gtrsim \Lambda_G$ is
physically unimportant, and in which these quantities can be reliably treated as
constants throughout the period in which $a_\lambda$ are oscillating.  One such  
situation arises in cases in which $t_\lambda > \tG$ for all of the
$a_\lambda$ which contribute meaningfully to the total dark-matter relic abundance, 
and therefore coherent axion oscillations do not occur until after $\mX(T)$ is 
effectively constant.  
This situation tends to arise either when $\Lambda_G$ is quite large, in which case
$\mX(T)$ attains its constant, late-time value very early, or else when
$\Lambda_G$ is fairly small, but $\fhatX$ is quite large, in which case $\mX$
itself is extremely small.  As we shall see in Sect.~\ref{sec:Abundances},
these turn out to be precisely the situations in which the total relic-abundance 
contribution from the ensemble of $a_\lambda$ successfully reproduces the observed
value for $\OmegaDM$ quoted in Eq.~(\ref{eq:OmegaWMAP}).  This retroactively justifies 
our use of the rapid-turn-on approximation.     
   
Our second comment concerns the assumption that $\Gamma_\lambda$ is sufficiently 
small that the solution of Eq.~(\ref{eq:TheDoubleDotEqnWithGammaLambda}) for any given 
$a_\lambda$ includes a period during which this equation of motion is effectively 
underdamped.  This is critical, since in the absence of such a period, 
coherent oscillations cannot occur.  Indeed, the energy density 
$\rho_\lambda$ stored in any $a_\lambda$ for which $\Gamma_\lambda \geq 2\lambda$ would
never scale in an appropriate manner for that field to behave like massive matter;
hence it would never contribute to $\OmegaDM$.  However,  
it is not difficult to demonstrate that this
situation essentially never arises in realistic bulk-axion scenarios, even 
for the most massive modes in a given tower.
For example, consider the case of a purely photonic axion with
$c_\gamma =1$.  In this case, it follows from Eq.~(\ref{eq:GammaDecayToPhotons})
that the solution for $a_\lambda$ becomes critically damped at a value
$\lambda_{\mathrm{cd}}$, which is determined by the condition 
\begin{equation}
  2\lambda_{\mathrm{cd}} ~=~ G_\gamma 
     \frac{\lambda_{\mathrm{cd}}^3}{\fhatX^2}(\wtl^2 A_\lambda)^2~.  
\end{equation} 
Solving this equation for $\lambda_{\mathrm{cd}}$, we find that
\begin{equation}
  \lambda_{\mathrm{cd}} ~=~ \frac{\fhatX}{\sqrt{2G_\gamma}}
  \biggg[1 + \sqrt{1+\frac{4G_\gamma \mX^2}{\fhatX^2}
    \left(1+\frac{\pi^2}{y^2}\right)}\biggg]^{1/2}~,
\end{equation}
which implies that $\lambda_{\mathrm{cd}} \geq \fhatX/\sqrt{2G_\gamma}$.
However, since the effective description of the theory in terms of a tower
of axion modes breaks down at the five-dimensional $U(1)_X$-breaking scale
$f_X \ll \fhatX$, we are assured that  
$\Gamma_\lambda \ll 2\lambda$ for all modes in such a tower.  Indeed, this 
qualitative result is not specific to a photonic axion, but applies
broadly to any axion field which couples to the SM fields with 
$\mathcal{O}(1)$ coupling coefficients.  As a corollary,  
this result also implies that the standard oscillation criterion 
$\lambda \sim 3H/2$ will always be met before the decay criterion 
$\Gamma_\lambda \sim H$.
This indicates that indeed $H$, rather than 
$\Gamma_\lambda$, sets the time scale at which oscillations begin, and that
any given $a_\lambda$ decays during a time frame in which its energy density 
can legitimately be described by Eq.~(\ref{eq:RhoOftEqnWithR}). 


\section{Characterizing the Ensemble:~ Total Abundances,\protect\\ Tower 
Fractions, and Equations of State\label{sec:Abundances}}


In the previous two sections, we derived expressions for the decay widths 
and relic abundances for the individual mass eigenstates $a_\lambda$ in a mixed tower of KK
axions.  We have shown that these quantities scale with $\lambda$ in an
appropriate, reciprocal manner for an ensemble of such states to serve as 
dynamical dark matter.  We are therefore finally equipped to address 
the dark-matter phenomenology of the ensemble as a whole.  

As discussed in  
Ref.~\cite{DynamicalDM1}, the crucial quantities which characterize a given 
dynamical dark-matter ensemble are the total relic 
abundance $\Omegatot$, the tower fraction $\eta$, and the effective 
equation-of-state parameter $w_{\mathrm{eff}}$.  In this section, we 
investigate how these three quantities depend on the scales  
$\fhatX$, $M_c$, and $\Lambda_G$ which characterize a given bulk-axion model 
and thereby assess which regions of parameter 
space are interesting from a dynamical dark-matter perspective.  
In the next section,
we discuss the applicable phenomenological constraints on the model and demonstrate
that substantial regions of parameter space exist within which all such constraints 
are satisfied.   

\subsection{General Definitions}

The first of the three principal quantities mentioned above which characterize
any given dynamical dark-matter ensemble is $\Omegatot$.  This 
is simply the total contribution to $\OmegaDM$ from
all constituent modes in the ensemble which have already begun 
oscillating: 
\begin{equation}
  \Omegatot ~ \equiv~ \sum_\lambda \Omega_\lambda~.
\end{equation}
The second is the so-called ``tower fraction'' $\eta$, which is a measure of
how the total abundance $\Omegatot$ is distributed across the ensemble.  
Specifically, $\eta$ is defined for a given dynamical dark-matter ensemble to be the 
fraction of $\Omegatot$ provided by all of the oscillating components of that ensemble 
except for the one which yields the largest individual contribution.  Explicitly,
\begin{equation}
  \eta ~\equiv~ 1 - \frac{\Omega_{\mathrm{max}}}{\Omegatot}~,
  \label{eq:DefOfTowerFrac} 
\end{equation}
where $\Omega_{\mathrm{max}} \equiv \max_\lambda\{\Omega_\lambda\}$.  After
all the $a_\lambda$ have begun oscillating, the lightest mass eigenstate $a_{\lambda_0}$
always yields the largest individual relic abundance $\Omega_{\lambda_0}$, 
and therefore $\Omega_{\mathrm{max}} = \Omega_{\lambda_0}$.  When $\eta \ll 1$, 
essentially the entirety of $\Omegatot$ is provided by a single field, as in most 
traditional dark-matter models.  By contrast, having $\eta \sim \mathcal{O}(1)$ 
signals a departure from this traditional setup, which is indeed one of the 
hallmarks of dynamical dark matter. 
(Note that when we say that $\eta$ should differ significantly
from zero, we are willing to accept $\eta\sim 0.1$, as such values could 
result in observable differences from traditional models, but not, for example, 
$\eta\sim 10^{-3}$.)  

In our dynamical dark-matter
framework, both $\Omegatot$ and $\eta$ are 
intrinsically dynamical quantities, with non-trivial time dependences.
For this reason, we will designate their present-day values as
$\Omegatotnow \equiv \Omegatot(\tnow)$ and $\etanow\equiv\eta(\tnow)$ in what
follows.

As discussed in Ref.~\cite{DynamicalDM1}, a given dynamical 
dark-matter ensemble as a whole can also be described in terms of
a single, effective equation-of-state parameter $w_{\mathrm{eff}}$:
\begin{equation}
  w_{\mathrm{eff}}~\equiv~ -\left(\frac{1}{3H}
    \frac{d\ln\rho_{\mathrm{tot}}}{dt}+1\right)~,
\end{equation}
where $\rho_{\mathrm{tot}} \equiv \Omegatot\rhocrit$.
Indeed, one of the hallmarks of this framework is that even
$w_{\mathrm{eff}}$ itself is continually changing in time.  
We shall therefore define $w_\ast \equiv w_{\mathrm{eff}}(\tnow)$.  As 
discussed in Ref.~\cite{DynamicalDM1}, this quantity is given by
\begin{equation}
   w_\ast ~ = ~ 
       \frac{AB}{2\Omegatotnow \tnow^{1+\alpha+\beta}}
   \label{eq:wstar}
\end{equation}   
for any given dynamical dark-matter ensemble in which the widths, abundances,
and densities of states obey the approximate scaling relations
$\Omega \approx A\Gamma^\alpha$ and $n_\Gamma \approx B \Gamma^\beta$, where 
$n_\Gamma$ denotes the density of states per unit decay width.   

Taken together, $\Omegatot$, $\eta$, and $w_{\mathrm{eff}}$ serve to characterize
any given dynamical dark-matter ensemble.
For the remainder of this section, then, our task will be to investigate how 
these three quantities depend on the parameters $\fhatX$, $M_c$, and $\Lambda_G$ 
which characterize our bulk-axion model.  Also recall that in Eq.~(\ref{eq:DefsOfyandmPQ}) 
we defined the mixing parameter 
\begin{equation}
  y ~=~ \frac{4\sqrt{2}\pi}{g_G\xi}\frac{\fhatX M_c}{\Lambda_G^2}~,
\end{equation}   
which quantifies the 
extent to which the different modes in the KK tower mix with each other.  We will also
therefore keep track of the corresponding values of $y$ in our analysis.
Moreover, as we have seen in Sect.~\ref{sec:Production},
our results will also depend on the cosmological framework 
adopted.  We shall therefore derive results in the context of the 
standard and LTR cosmologies independently.       
However, as discussed above, constraints on $T_\ast$ in theories with large, flat 
extra dimensions provide a strong motivation for working within the context of an 
LTR cosmology with a reheating temperature $\TRH\sim \mathcal{O}(\mathrm{MeV})$.  For 
this reason, the results obtained for the LTR cosmology are more likely to 
be realistic. 

Needless to say, phenomenological consistency imposes certain constraints on the 
parameters $\Omegatot$, $\eta$, and $w_{\mathrm{eff}}$.  For example, WMAP data 
require that $\Omegatotnow \approx \OmegaDM$; likewise, $w_\ast$ should not differ
too significantly from zero.  Beyond this, however, $\eta$ and $w_{\mathrm{eff}}$
are fairly unconstrained by data.  Nevertheless, while any values for $\eta$ 
and $w_{\mathrm{eff}}$ can be realized within the general dynamical dark-matter framework, 
we are particularly interested in situations in which $\etanow$ is also significantly 
different from zero, for these are the situations in which our dynamical 
dark-matter ensemble represents a significant departure from traditional, single-component 
models of dark matter.

\subsection{Dark Towers: Relic Abundances and Tower Fractions}

\begin{figure}[p]
~\vskip 0.5 truein
\begin{center}
  \epsfxsize 2.25 truein \epsfbox {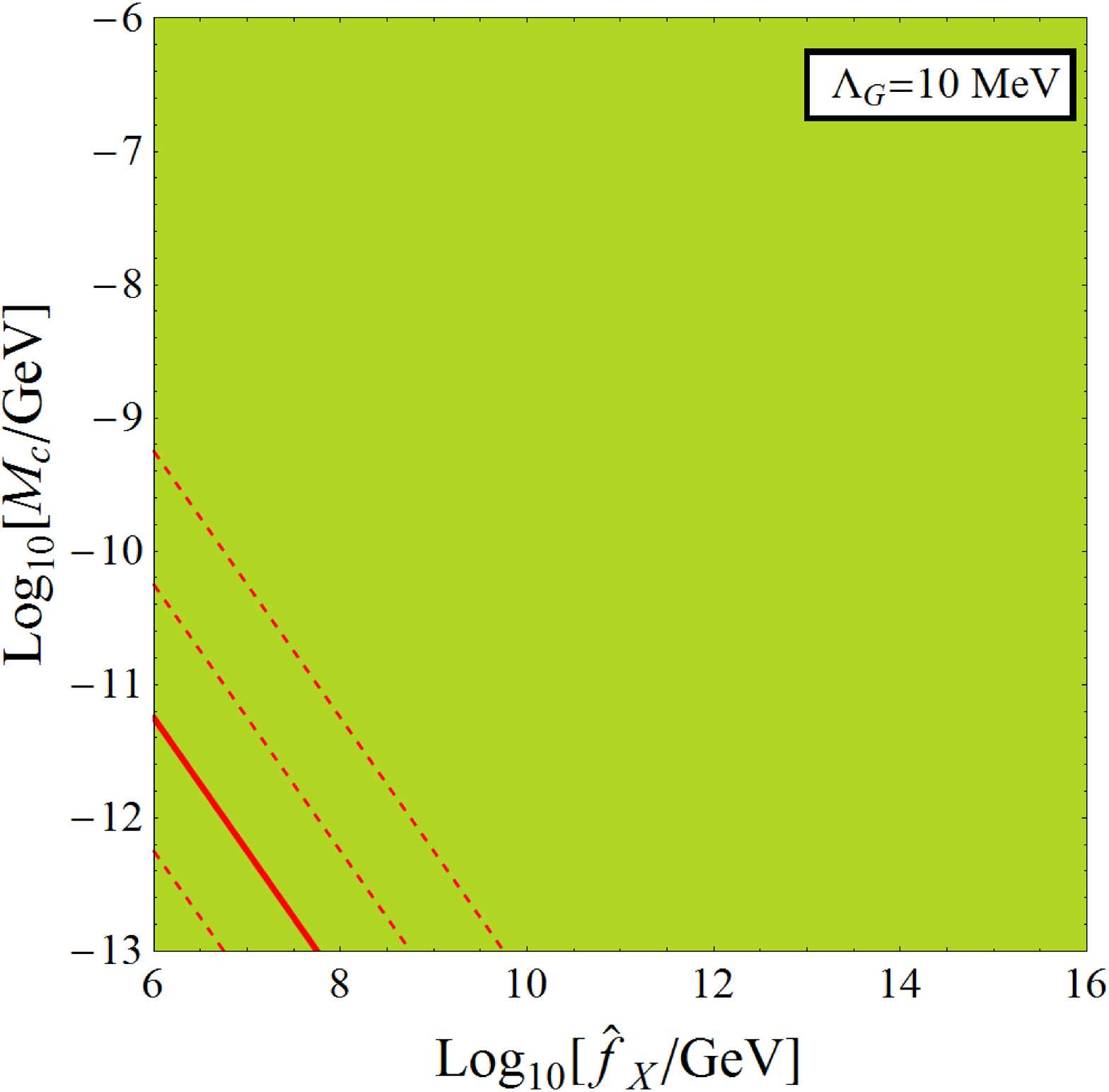} 
  \epsfxsize 2.25 truein \epsfbox {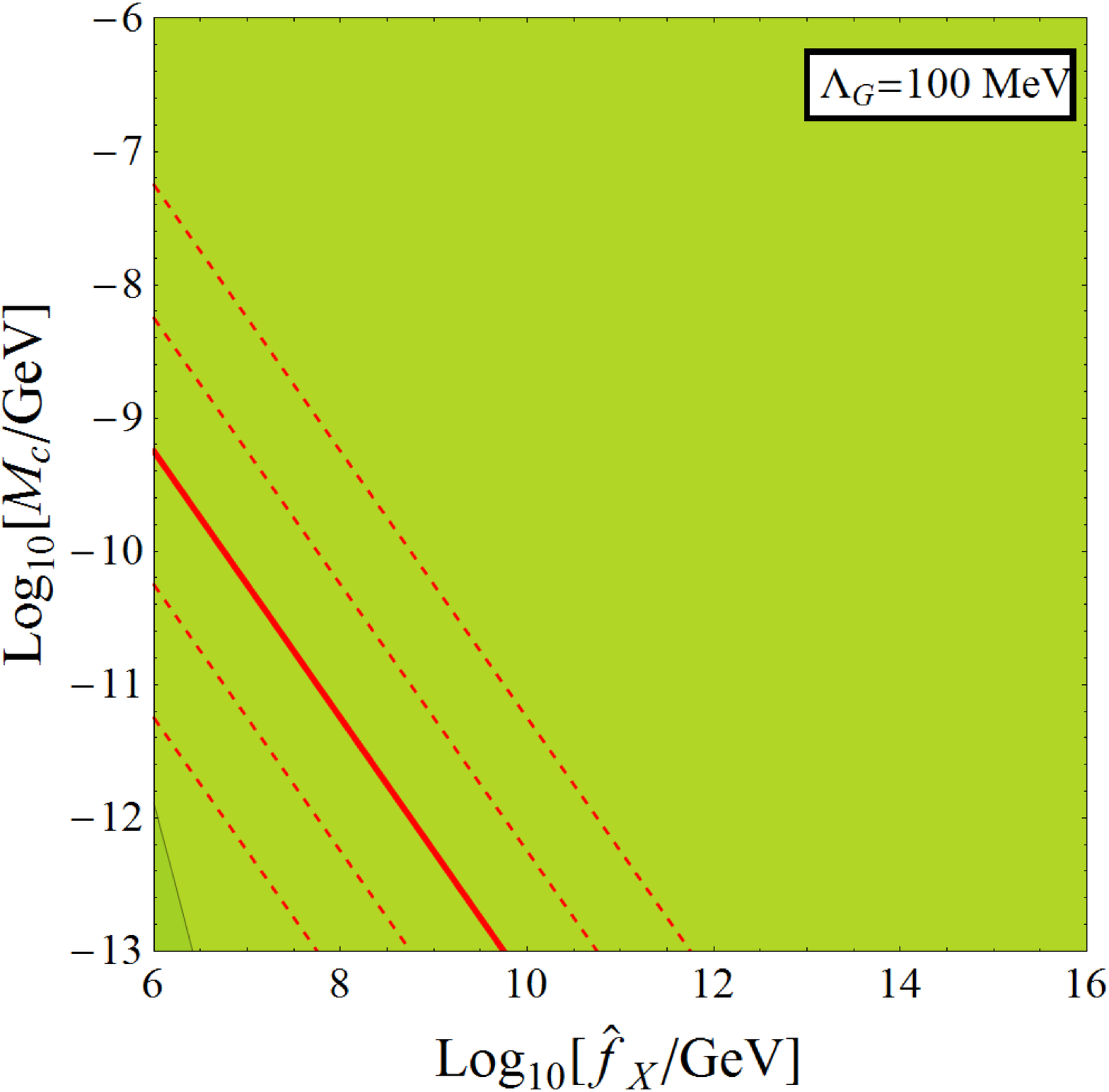}
  \epsfxsize 2.25 truein \epsfbox {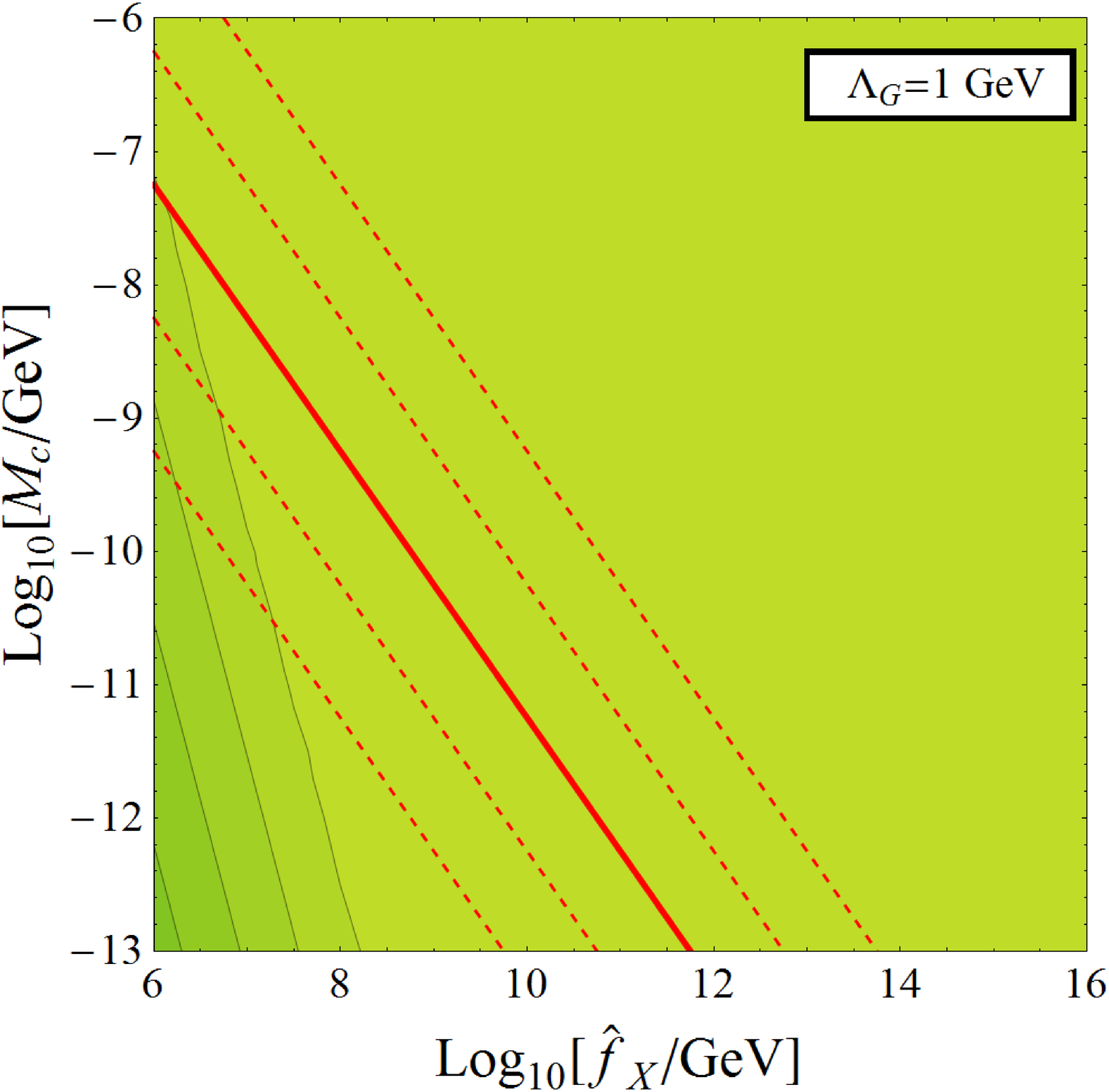}\\
  \epsfxsize 2.25 truein \epsfbox {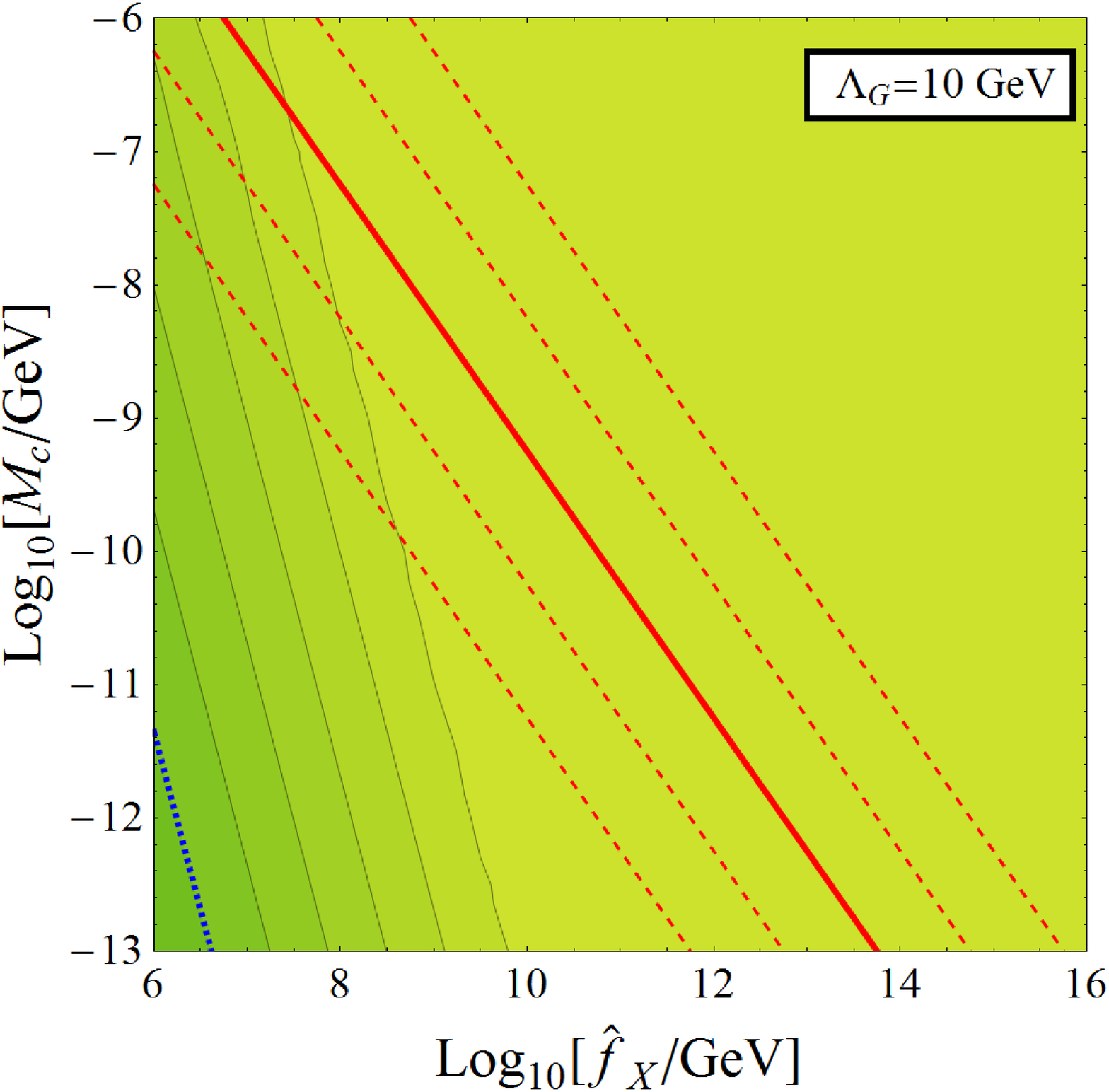}
  \epsfxsize 2.25 truein \epsfbox {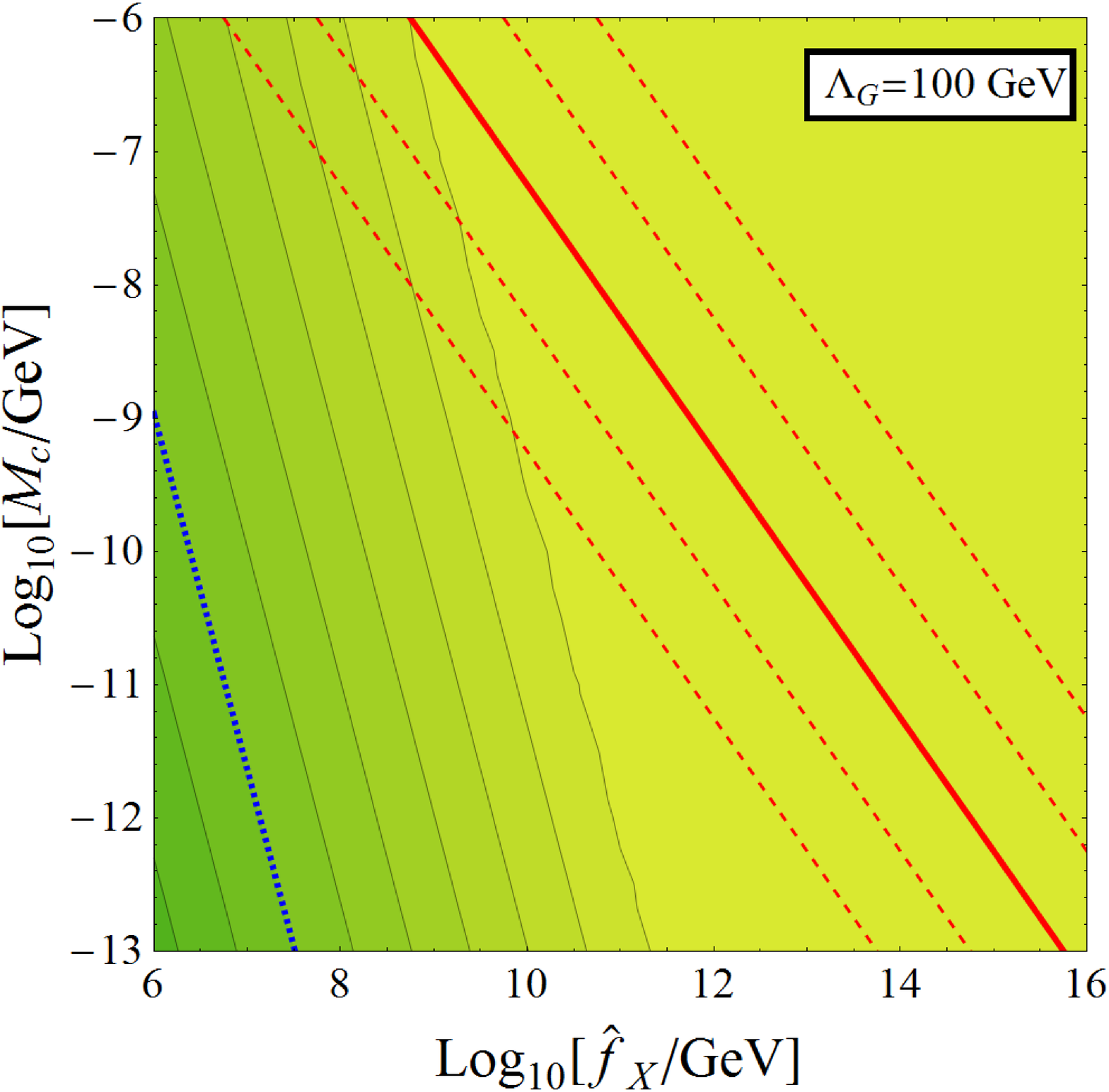}
  \epsfxsize 2.25 truein \epsfbox {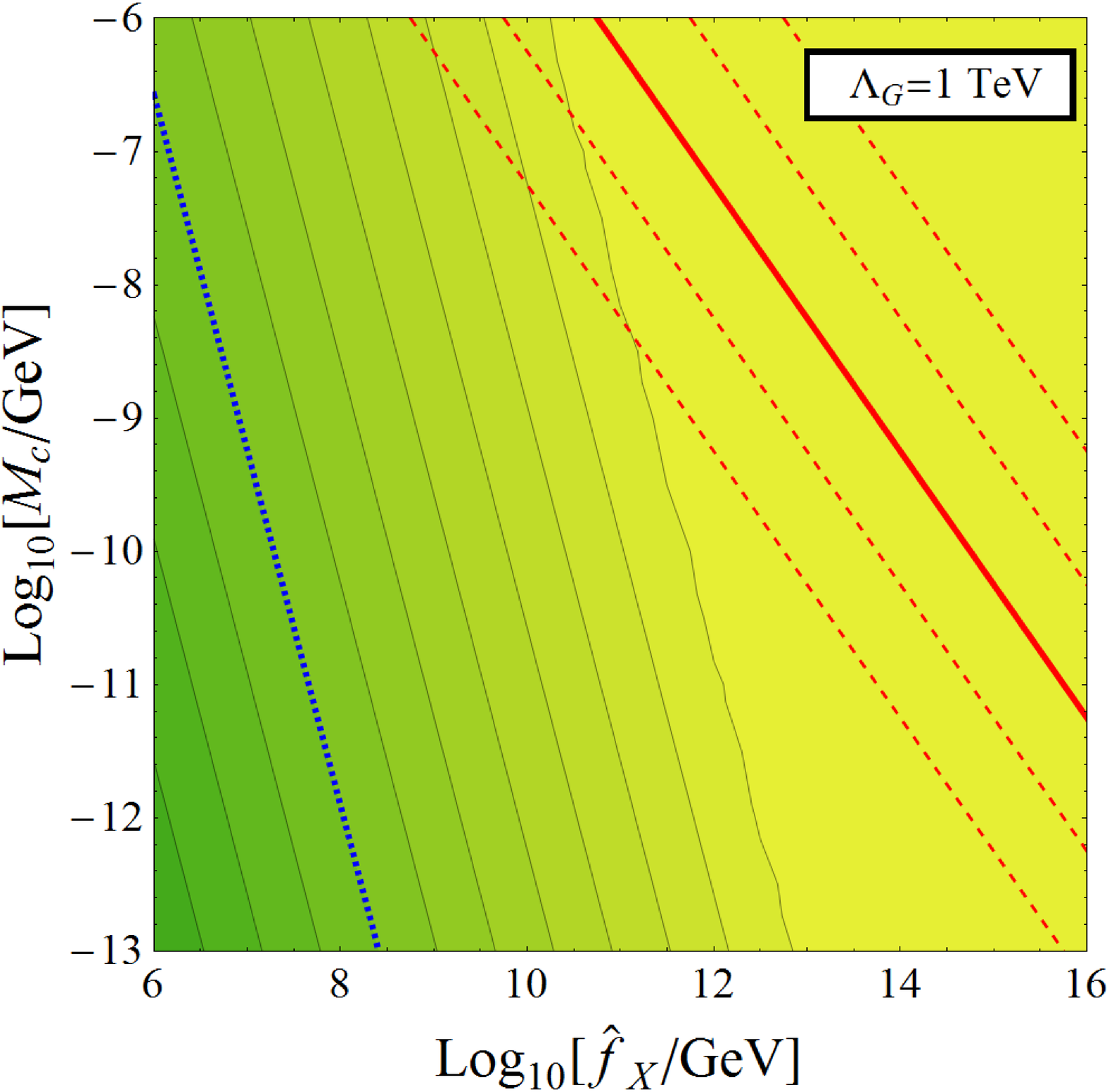}\\
  \epsfxsize 2.25 truein \epsfbox {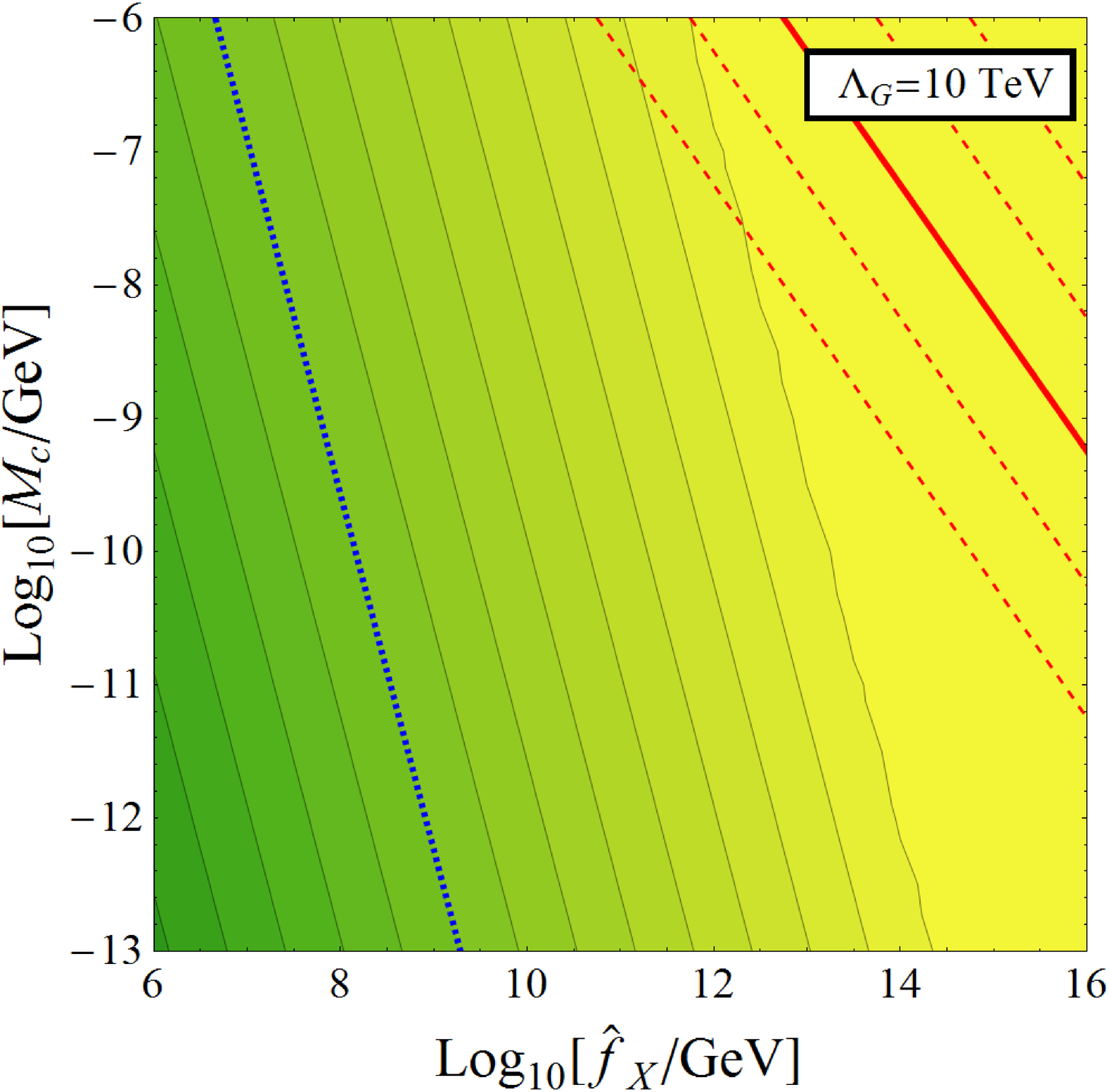}
  \epsfxsize 2.25 truein \epsfbox {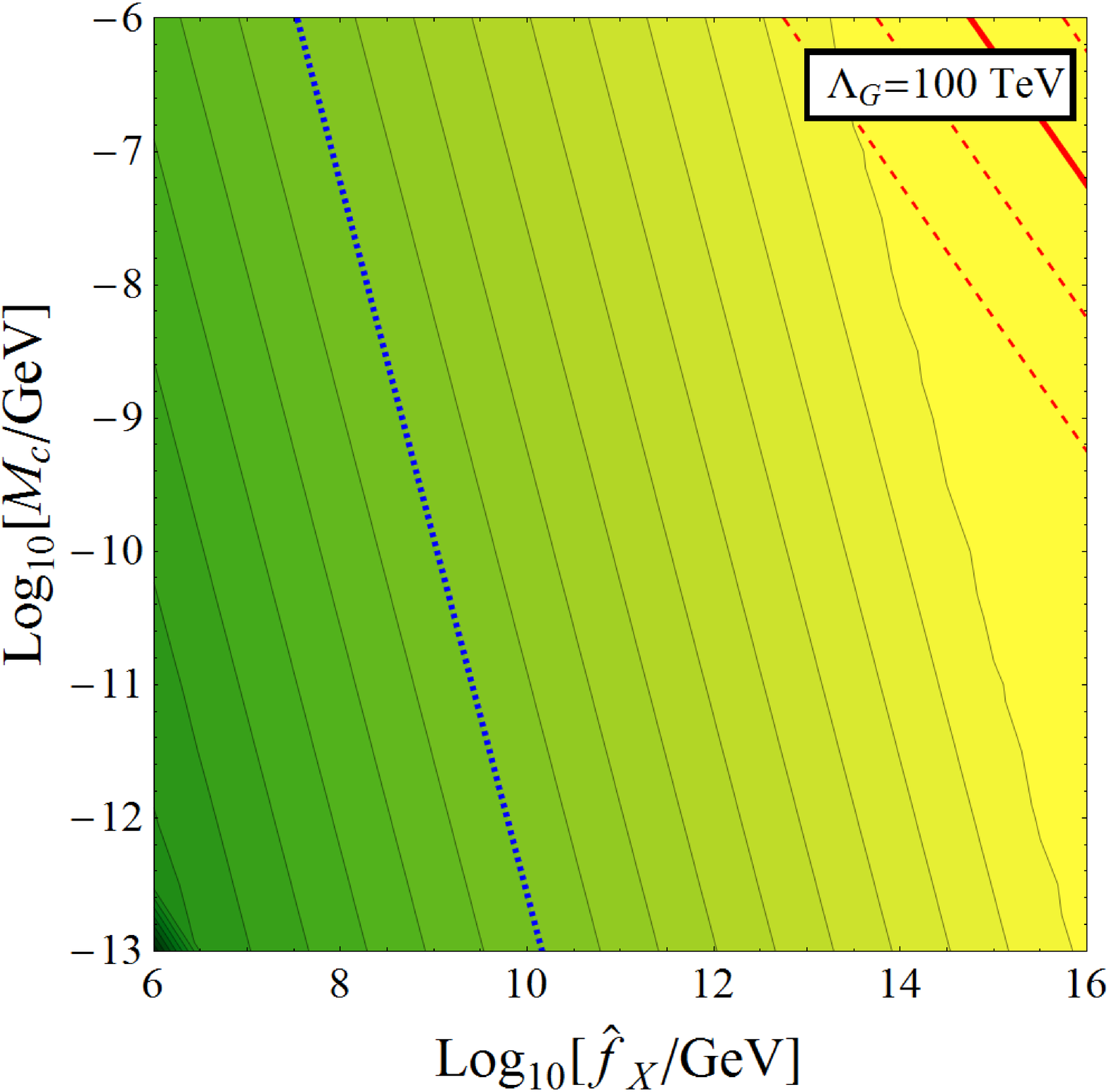}\\
  \raisebox{0.3cm}{\large$\Omegatotnow$}\epsfxsize 6.00 truein \epsfbox {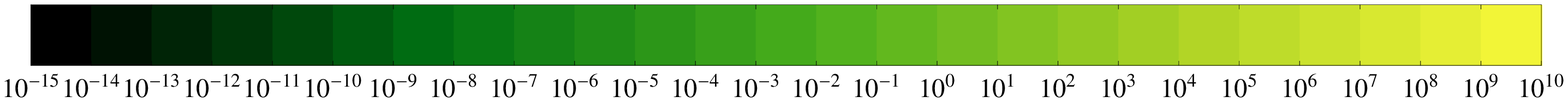}
\end{center}
\caption{Contours of the total contribution $\Omegatotnow$ from a KK tower of
general axions, plotted in $(\fhatX, M_c)$ space,
assuming the standard cosmology.  Each panel
corresponds to a different choice of $\Lambda_G$ ranging from 10~MeV to 100~TeV.~  
In each case, we have taken $\xi = g_G = \theta = 1$ and set $H_I = 10^{-7}$~GeV.~ 
Note that the contour corresponding to $\Omegatotnow = 1$ 
has been highlighted with a dotted blue line in each panel for clarity.  
The solid oblique red line appearing in each panel indicates 
where $y=1$, and proceeding from left to right, 
the dashed red lines correspond to $y = \{0.01,0.1,10,100\}$.   
\label{fig:OmegaTotPanelsStd}}
\end{figure}

\begin{figure}[p]
~\vskip 0.5 truein
\begin{center}
  \epsfxsize 2.25 truein \epsfbox {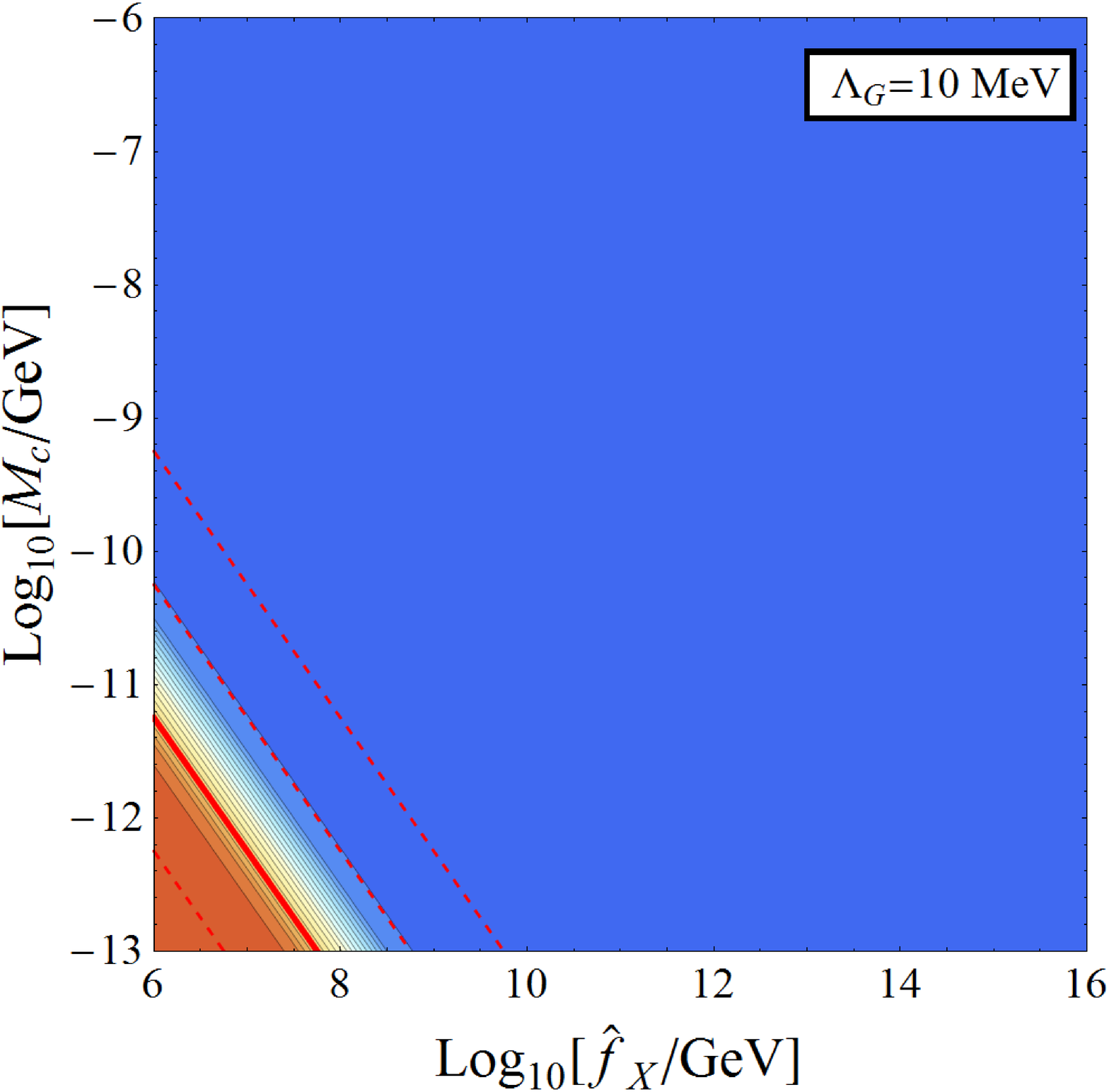} 
  \epsfxsize 2.25 truein \epsfbox {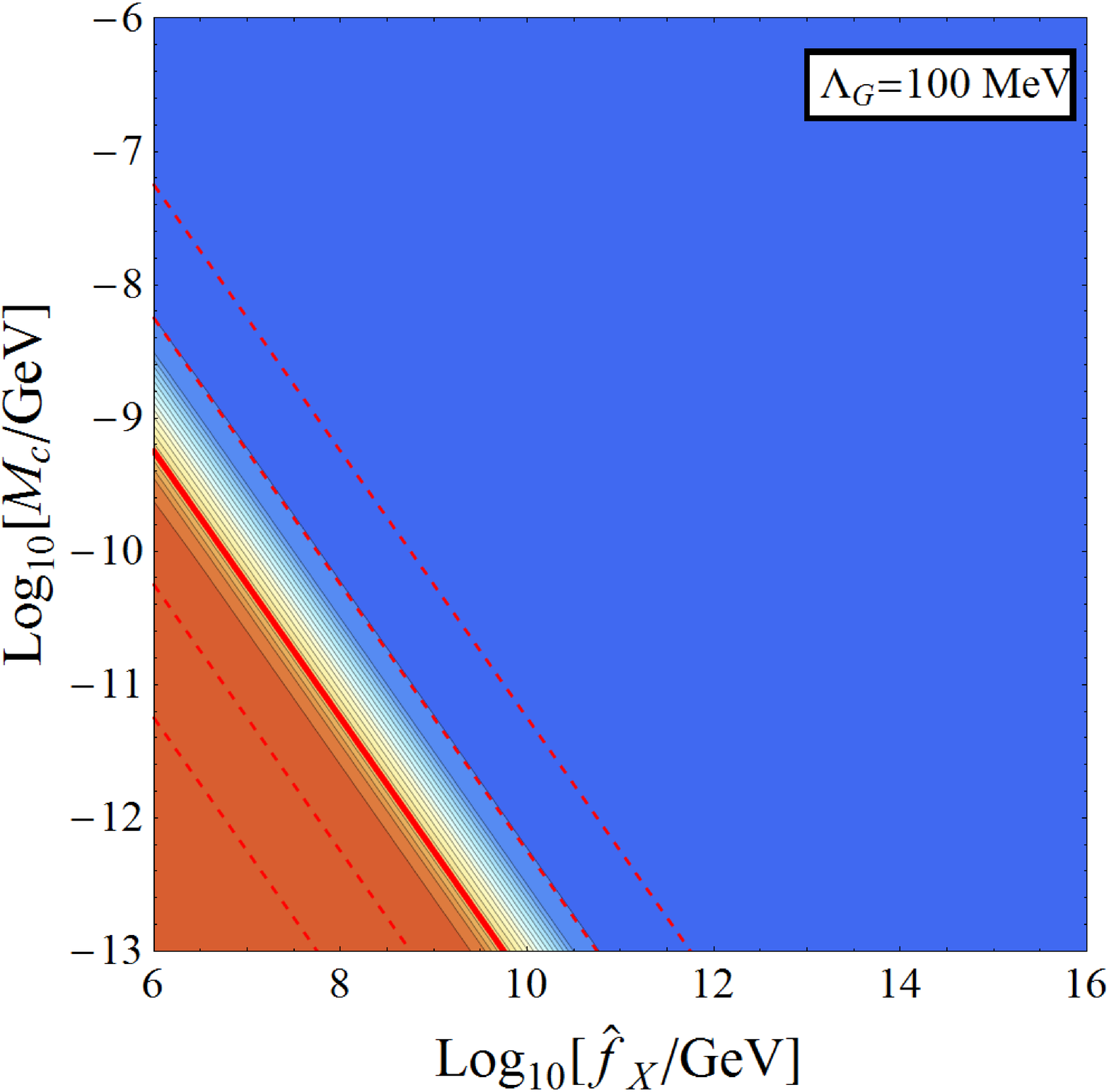}
  \epsfxsize 2.25 truein \epsfbox {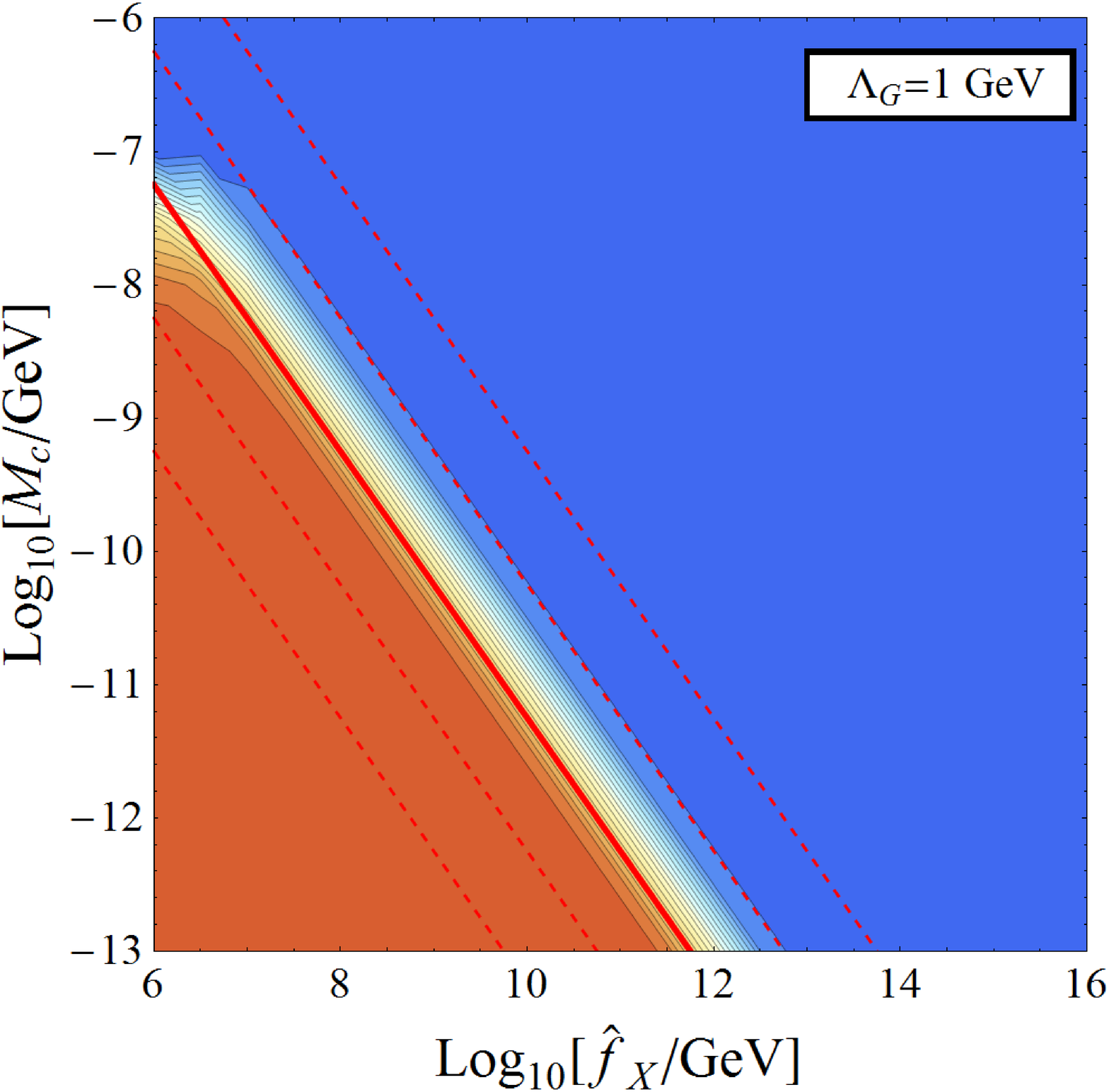}\\
  \epsfxsize 2.25 truein \epsfbox {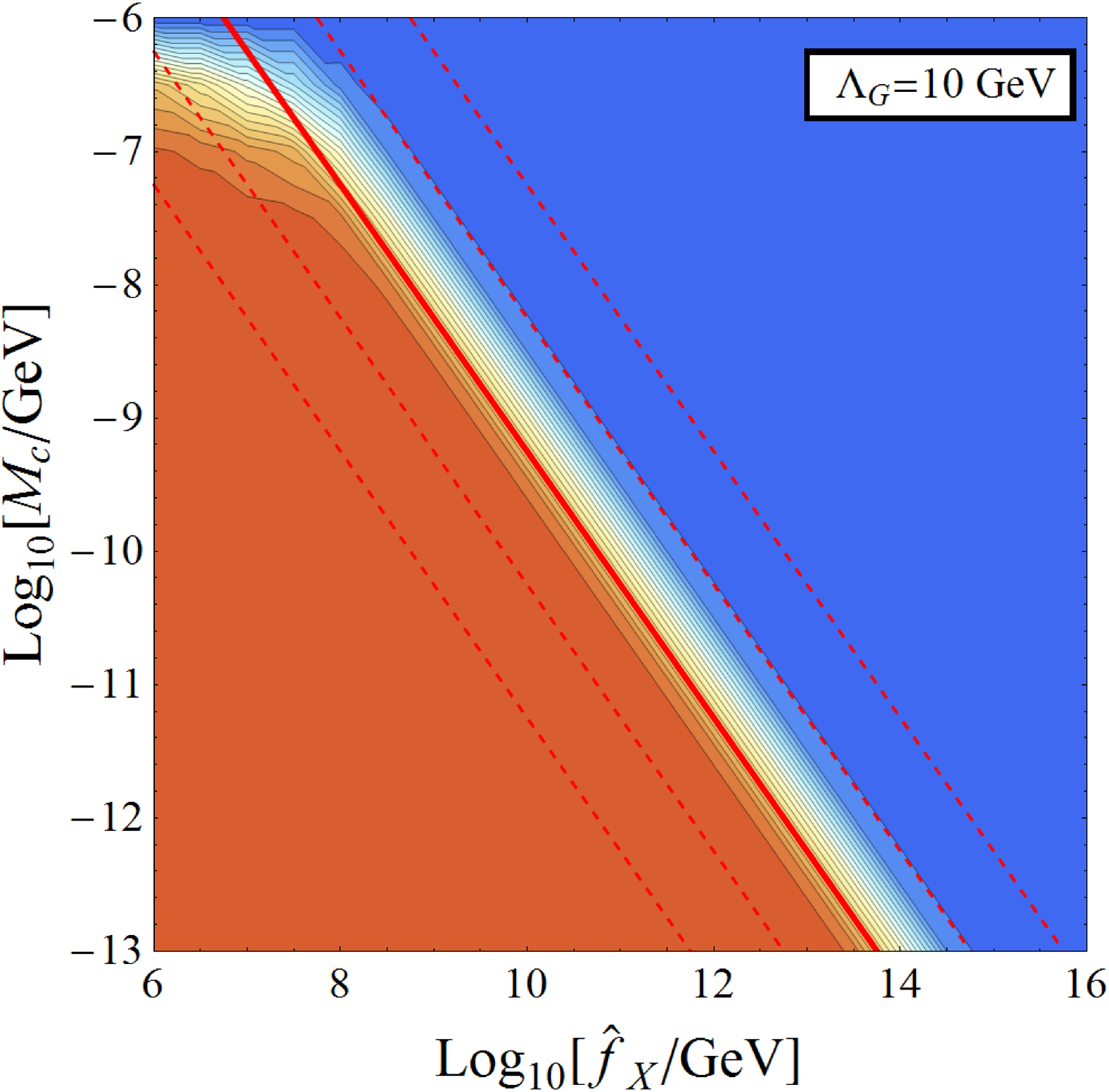}
  \epsfxsize 2.25 truein \epsfbox {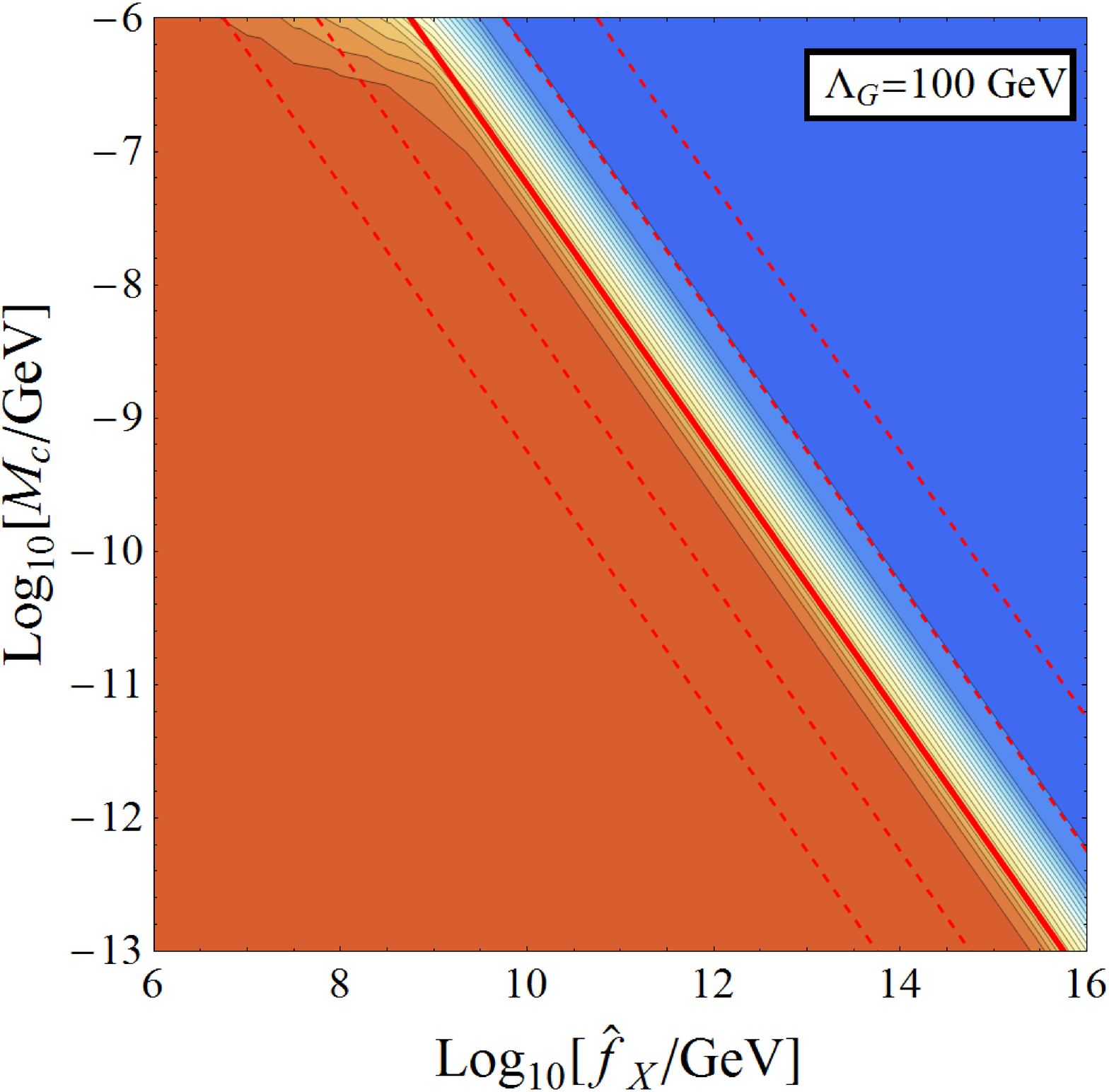}
  \epsfxsize 2.25 truein \epsfbox {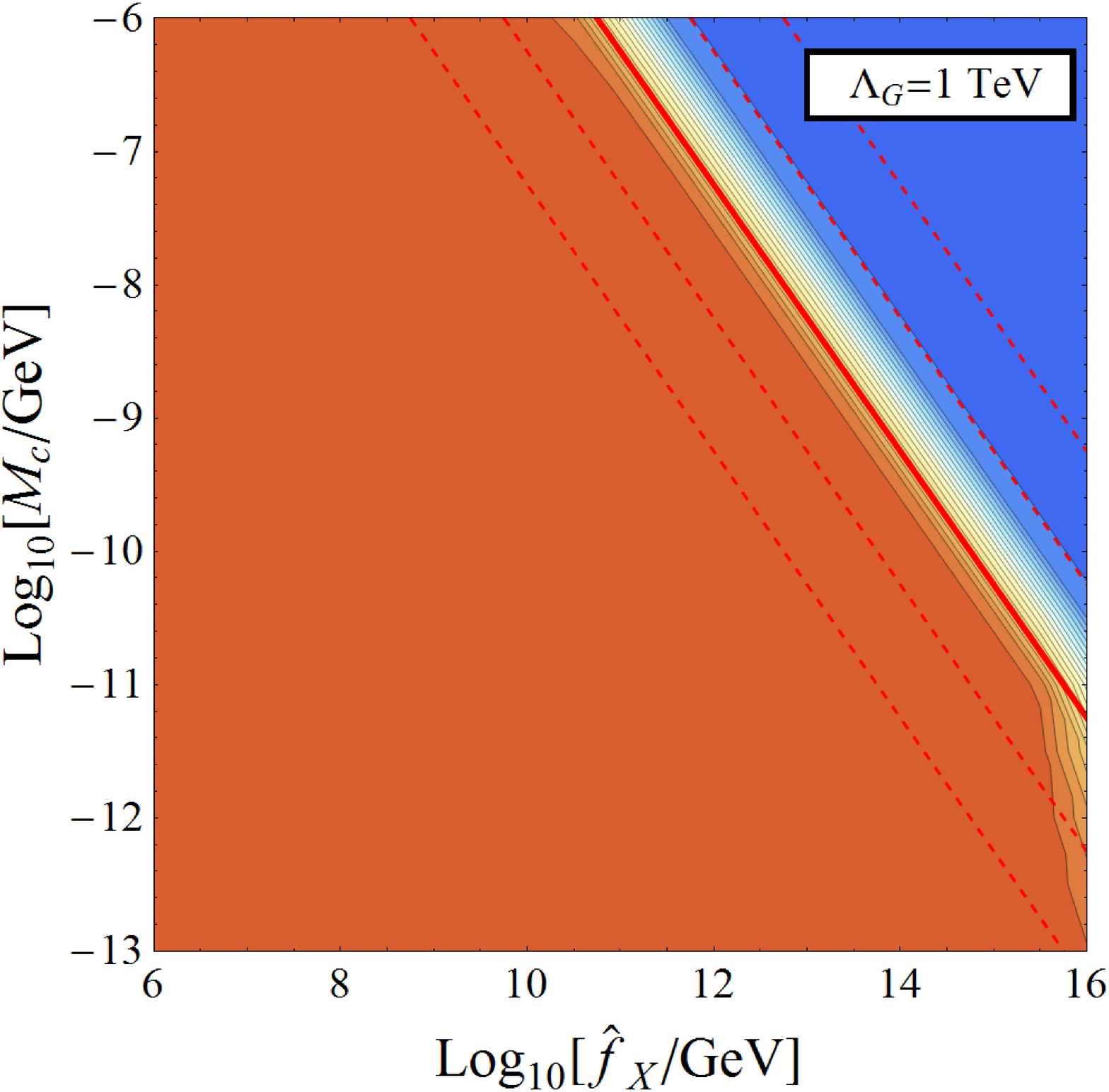}\\
  \epsfxsize 2.25 truein \epsfbox {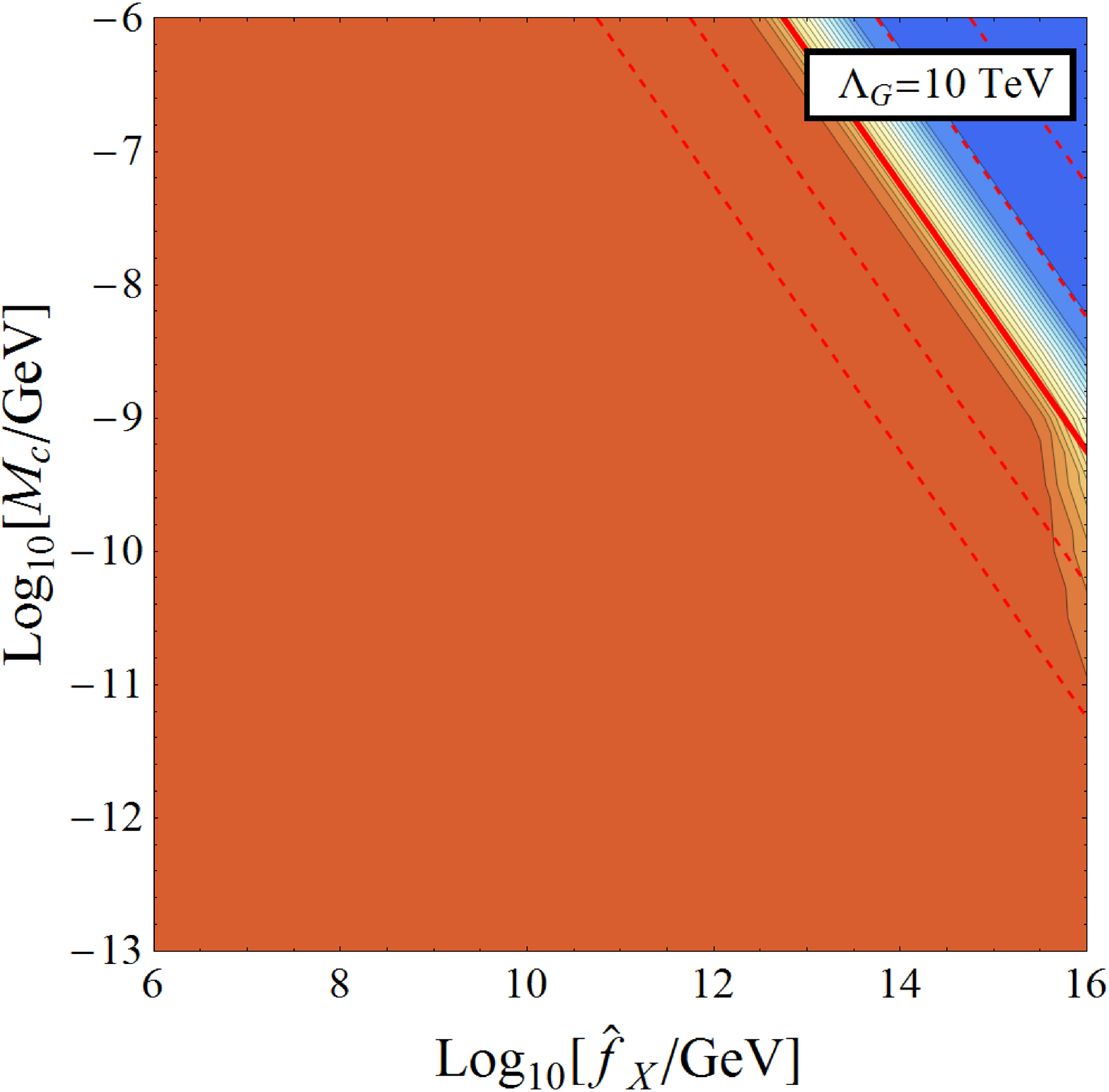}
  \epsfxsize 2.25 truein \epsfbox {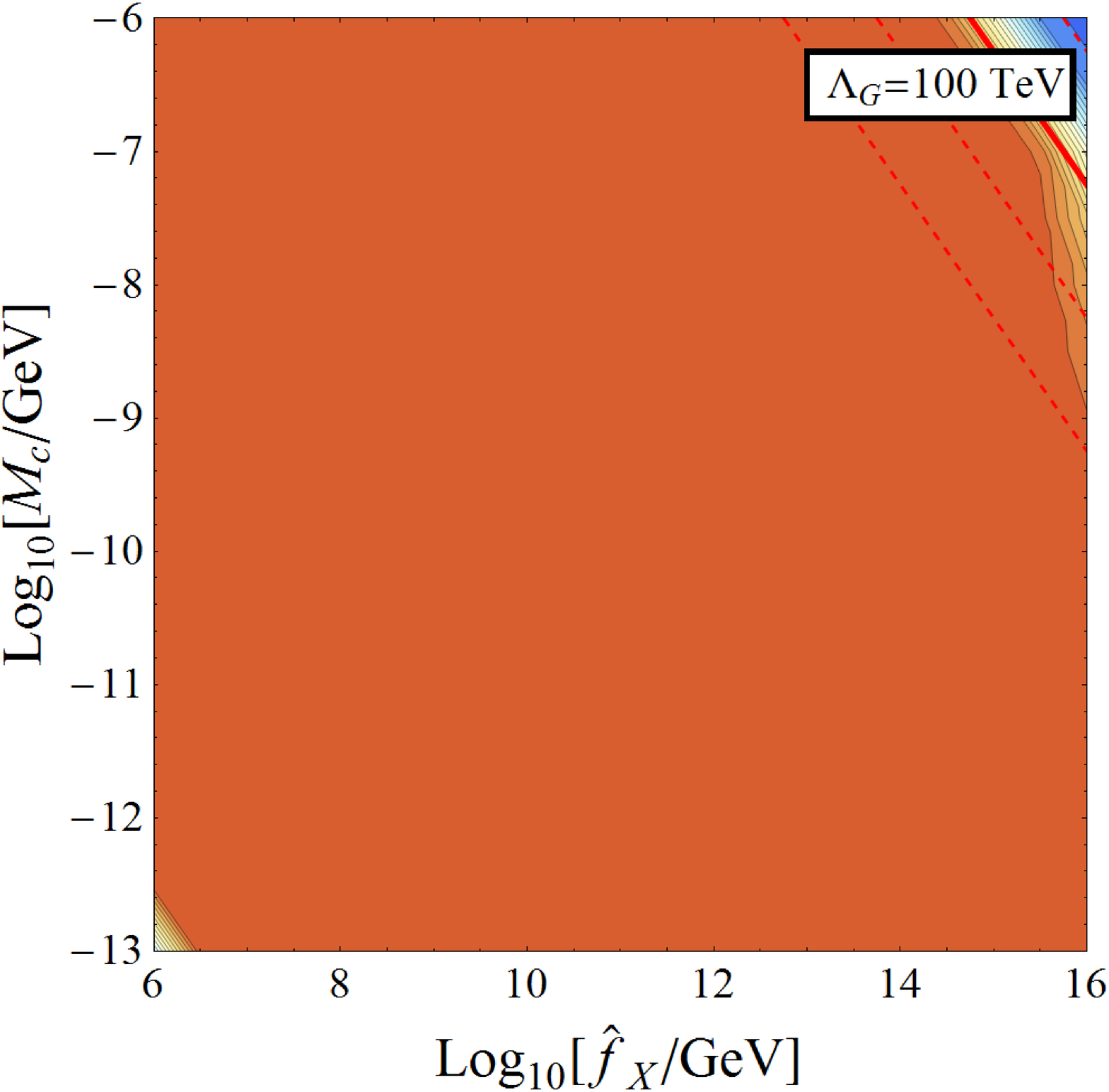}\\
  \raisebox{0.3cm}{\large $\etanow$}~\epsfxsize 5.00 truein \epsfbox {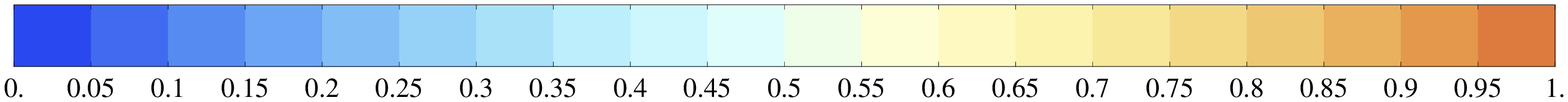}
\end{center}
\caption{
Contours of the tower fraction $\etanow$ from a KK tower of
general axions, plotted in $(\fhatX, M_c)$ space,
assuming the standard cosmology.  As in Fig.~\protect\ref{fig:OmegaTotPanelsStd}, each panel
corresponds to a different choice of $\Lambda_G$ ranging from 10~MeV to 100~TeV; likewise,  
we have taken $\xi = g_G = \theta = 1$ and set $H_I = 10^{-7}$~GeV.~   
Once again, the solid oblique red line appearing in each panel indicates 
where $y=1$, and proceeding from left to right, the dashed red lines 
correspond to $y = \{0.01,0.1,10,100\}$. 
\label{fig:EtaPanelsStd}}
\end{figure}

\begin{figure}[p]
~\vskip 0.75 truein
\begin{center}
  \epsfxsize 2.25 truein \epsfbox {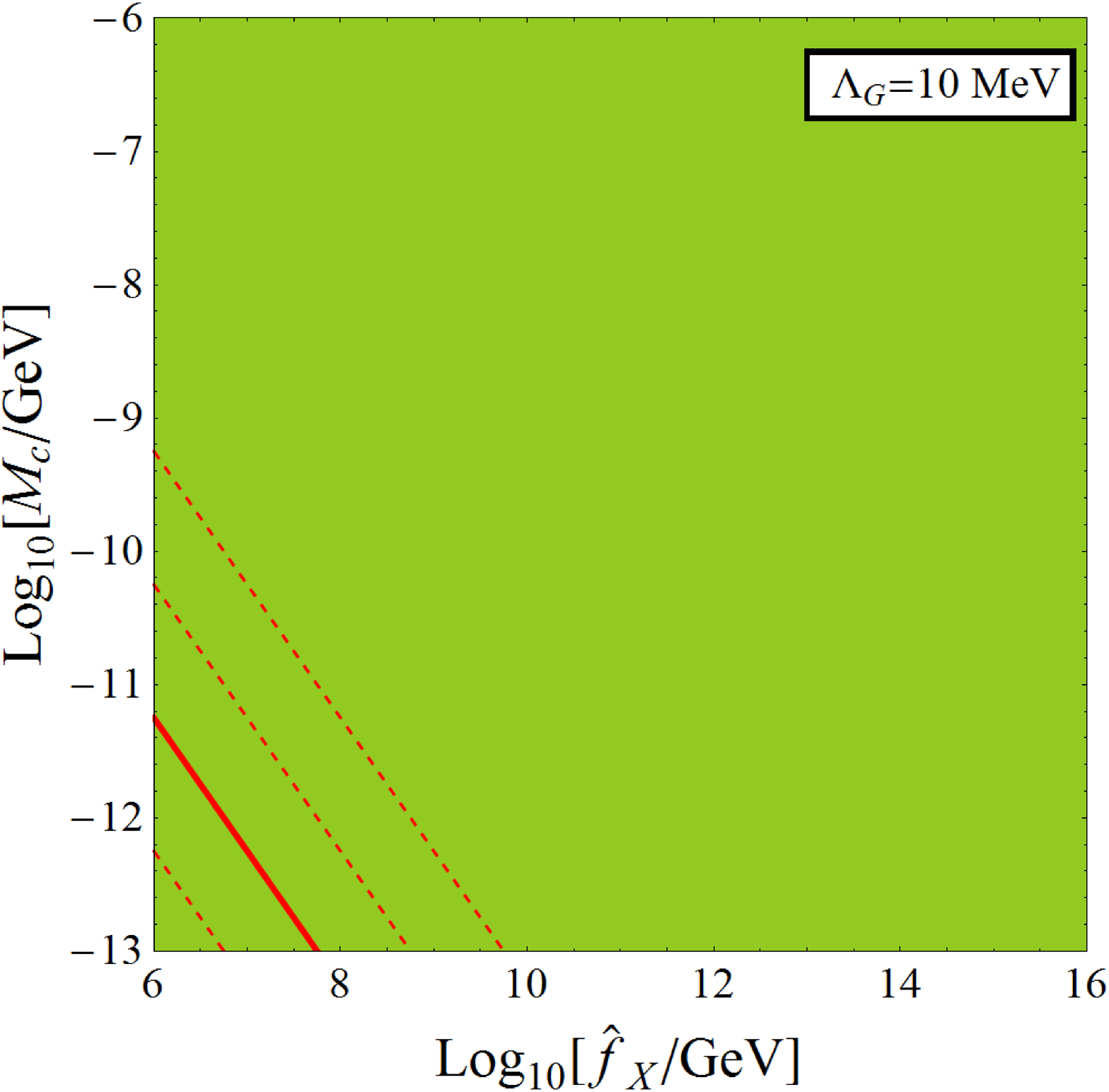} 
  \epsfxsize 2.25 truein \epsfbox {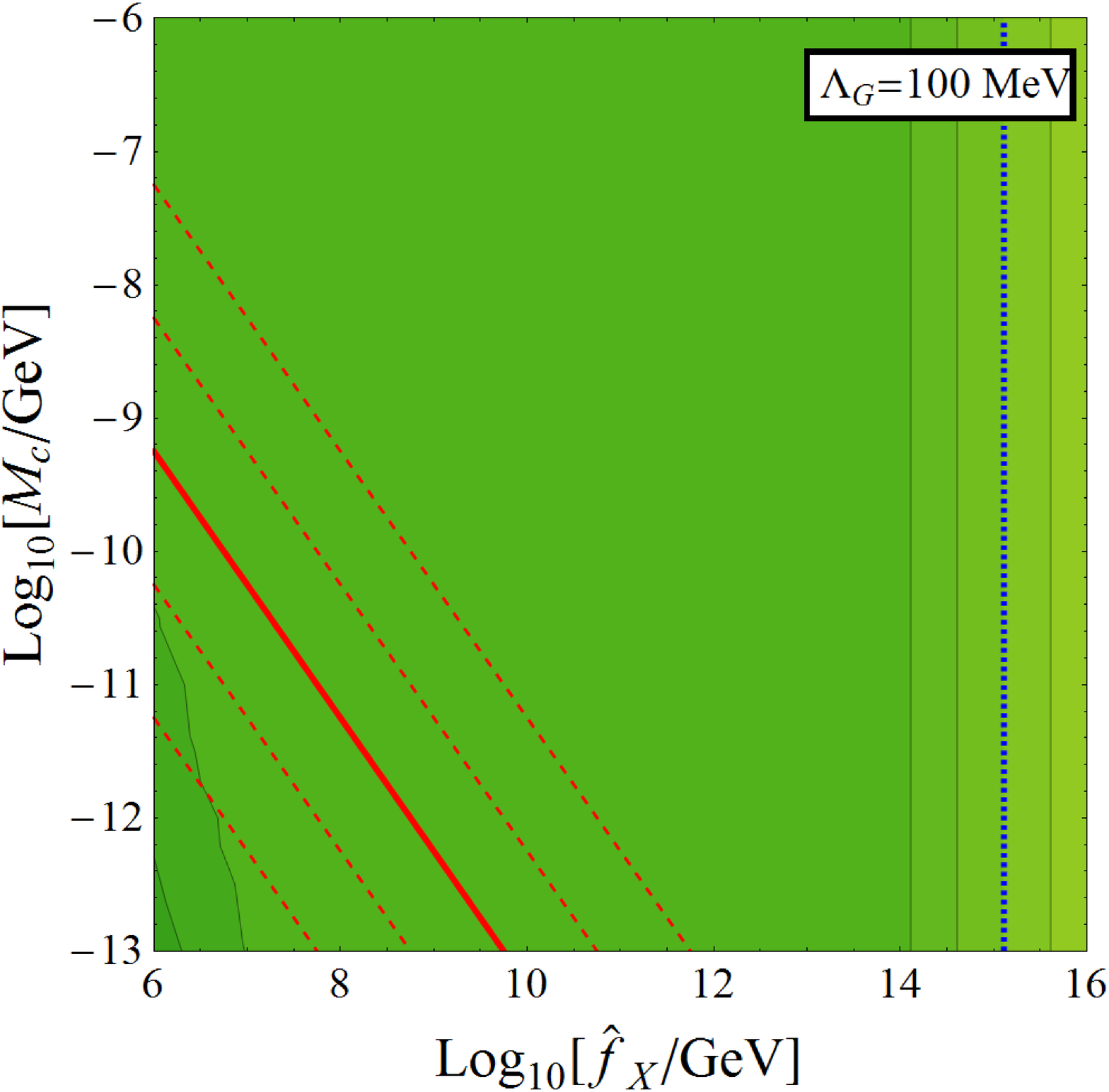}
  \epsfxsize 2.25 truein \epsfbox {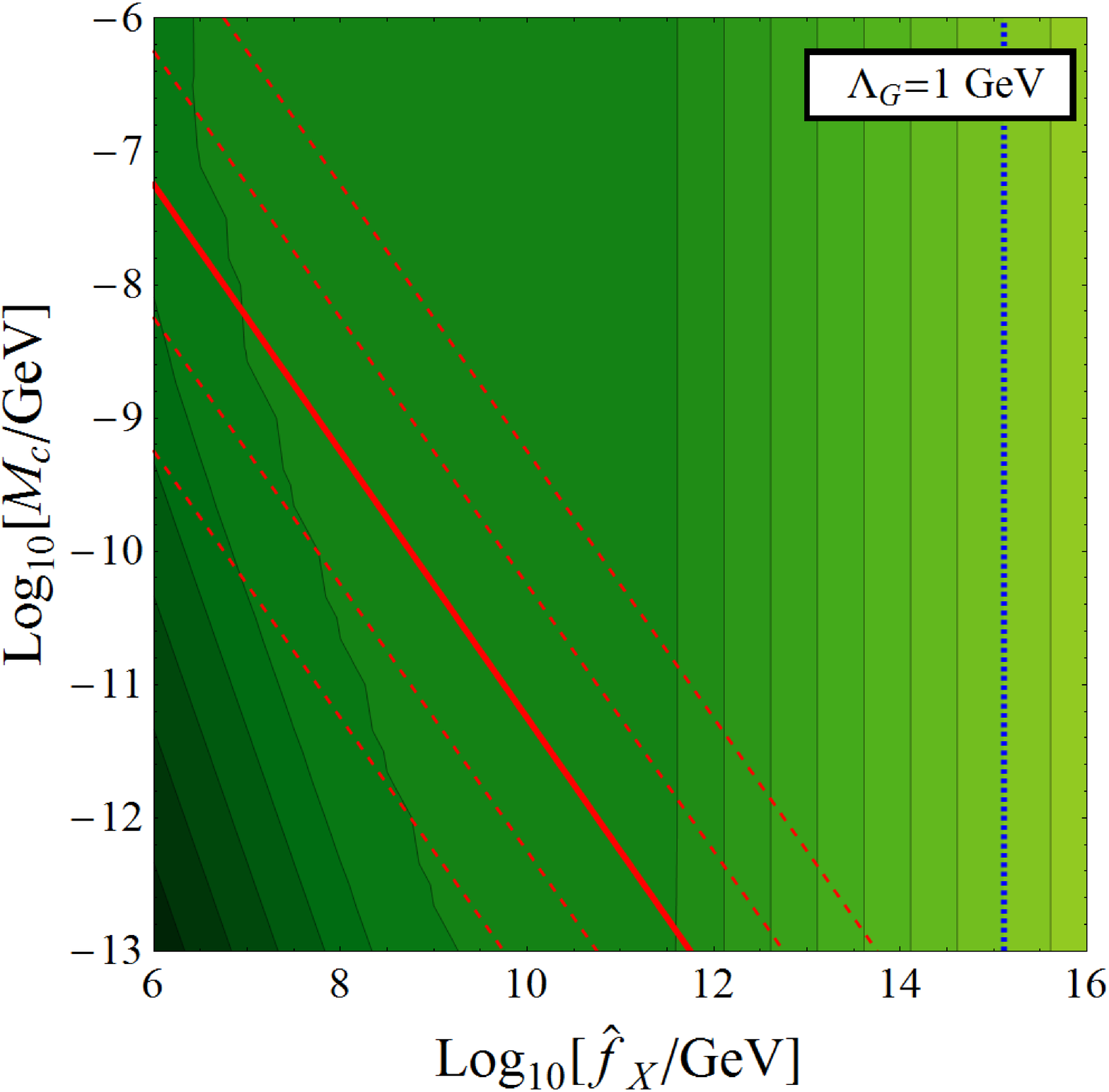}\\
  \epsfxsize 2.25 truein \epsfbox {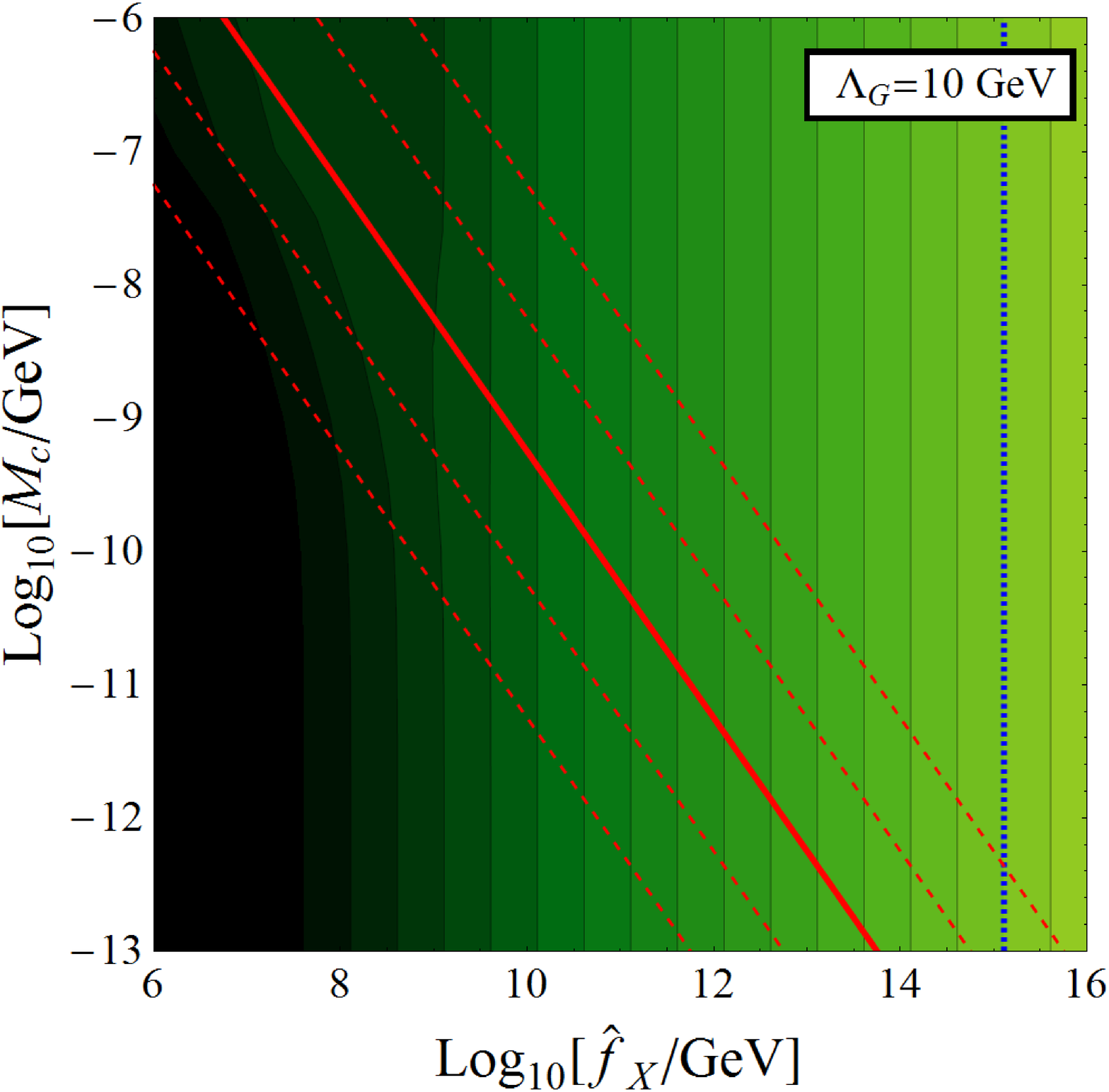}
  \epsfxsize 2.25 truein \epsfbox {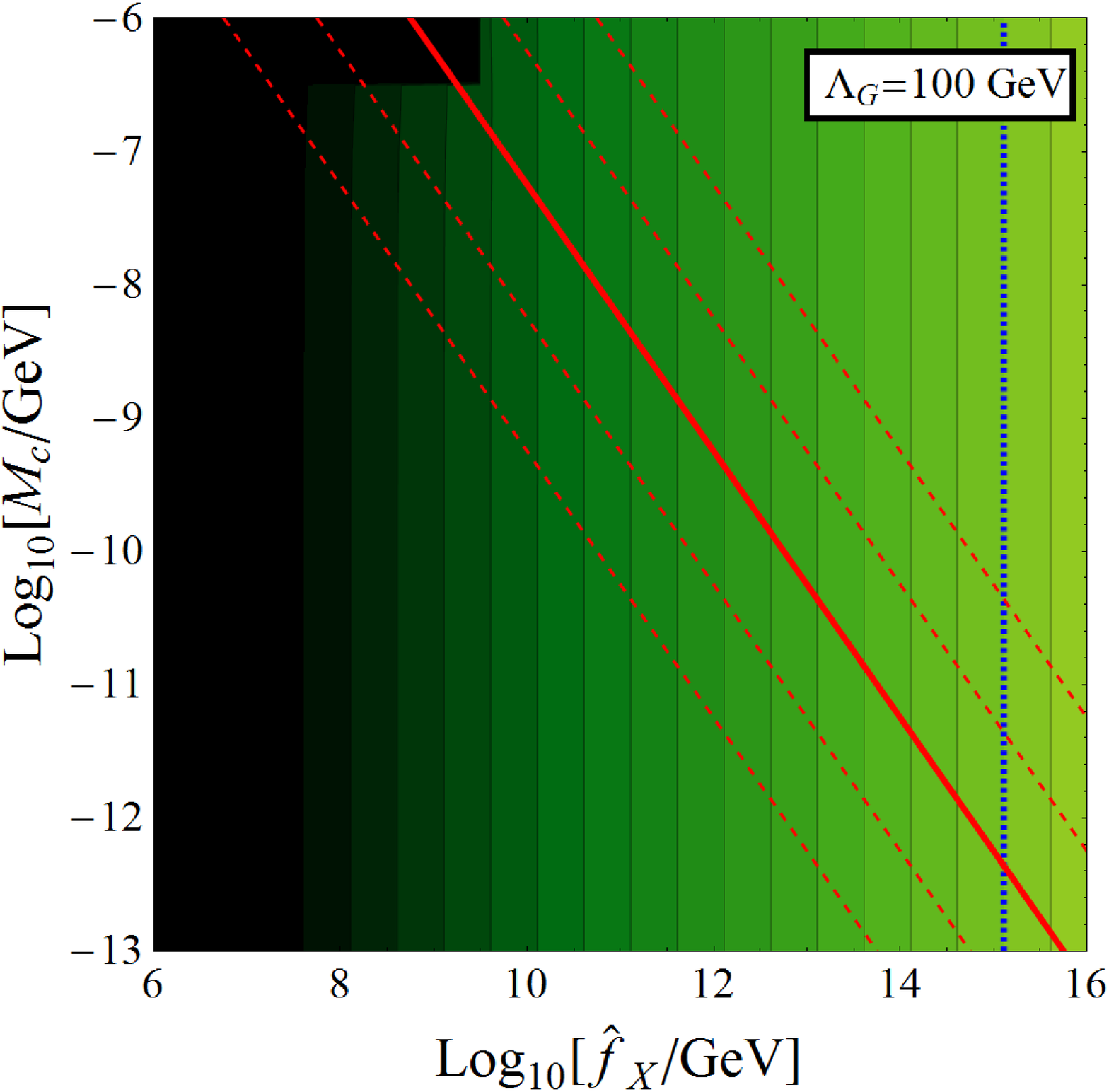}
  \epsfxsize 2.25 truein \epsfbox {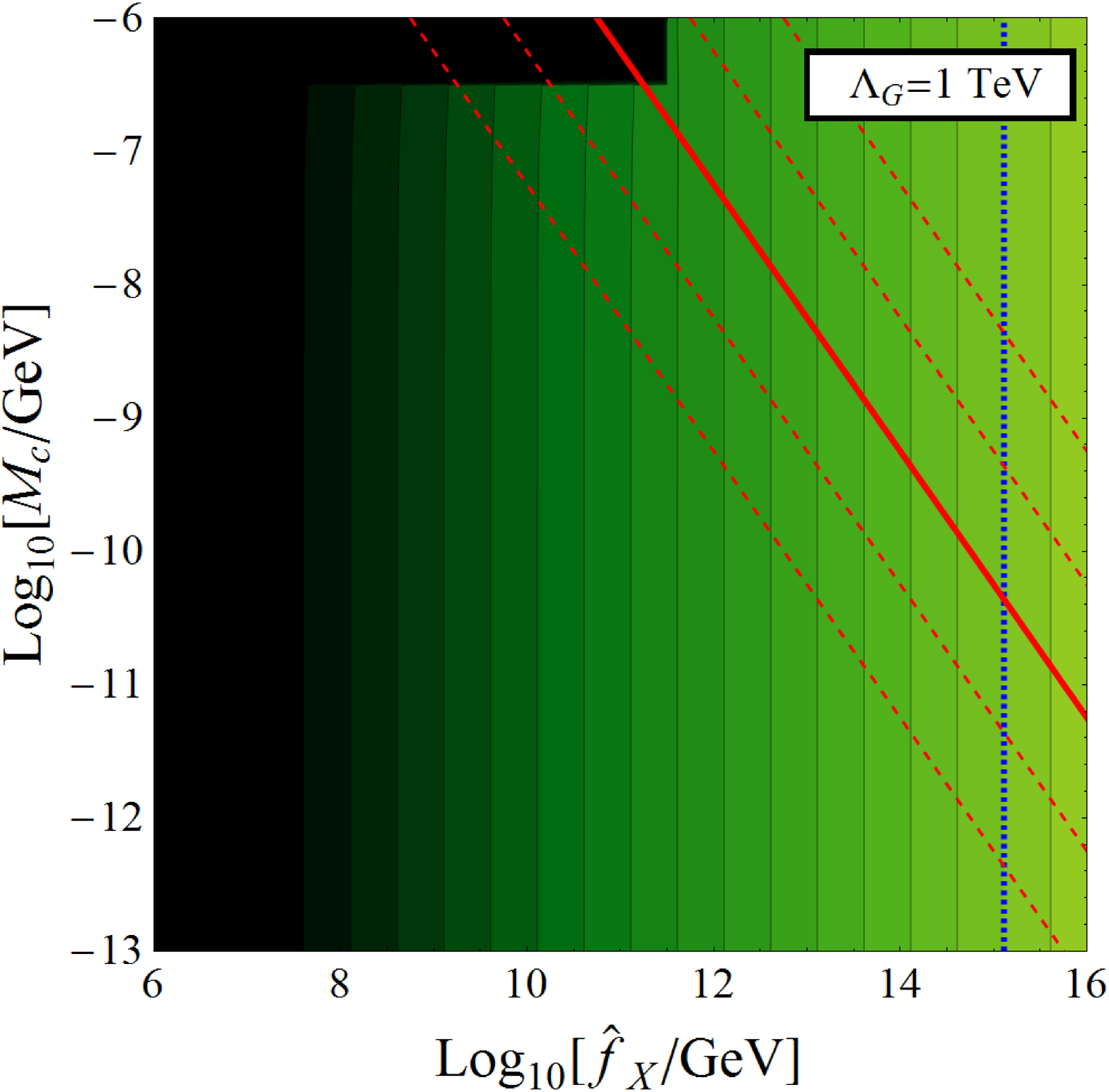}\\  
  \epsfxsize 2.25 truein \epsfbox {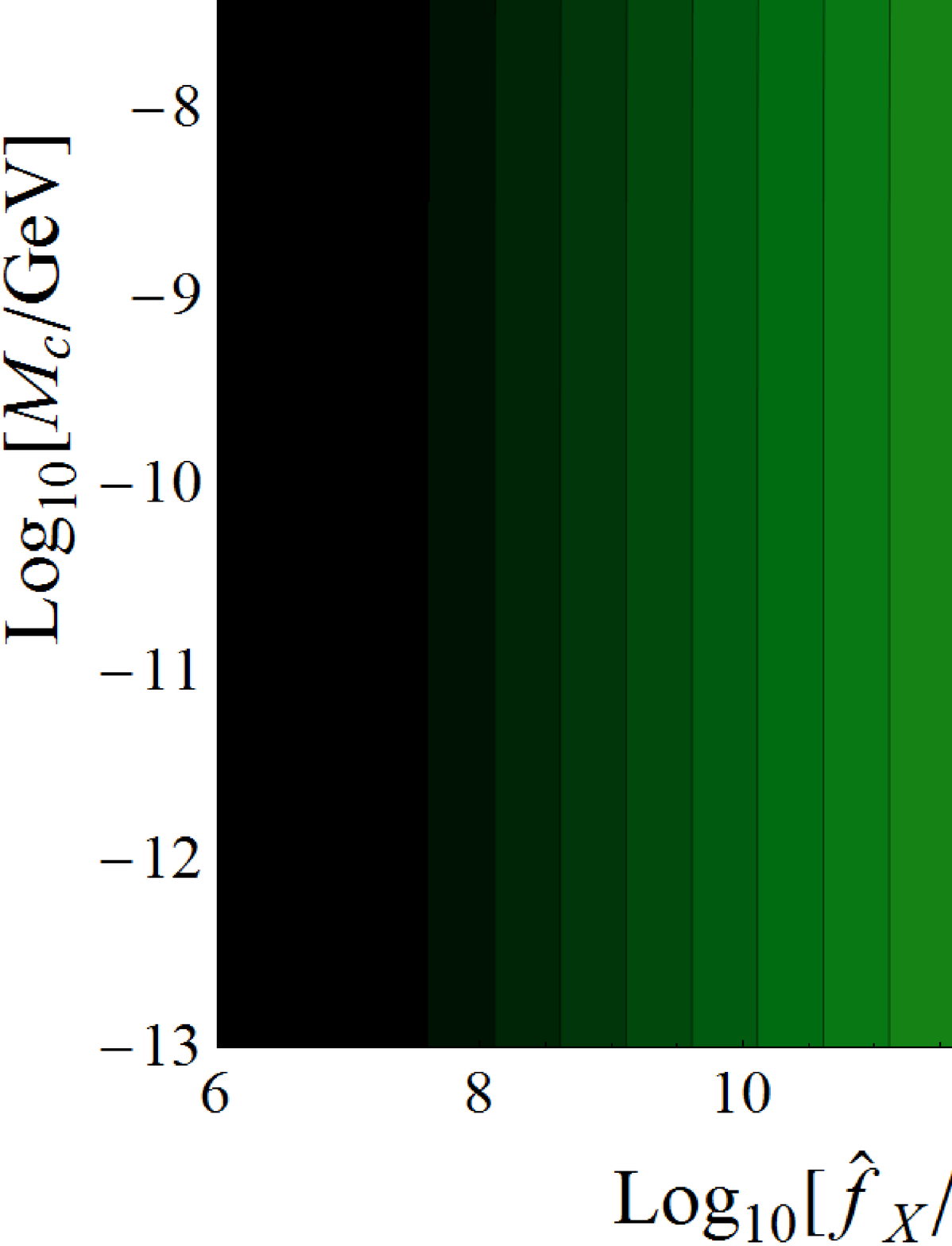}
  \epsfxsize 2.25 truein \epsfbox {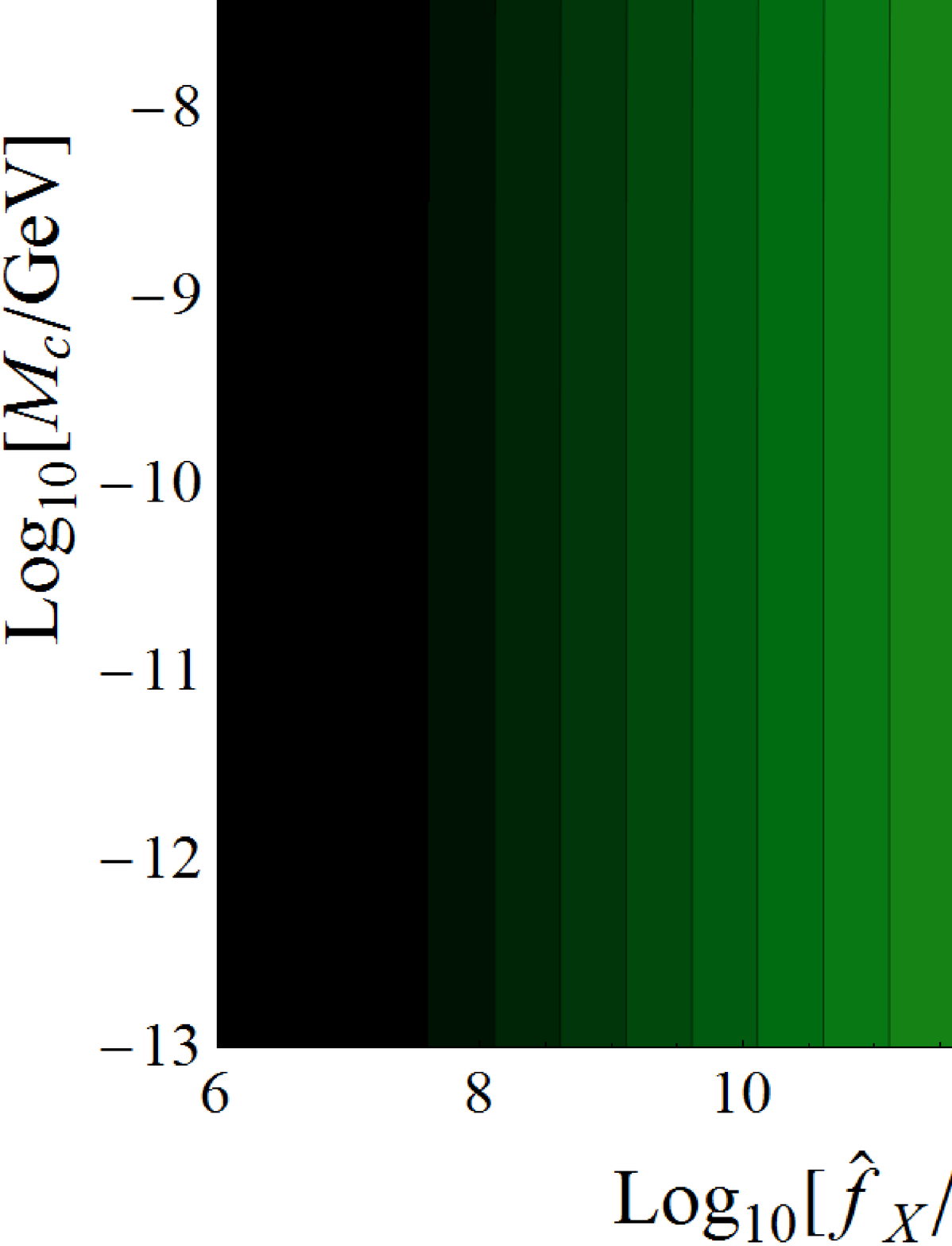}\\
  \raisebox{0.3cm}{\large$\Omegatotnow$}\epsfxsize 5.00 truein \epsfbox {ColorBarOmega.eps}
\end{center}
\caption{Same as in Fig.~\protect\ref{fig:OmegaTotPanelsStd}, but for the LTR cosmology
rather than the standard cosmology. 
\label{fig:OmegaTotPanelsLTR}}
\end{figure}

\begin{figure}[p]
~\vskip 0.75 truein
\begin{center}
  \epsfxsize 2.25 truein \epsfbox {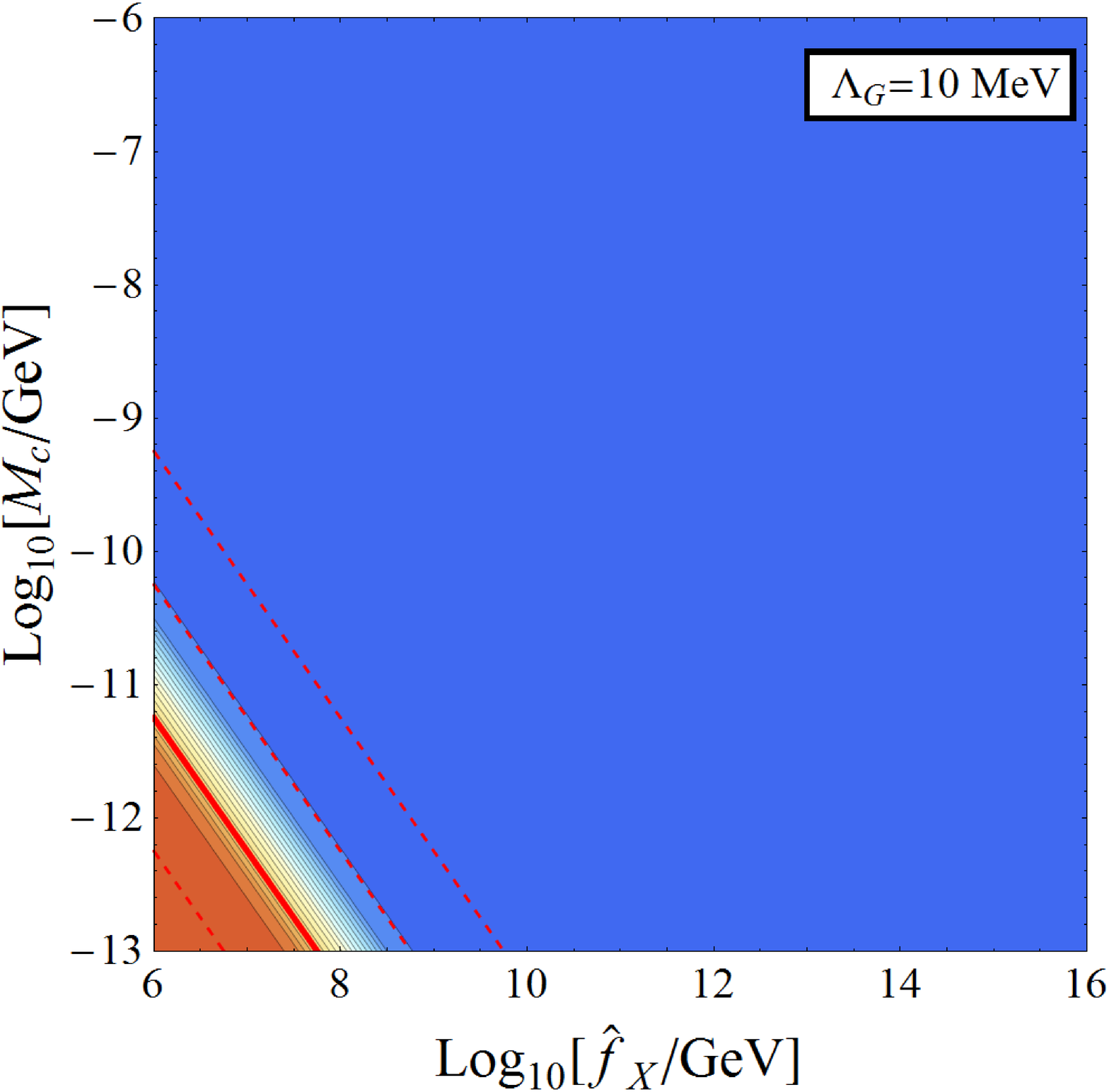} 
  \epsfxsize 2.25 truein \epsfbox {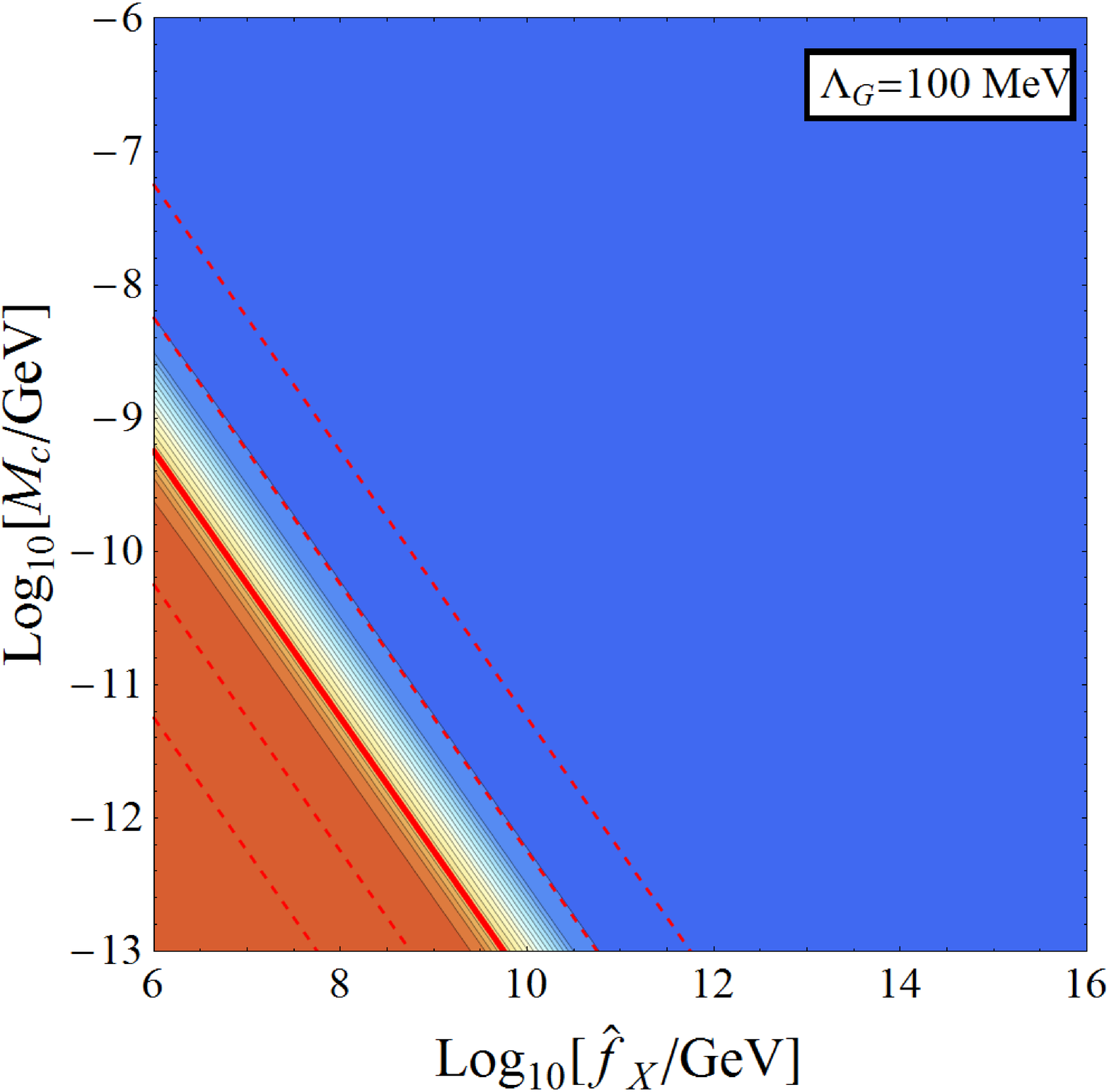}
  \epsfxsize 2.25 truein \epsfbox {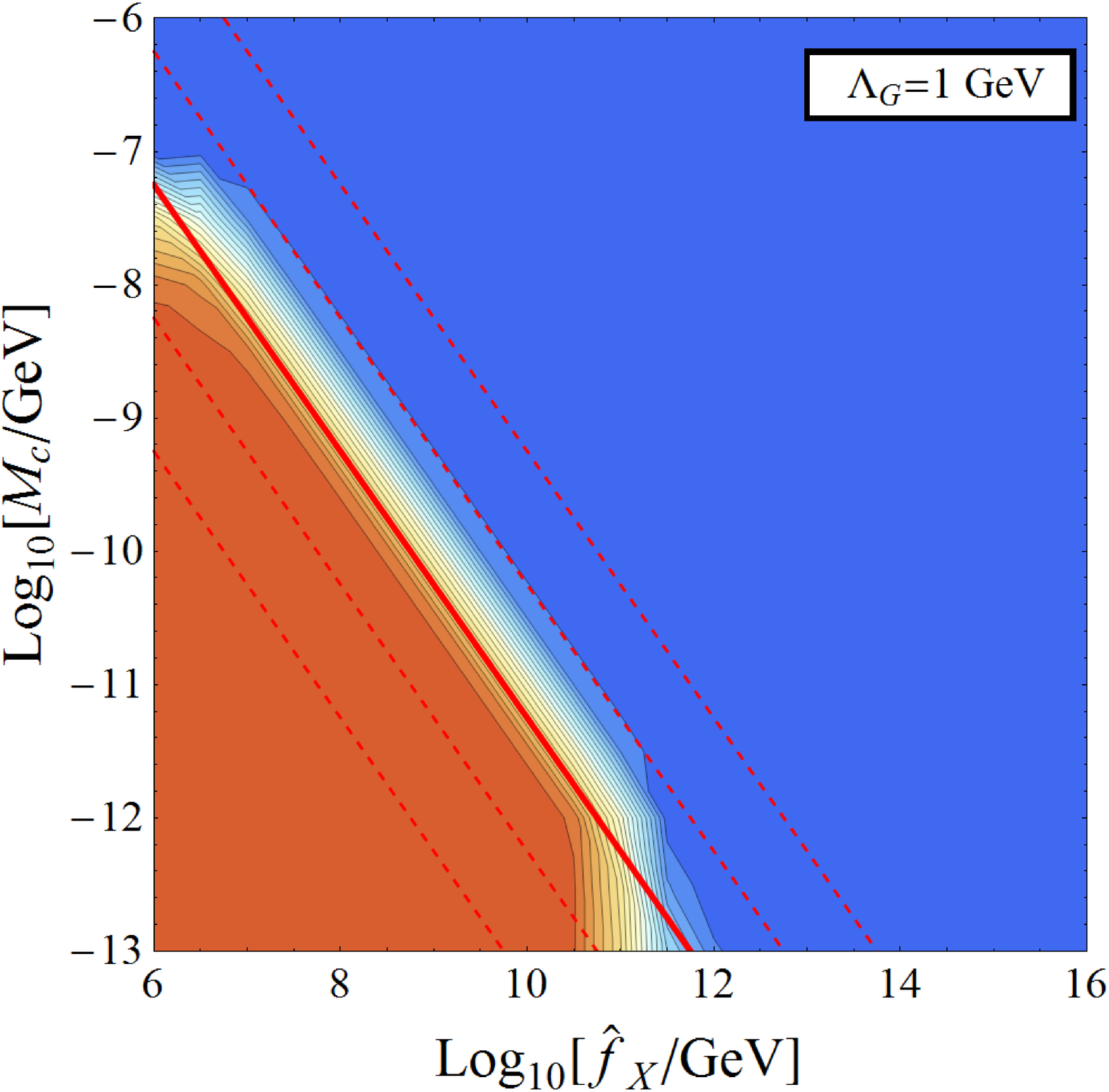}\\
  \epsfxsize 2.25 truein \epsfbox {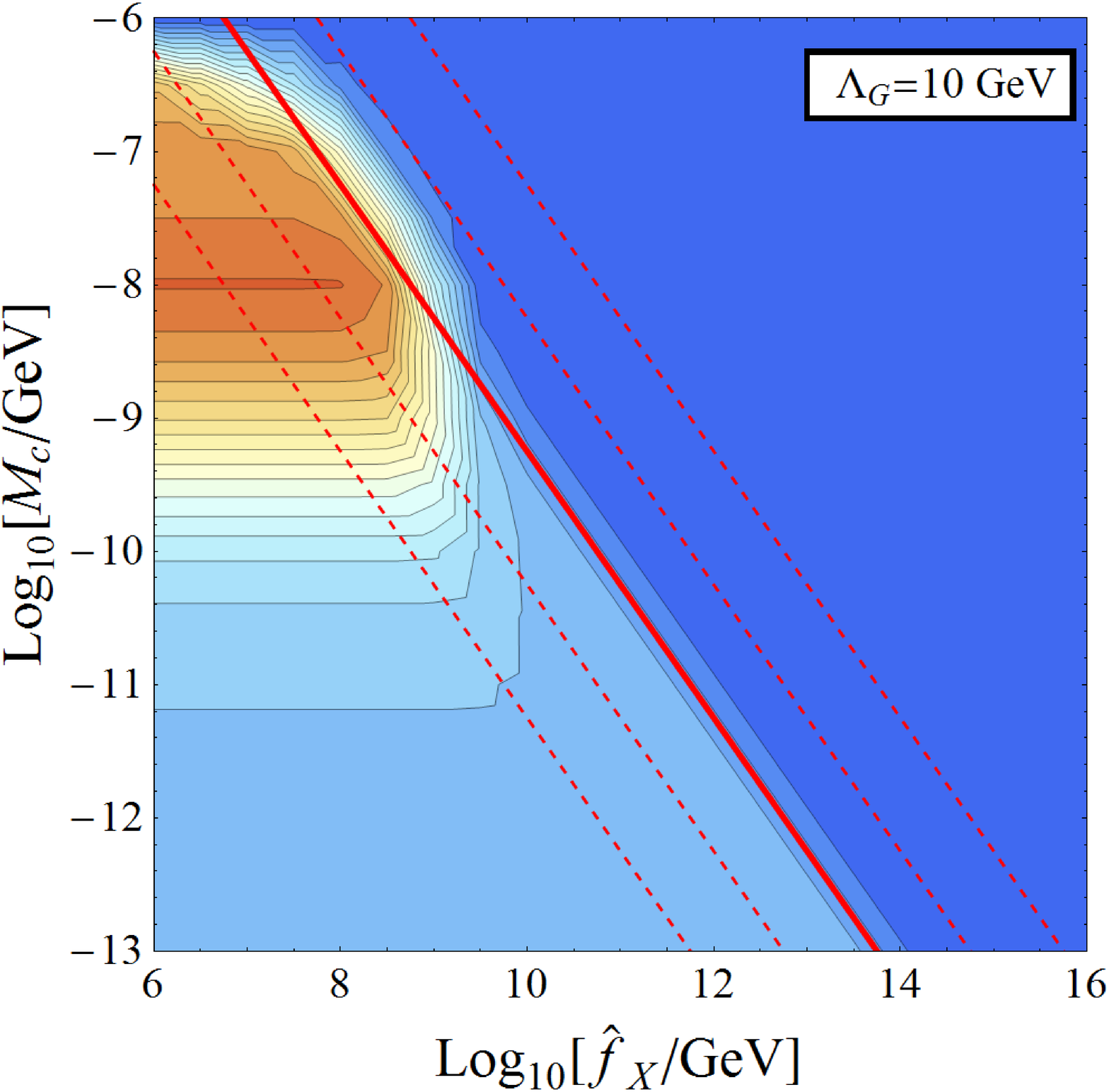}
  \epsfxsize 2.25 truein \epsfbox {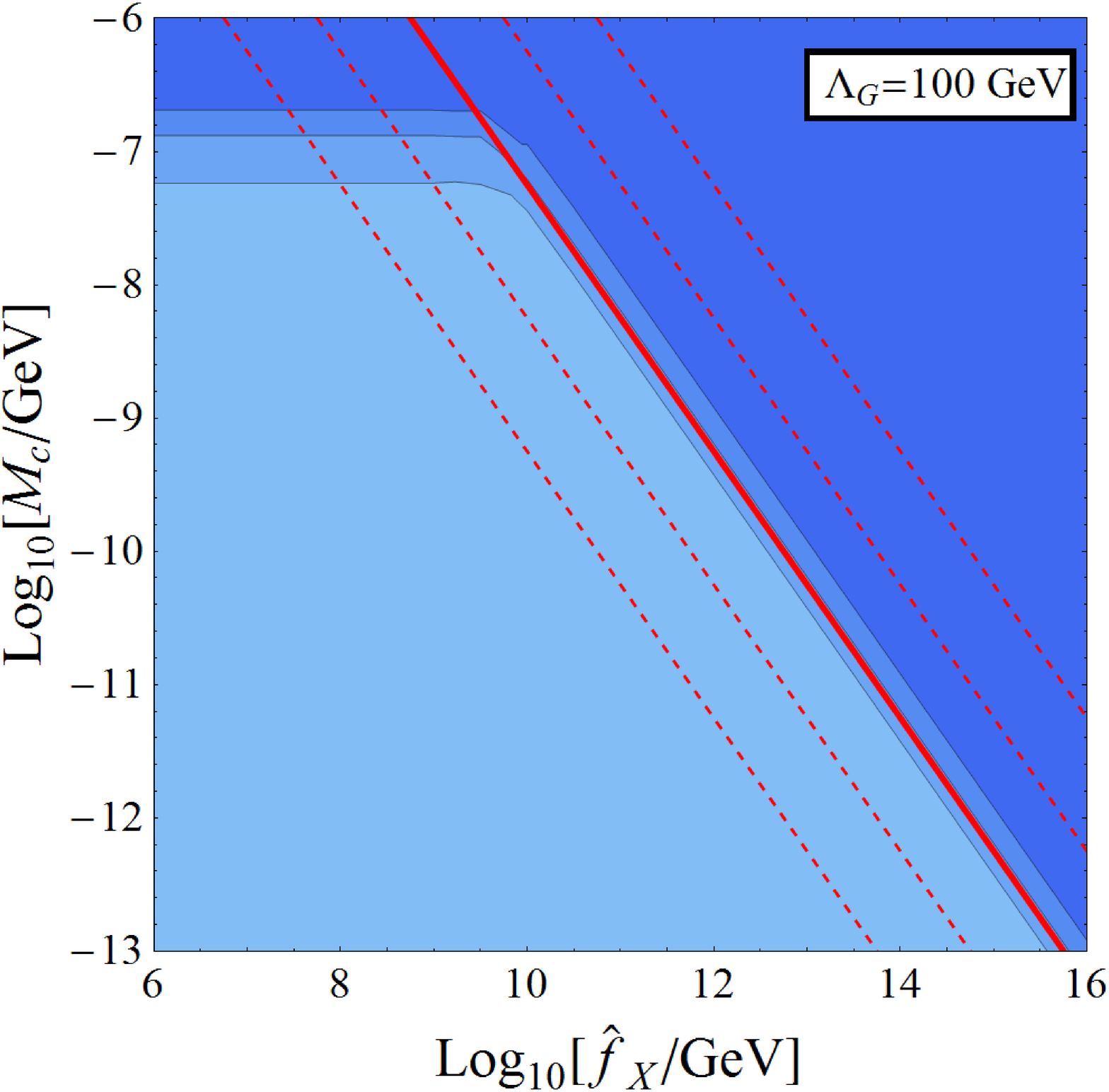}
  \epsfxsize 2.25 truein \epsfbox {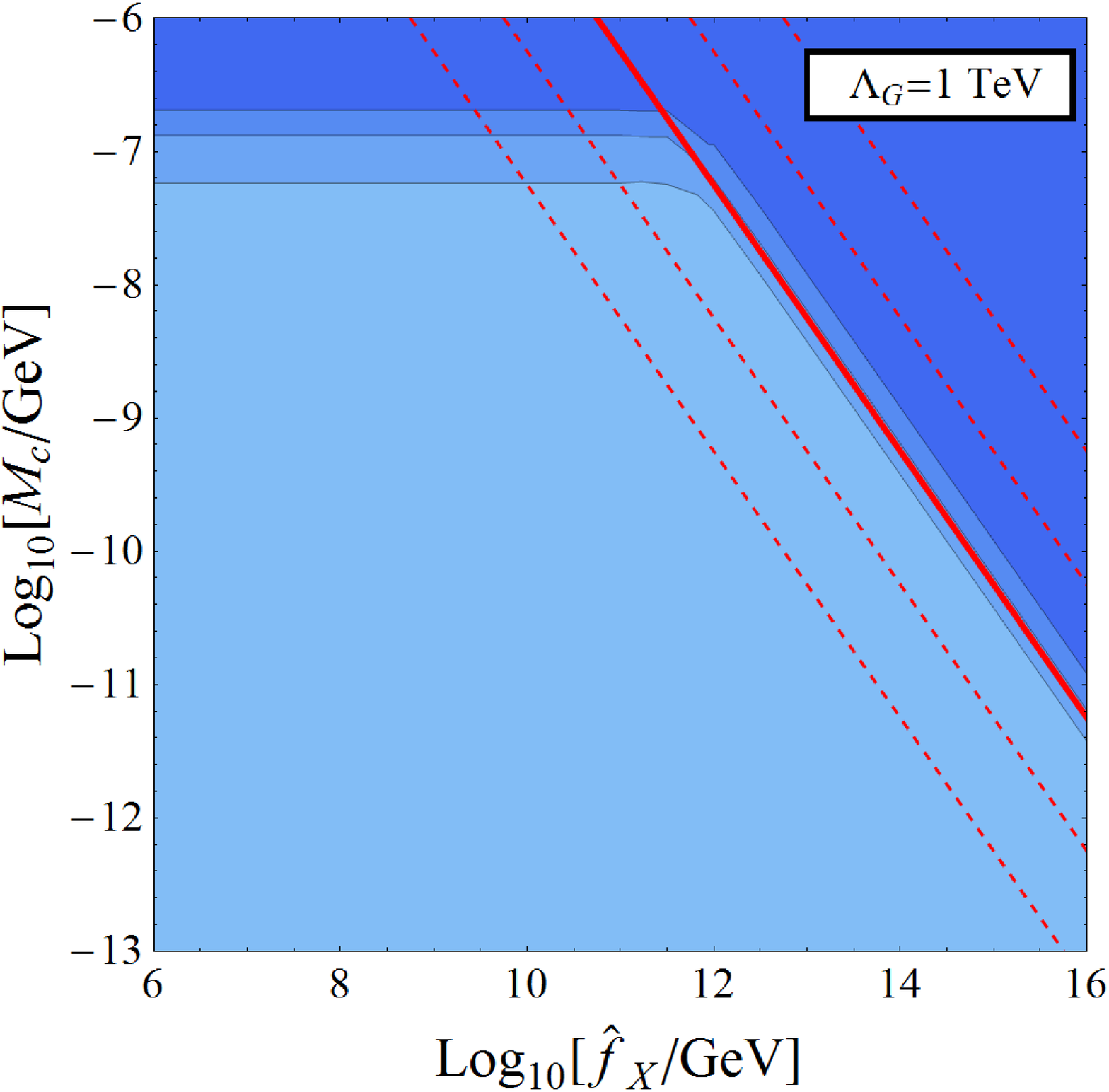}\\
  \epsfxsize 2.25 truein \epsfbox {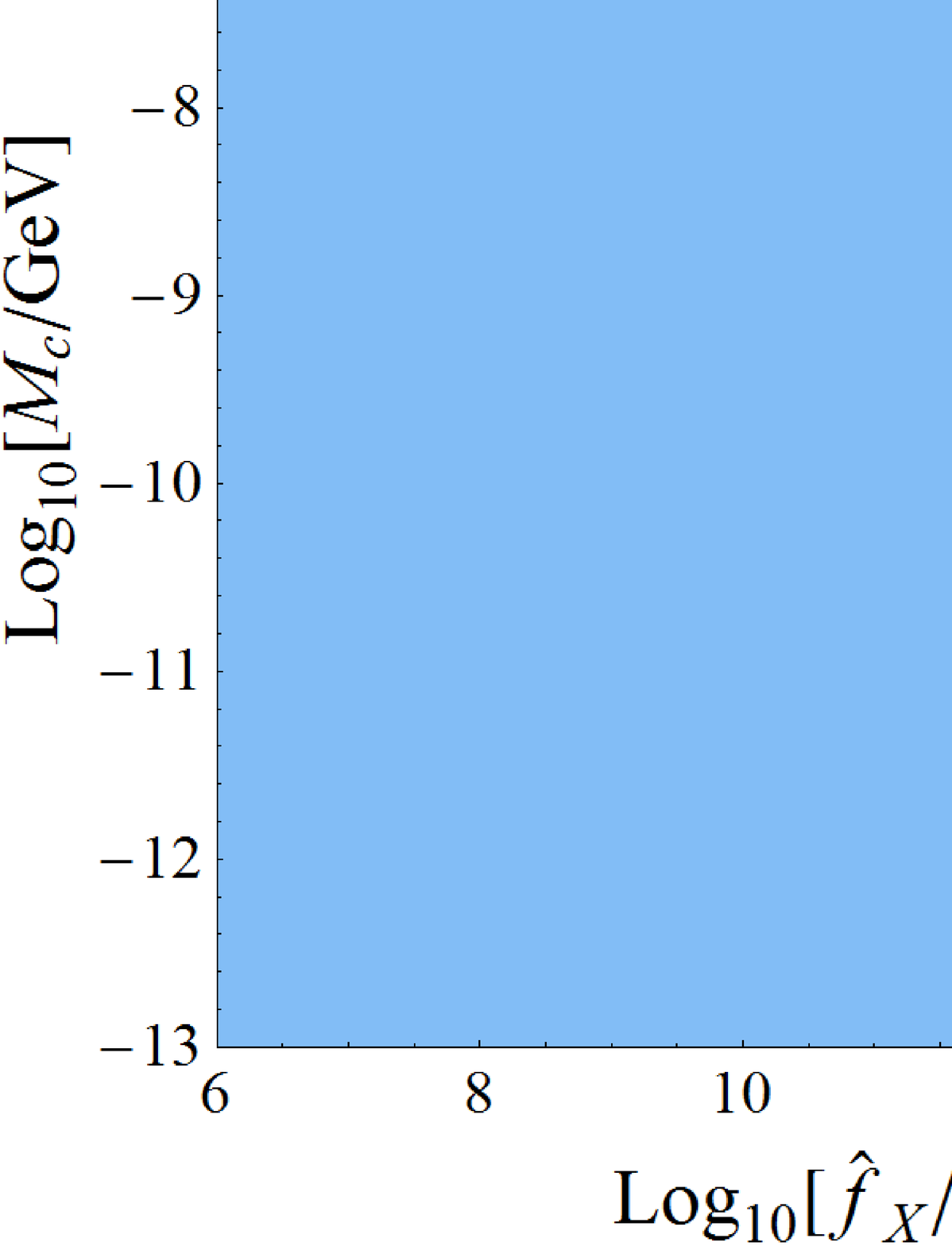}
  \epsfxsize 2.25 truein \epsfbox {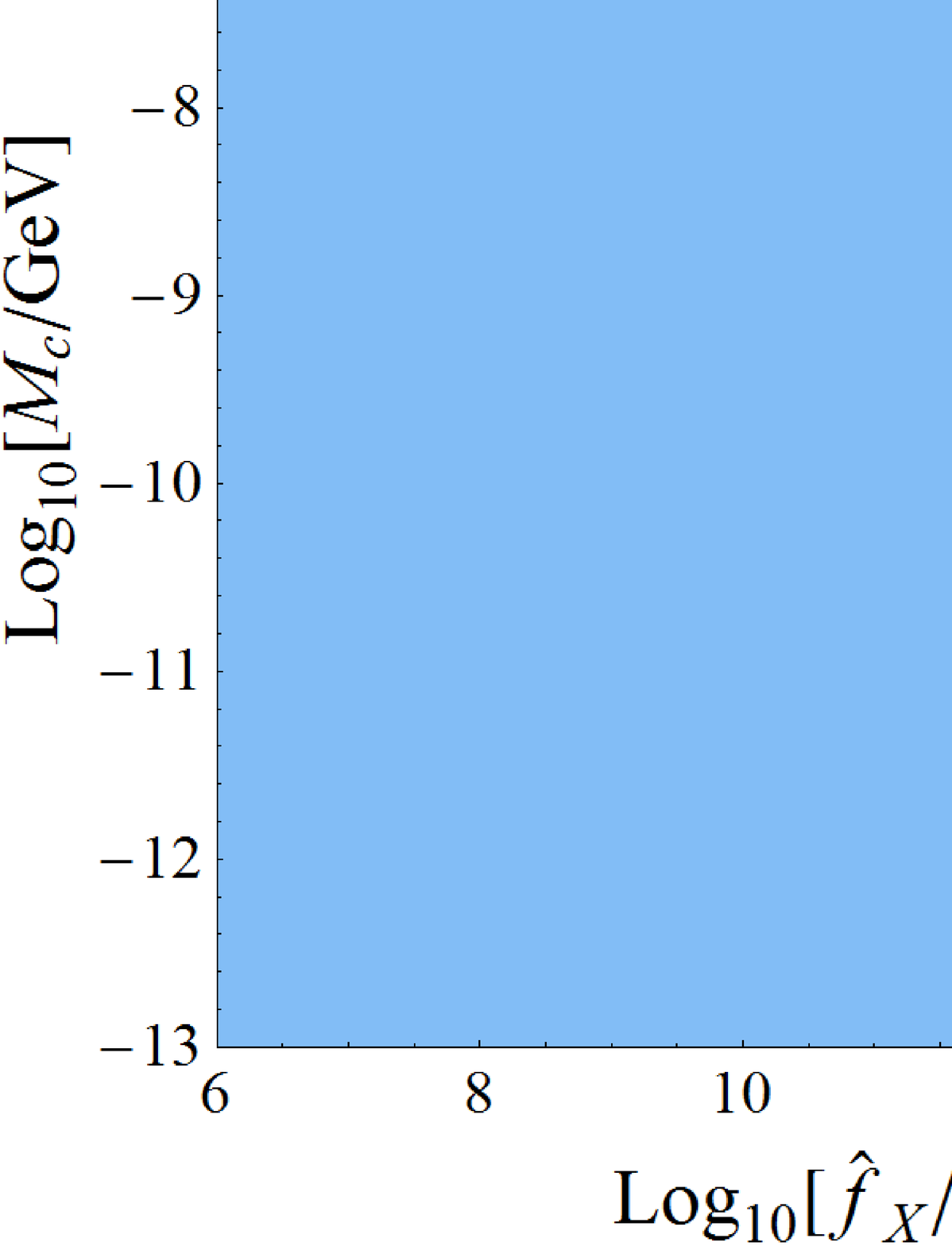}\\
  \raisebox{0.3cm}{\large$\etanow$}~\epsfxsize 5.00 truein \epsfbox {ColorBarEta.eps}
\end{center}
\caption{Same as in Fig.~\protect\ref{fig:EtaPanelsStd}, but for the LTR cosmology
rather than the standard cosmology.
\label{fig:EtaPanelsLTR}}
\end{figure}

In Fig.~\ref{fig:OmegaTotPanelsStd}, we show how 
the total present-day dark-matter relic abundance $\Omegatotnow$ depends
on $\fhatX$, $M_c$, and $\Lambda_G$ in the standard cosmology, assuming a
photonic axion with $c_\gamma = 1$. 
Each panel in the figure displays contours of $\Omegatotnow$ for a different 
choice of $\Lambda_G$.  A dashed blue line highlighting the 
$\Omegatotnow = 1$ contour has also 
been included in each panel.  The red lines
are contours of $y$: the solid red line corresponds to $y=1$, which roughly 
indicates the transition point between the strongly-mixed regime (below and to the left
of the contour) and the weakly-mixed regime (above and to the right of the contour).
Proceeding from left to right, the dotted lines correspond to the values 
$y=\{0.01,0.1,10,100\}$.  In Fig.~\ref{fig:EtaPanelsStd}, we present 
the corresponding contour plots for the tower fraction $\etanow$.
Moreover, to complement the results shown in Figs.~\ref{fig:OmegaTotPanelsStd} 
and~\ref{fig:EtaPanelsStd} for the standard cosmology, we present the 
corresponding results for the LTR cosmology in 
Figs.~\ref{fig:OmegaTotPanelsLTR} and~\ref{fig:EtaPanelsLTR}. 

As Figs.~\ref{fig:OmegaTotPanelsStd} and~\ref{fig:OmegaTotPanelsLTR} illustrate,
the dependence of $\Omegatotnow$ on the model parameters $\fhatX$, $M_c$, 
and $\Lambda_G$ is somewhat complicated, and the results displayed therein 
clearly warrant detailed explanation.  Perhaps the most intuitive way of 
understanding these results is to begin by examining them in certain limiting regimes.
For example, consider a situation in which $\Lambda_G$ is relatively small.  In this 
case, the confinement time scale $\tG$ is relatively late.
If $\tG$ is sufficiently late that all modes in the tower begin oscillating 
immediately at $\tG$, each $\Omega_\lambda$ is
given in the standard
cosmology by Eq.~(\ref{eq:OmegaLambdaOftEqntG}) for all $\lambda$ and by Eq.~(\ref{eq:OmegaLambdaOftEqnLTRtG}) in the LTR cosmology.  In addition,
let us assume that $\fhatX$ is large enough that decays can be neglected, and 
that $H_I$ is at least moderately large, so that essentially all of the
$a_\lambda$ which contribute meaningfully to $\Omegatotnow$ survive inflation.
In this special case, we can explicitly sum the contributions $\Omega_\lambda$ to
obtain the result  
\begin{equation}
  \Omegatotnow ~\approx~
    \frac{3}{256\pi^2}(g_G\xi)^2\left(\frac{\theta \Lambda_G^2}{M_P}\right)^2
      \tG^{3/2}\tMRE^{1/2}\times
      \begin{cases} \vspace{0.25cm}
      \displaystyle  1 & \mathrm{standard~cosmology}\\
      \displaystyle  (\tG/\tRH)^{1/2} & \mathrm{LTR~cosmology}~.
      \end{cases}
  \label{eq:OmegaLambdaShelfLimit}
\end{equation} 
In deriving this result, we have used the second identity in Eq.~(\ref{eq:AlambdaSqdID}).
Note that Eq.~(\ref{eq:tGtlambdaRegimes}) implies that $t_G < \tRH$; hence $\Omegatotnow$
is suppressed in the LTR cosmology relative to the standard cosmology by a factor which 
can be quite significant.  Indeed, this suppression factor in $\Omegatotnow$ is due to 
the uniform suppression of each individual contribution $\Omega_\lambda$ in this 
regime by the factor $\mathcal{E}_{\mathrm{LTR}}$ given for the 
$\lambda \geq 1/t_G$ case in Eq.~(\ref{eq:LTRSuppressionFactor}). 

Perhaps the most interesting aspect of this result is that it depends
only on $\Lambda_G^4$, and is independent of both $M_c$ and $\fhatX$.   
This is quite surprising indeed, for it
indicates that in this regime, no matter how many of the $a_\lambda$ contribute 
significantly to $\Omegatot$, the total contribution to the dark-matter relic
abundance is the same.  For example, a strongly-mixed scenario with $y \ll 1$ and
a vast number of modes contribute more or less democratically to $\Omegatot$ 
will yield the same abundance as a weakly-mixed scenario with $y\gg 1$ in which 
a single, light axion accounts for essentially the entirety of the dark matter.
This situation is realized in the $\Lambda_G = 10$~MeV and $\Lambda_G = 100$~MeV 
panels in Figs.~\ref{fig:OmegaTotPanelsStd} and~\ref{fig:OmegaTotPanelsLTR}, 
in which the value of $\Omegatotnow$
remains essentially constant throughout the region of parameter space shown.

At an algebraic level, the fact that $\Omegatotnow$ is a constant throughout 
substantial regions of $(\fhatX, M_c)$ space for small $\Lambda_G$ is a 
reflection of the fact that the identity in 
Eq.~(\ref{eq:AlambdaSqdID}) holds regardless of the value of $y$.  Of course,
this identity requires that the sum over $\Omega_\lambda$ be taken over the 
entire KK tower, from the lowest mass eigenstate up to infinity.
At a physical level, this is the appropriate sum to take for small $\Lambda_G$,
because all of the modes begin oscillating at a common time $t_G$, 
and because the full structure of the tower is undisturbed by the decay of 
any modes which contribute meaningfully in the sum.  Indeed, as we have seen from
the panels of Figs.~\ref{fig:OmegaTotPanelsStd} and~\ref{fig:OmegaTotPanelsLTR},
this result is characteristic of situations in which $\Lambda_G$ is small.

However, as we increase $\Lambda_G$, three effects can begin to alter
this picture and thereby destroy the uniformity of $\Omegatotnow$:
\begin{itemize}

\item First, $t_G$ becomes smaller and smaller, and consequently $t_\lambda$ can
begin to exceed $t_G$ for the lower modes in the KK tower.  In other words, 
these lower modes may begin to experience oscillations with staggered onset 
times, a phenomenon which begins with the lowest-lying modes in the tower and 
ultimately affects higher and higher modes as $\Lambda_G$ increases. 

\item Second, $y$ decreases with increasing $\Lambda_G$ (for fixed $\fhatX$ and $M_c$), 
and consequently more and more of the excited $a_\lambda$ contribute significantly 
to $\Omegatotnow$.  Although the lifetimes $\tau_\lambda$ of these modes also 
increase with increasing $\Lambda_G$, they do so at a slower rate.  As
a result, a larger and larger fraction of the contributing portion of the KK tower is 
effectively truncated by decays.  This effect can therefore lead to a reduction 
in $\Omegatotnow$, especially in the $y\ll 1$ regime.

\item   Third, as $\Lambda_G$ increases,
$t_G$ can be pushed back into the inflationary era.  The contributions from those 
modes which begin oscillating prior to or during inflation will therefore be
inflated away.  This too can result in a reduction of $\Omegatotnow$. 

\end{itemize}  

Of course, which of these effects happens to be relevant in any given situation 
ultimately depends on the parameters in question, and whether we are
working in the standard cosmology or an LTR cosmology.
Let us therefore begin by examining the situation in the standard cosmology, as 
shown in Fig.~\ref{fig:OmegaTotPanelsStd}.  As we increase $\Lambda_G$ from 
$10$~MeV to $100$~TeV, we see that a series of contours with smaller and 
smaller values of $\Omegatotnow$ emerges in the region of parameter space where 
$y\ll 1$ and gradually spreads over a substantial region of $(\fhatX,M_c)$ space.   
This is the effect of decays truncating the contributions from the
higher modes in the tower, as discussed above.  Note that for the regions of  
parameter space shown in Fig.~\ref{fig:OmegaTotPanelsStd}, neither of the other two 
effects outlined above is apparent.  
In particular, staggering effects only occur within regions of parameter space for 
which $t_G < t_{\lambda_0}$, where $t_{\lambda_0}$ is the 
time at which the lightest mode in the tower begins oscillating.
This criterion can be rephrased as a condition on the model parameters 
$\fhatX$, $M_c$, and $\Lambda_G$ by substituting $\Lambda_G$ for $T$ in
the middle line of Eq.~(\ref{eq:tTempRelLTR}).  In the $y\gg 1$ and $y\ll 1$ 
regimes, we can approximate $\lambda_0 \mX$ and $\lambda_0 \approx M_c/2$ 
respectively to obtain
\begin{equation}
  \mathrm{standard~cosmology}:~
  \begin{cases}\vspace{0.25cm}
  \displaystyle
  \fhatX \gtrsim (1.38\times 10^{17}\mathrm{~GeV}) \times 
    g_G\,\xi \left[g_\ast^{-1/2}(\Lambda_G)\right] ~~~
    & y \gg 1 \\
  \displaystyle
  M_c \lesssim (8.18\times 10^{-19}\mathrm{~GeV})\times
   \left[g_\ast^{1/2}(\Lambda_G)\right]
   \left(\frac{\Lambda_G}{\mathrm{GeV}}\right)^2~~~
    & y \ll 1~.
  \end{cases}
  \label{eq:fhatXCriticalAllStaggeredStd}
\end{equation}
Given these results, it is clear that staggering effects will not be visible in
Fig.~\ref{fig:OmegaTotPanelsStd}: for the $y\gg 1$ case, extremely large values
of $\fhatX$ are required, regardless of the value of $\Lambda_G$; for the 
$y\ll 1$ case, $M_c \lesssim 10^{13}$~GeV or $\Lambda_G \gtrsim 100$~TeV is 
required for these effects to be apparent.
 
The situation is quite different in the LTR cosmology, as shown in
Fig.~\ref{fig:OmegaTotPanelsLTR}.  Indeed, as we increase $\Lambda_G$, 
all three of the above effects begin
to become relevant.  First, we observe the same effect of decaying
$a_\lambda$ modes in the $y\ll 1$ region that we saw in 
Fig.~\ref{fig:OmegaTotPanelsStd}.  This effect is particularly
evident in the $\Lambda_G = 1$~GeV panel of Fig.~\ref{fig:OmegaTotPanelsLTR}.  
However, in the LTR case, we also observe effects 
due to staggering, which begin to appear in the large-$\fhatX$ region.  
Indeed, following the same procedure applied above for the standard 
cosmology but using the top line in Eq.~(\ref{eq:tTempRelLTR}), we 
find that these effects emerge in regions of parameter space where
\begin{equation}
  \mathrm{LTR~cosmology}:~
  \begin{cases}\vspace{0.25cm}
  \displaystyle
  \fhatX \gtrsim (1.03\times 10^{11}\mathrm{~GeV}) \times g_G\,\xi
  \left[\frac{g_\ast^{1/2}(\TRH)}{g_*(\Lambda_G)}\right]
  \left(\frac{\TRH}{\mathrm{MeV}}\right)^2
  \left(\frac{\Lambda_G}{\mathrm{GeV}}\right)^{-2} ~~~
    & y \gg 1 \\
  \displaystyle
  M_c \lesssim (2.73 \times 10^{-13}\mathrm{~GeV}) \times
   \left[\frac{g_\ast(\Lambda_G)}{g_\ast^{1/2}(\TRH)}\right]
   \left(\frac{\TRH}{\mathrm{MeV}}\right)^{-2}
   \left(\frac{\Lambda_G}{\mathrm{GeV}}\right)^4
    & y \ll 1~.
  \end{cases}
  \label{eq:fhatXCriticalAllStaggeredLTR}
\end{equation} 
The first of these limiting forms accounts for the vertical
contours which appear on the right side of the $\Lambda_G = 100$~MeV
panel in Fig.~\ref{fig:OmegaTotPanelsLTR} and encroach
further and further to the left as $\Lambda_G$ increases.  

Finally, as $\Lambda_G$ grows beyond $100$~GeV, we see the third effect 
emerging: the inflating away of heavy KK modes.  
In particular, since we have chosen $H_I = 10^{-7}$~GeV in this plot, the 
relic-abundance contributions from all modes with $\lambda\gtrsim 3H_I/2$ will 
be inflated away.  Indeed, we see that when $y\ll 1$ (which implies that 
$\lambda_0 \approx M_c/2$) and $M_c \gtrsim 3\times 10^{-7}$~GeV, the entire 
tower is inflated away, yielding $\Omegatotnow = 0$. 
Indeed, since $y$ increases with $\Lambda_G$ for fixed $\fhatX$ and $M_c$, 
this effect spreads across a wider region as $\Lambda_G$ increases.

Ultimately, for large $\Lambda_G$, those modes which have not inflated away 
exhibit a completely staggered behavior in the LTR cosmology.  This limit may be 
regarded as the converse of the ``instantaneous turn-on'' limit taken in 
Eq.~(\ref{eq:OmegaLambdaShelfLimit}) for small $\Lambda_G$: indeed, we 
now have $t_\lambda > t_G$ for {\it all} of the modes which 
contribute significantly to $\Omegatotnow$.  Moreover, in this case the 
sum over $\Omega_\lambda$ can be explicitly evaluated 
using the first identity in Eq.~(\ref{eq:AlambdaSqdID}), allowing us 
to obtain explicit results for $\Omegatotnow$ in
the completely staggered limit for both the standard and LTR cosmologies:        
 \begin{eqnarray}
  \mathrm{standard~cosmology}:~ & & \vspace{0.25cm}
    \Omegatotnow \approx
  \frac{3^{5/4}}{2^{29/4}\pi^{1/2}}(g_G\xi)^{1/2}
     \left(\frac{\theta}{M_P}\right)^2\tMRE^{1/2}\,\fhatX^{3/2}\Lambda_G\,C(y)  
  \nonumber\\
  \mathrm{LTR~cosmology}:~ & & 
    \Omegatotnow \approx \frac{3}{8} \left(\frac{\theta}{M_P}\right)^2
     \left(\frac{\tMRE}{\tRH}\right)^{1/2} \fhatX^2~,
  \label{eq:OmegaLambdaStaggeredLimit}
\end{eqnarray}  
where $C(y) \equiv \sum_{\lambda}\wtl^{1/2}A_\lambda^2$ for the 
standard-cosmology case.  Note that the $y$-dependence of $C(y)$ is   
illustrated in Fig.~4 of Ref.~\cite{DynamicalDM1}. 
In sharp contrast with the results
obtained in Eq.~(\ref{eq:OmegaLambdaShelfLimit}), we see that in 
this staggered regime, the expressions for $\Omegatotnow$
in the standard and LTR cosmologies differ significantly.  In the 
standard cosmology, $\Omegatotnow$ depends non-trivially on $\fhatX$,
$\Lambda_G$, and $M_c$ (through its dependence on $y$).
By contrast, in the LTR cosmology, $\Omegatotnow$ depends on $\fhatX$
in this staggered-oscillation regime, but not on $\Lambda_G$ or $M_c$.

Following similar reasoning, we can also understand the behavior of
the tower fraction $\etanow$, beginning with the results 
displayed in Fig.~\ref{fig:EtaPanelsStd} for the standard cosmology.
These results nicely illustrate an important general property of 
$\etanow$ in brane/bulk models of dynamical dark matter discussed in
Ref.~\cite{DynamicalDM1}, which is 
that the behavior of $\etanow$ is strongly correlated with the  
value of $y$.  In particular, when $y \gg 1$, the mass
of the lightest mode in the KK tower is proportionally far lighter than those
all of the excited modes, and consequently its contribution $\Omega_{\lambda_0}$ 
to the total abundance will be much larger than the contributions from all of 
those other modes combined.  Indeed, this is nothing but the four-dimensional 
limit of the KK theory.  This property of $\etanow$ is independent of both 
the specific cosmological context and whether or not any of the modes in the 
tower have staggered oscillation onset times, as can be seen from 
Eqs.~(\ref{eq:OmegaLambdaOftEqntG}) through~(\ref{eq:OmegaLambdaOftEqnLTRtlambda}).     
By contrast, when $y \ll 1$, each mode with a mass 
$\lambda \ll \lambdatrans$ contributes essentially equally toward 
$\Omegatotnow$ when $t_\lambda = t_G$ for all $a_\lambda$, 
and hence $\etanow \approx 1$.  This behavior is manifest in the 
various panels of Fig.~\ref{fig:EtaPanelsStd}, in which $\etanow$
rapidly transitions from nearly zero to nearly unity as one crosses 
the $y=1$ contour.

In Fig.~\ref{fig:EtaPanelsLTR}, we display the behavior of $\etanow$ in the
LTR cosmology.  For small $\Lambda_G$, the situation is very similar to that
in the standard cosmology: all modes in the tower begin oscillating at $t_G$,
and hence $\eta \approx 0$ for $y \gg 1$, while $\eta \approx 1$ for $y \ll 1$.
However, as $\Lambda_G$ increases, the oscillation onset times for more and more of 
the lighter modes in the tower become staggered in time.  In this regime,  
the energy densities $\rho_\lambda$ associated
with the lighter modes in the tower scale like vacuum energy (and are hence 
unaffected by Hubble dilution) for a longer time before coherent oscillations set in 
and they begin to scale like massive matter, as illustrated in Fig.~1 of 
Ref.~\cite{DynamicalDM1}.  It follows that in this regime, the
lighter modes account for a greater fraction of $\Omegatotnow$.  For this reason, 
as discussed in Ref.~\cite{DynamicalDM1}, $\etanow$ no longer approaches 
unity for $y \ll 1$ in those regions of parameter space in which oscillation onset 
times are staggered, but instead asymptotes to\footnote{
Note that there are two (ultimately equivalent) ways to derive this result,
corresponding to two different methods of taking the $y\to 0$ limit.
In Ref.~\cite{DynamicalDM1}, we recognized that $\lambda_n\approx (n+1/2)M_c$
as $y\to 0$.   Since the $y\to 0$ limit also implies  
that $\lambda\ll \lambda_{\rm trans}$ for all $\lambda$,
we can similarly approximate $A_\lambda\sim 1/\lambda$ in this limit.
We then have $\Omega_\lambda\sim A_\lambda^2\sim 1/\lambda^2$, whereupon
it follows that $\Omega_{\lambda_0}/\Omegatot = 4/\sum_n (n+1/2)^{-2}= 8/\pi^2$,
or equivalently $\eta_{\rm max}= 1-8/\pi^2$.  However, it is also possible to 
retain the exact form $\Omega_\lambda\sim A_\lambda^2$, whereupon we see that
$\Omega_{\lambda_0}/\Omegatot=A_{\lambda_0}^2$ where $A_{\lambda_0}$ is the value of 
$A_\lambda$ for the lightest eigenvalue $\lambda_0$ and where we have used the identity 
$\sum_\lambda A_\lambda^2=1$ to perform the sum over KK modes.  Note that this 
result is exact and valid for all $y$.  However, it is easy to verify that 
$A_{\lambda_0}\to 2\sqrt{2}/\pi$ as $y\to 0$.  We thus again 
find that $\eta_{\rm max}= 1-8/\pi^2$.} 
\begin{equation}
  \eta_{\mathrm{\max}} ~\equiv~ 1 - \frac{8}{\pi^2} ~\approx~ 0.189~,
  \label{eq:EtaLimitInStaggeredRegime}
\end{equation}
for the case of the LTR cosmology.  Roughly speaking, the regions of 
parameter space in which this occurs are those in which the staggered-onset criteria in 
Eq.~(\ref{eq:fhatXCriticalAllStaggeredLTR}) are satisfied.  Indeed, this effect
first becomes apparent in the $\Lambda_G = 1$~GeV panel of Fig.~\ref{fig:EtaPanelsLTR}
and becomes increasingly significant as $\Lambda_G$ increases and staggering
effects become relevant for smaller and smaller $\fhatX$ and larger and larger $M_c$.
By $\Lambda_G = 10$~GeV, these staggering effects are realized over nearly the entirety
of $(\fhatX,M_c)$ space shown, leaving only a narrow strip in which all modes still
begin oscillating at $t_G$, and by $\Lambda_G = 100$~GeV, even this strip vanishes.  
Note also that the effect on $\etanow$ 
of modes being inflated away is apparent in the upper left of those panels
in Fig.~\ref{fig:EtaPanelsLTR} for which $\Lambda_G \geq 10$~GeV.  While $\eta$ is 
technically undefined in this region of parameter space because $\Omegatotnow = 0$,
we have set $\etanow = 0$ within this region to illustrate where this effect is 
relevant.  

As discussed in the beginning of this section, the interesting regions of parameter 
space for dynamical dark matter are ultimately those in which 
$\Omegatotnow \approx \OmegaDM$, while at the same time $\etanow$ differs 
significantly from zero.  Given the results in Figs.~\ref{fig:OmegaTotPanelsStd}
through~\ref{fig:EtaPanelsLTR}, we can now determine whether this situation ever
actually arises in our model.  Comparing the results in Figs.~\ref{fig:OmegaTotPanelsStd}
and~\ref{fig:EtaPanelsStd}, we see that this occurs in the standard cosmology
for small values of $\fhatX$, within a diagonal stripe of parameter space slightly
to the left of the blue $\Omegatotnow = 1$ contour in each panel.  We also see that
this stripe moves to the right in $(\fhatX,M_c)$ space as $\Lambda_G$ increases.
By contrast, comparing the results in Figs.~\ref{fig:OmegaTotPanelsLTR} 
and~\ref{fig:EtaPanelsLTR}, we see that the above conditions are satisfied 
in the LTR cosmology in the region of parameter space where 
\begin{equation}
\mathrm{preferred~region~(LTR)}:~~
\begin{cases}
  \,\bullet~~\fhatX \sim 10^{14} - 10^{15}\mathrm{~GeV} \\
  \,\bullet~~\Lambda_G \gtrsim 100\mathrm{~GeV}\\
  \,\bullet~~M_c\mathrm{~small~enough~that~} y \lesssim 1~.
\end{cases}
\label{eq:PreferredRegion}
\end{equation}
This result is certainly intriguing, as 
it suggests that the preferred scale for $\Lambda_G$ in this model 
is roughly the TeV scale for the LTR cosmology ---  a scale at which 
there is good reason to expect new physics to appear.
 
The principal message of Figs.~\ref{fig:OmegaTotPanelsStd} 
through~\ref{fig:EtaPanelsLTR}, then, is that our bulk-axion model indeed satisfies
the conditions on $\Omegatotnow$ and $\etanow$ for dynamical dark 
matter within these regions of parameter space.  In other words, within these
regions, our axion ensemble reproduces the observed dark-matter relic abundance, and
does so in a non-trivial manner, with a substantial number of its constituents
contributing significantly to $\OmegaDM$.  Of course these alone are not 
sufficient conditions for a successful model of dynamical dark matter: such
a model must also not only have an appropriate present-day equation-of-state 
parameter $w_\ast$, but also satisfy all additional relevant phenomenological 
constraints.  In the remainder of this section, we will address the constraints 
on $w_\ast$; the rest of the applicable constraints will be addressed in 
Sect.~\ref{sec:Constraints}.

One particular ramification of these constraints, however, is appropriate to
mention before proceeding further. 
As discussed in Sect.~\ref{sec:Production}, certain
bounds which apply generically to models with large, flat extra dimensions 
strongly prefer the LTR cosmology over the standard cosmology.  For this reason, 
we will focus primarily on the LTR case from this point forward.     

\subsection{Dark Towers: Equations of State}

Having characterized the behavior of $\Omegatotnow$ and $\etanow$
over the parameter space of our bulk-axion model, we now proceed to discuss 
the third critical quantity which characterizes the dynamical dark-matter 
ensemble in this model: the present-day effective equation-of-state parameter 
$w_\ast$.  Since it is now clear which regions of model parameter
space are suitable for dynamical dark matter, we will not perform a general
survey of $w_\ast$ over the entirety of that parameter space, as we did with
$\Omegatotnow$ and $\etanow$, but instead focus on the preferred regions
indicated in Eq.~(\ref{eq:PreferredRegion}).

In order to calculate $w_\ast$ we need to know the values of the coefficients and
exponents $A$, $B$, $\alpha$, and $\beta$ appearing in Eq.~(\ref{eq:wstar}). This, in 
turn, requires knowledge of how our abundances and decay widths scale with $\lambda$. 
As in the previous subsection, we will assume that the abundances of the $a_\lambda$ 
result from misalignment production, and likewise we will assume that their decay 
widths $\Gamma_\lambda$ are those appropriate for 
a photonic axion with $c_\gamma = 1$.  Because the preferred region of parameter 
space specified in Eq.~(\ref{eq:PreferredRegion}) for our model is one which is 
well approximated by assuming staggered oscillation onset times for all relevant
modes, the correct expression for $\Omega_\lambda$ is the one given in
Eq.~(\ref{eq:OmegaLambdaOftEqnLTRtlambda}).  Likewise, the decay width for a
photonic axion is given by the expression in Eq.~(\ref{eq:GammaDecayToPhotons}).
We therefore find that the coefficients $A$ and $B$ appearing in 
Eq.~(\ref{eq:wstar}) are given respectively by    
\begin{equation}
  A ~ = ~ 3 \frac{\theta^2}{M_P^2}\left(\frac{\tMRE}{\tRH}\right)^2
     \times \begin{cases}
     \displaystyle\vspace{0.2cm} \frac{2^{4/3}G_\gamma^{4/3}\mX^4}{\fhatX^{2/3}}~~~~~
     &\displaystyle\lambda ~\gtrsim~ \frac{\pi \mX^2}{M_c} \\  
     \displaystyle \frac{2^{2/5}G_\gamma^{4/3}(\fhatX\mX)^{6/5}}
     {(1+\pi^2/y^2)^{7/5}}~~~~~
     &\displaystyle\lambda ~\lesssim~ \frac{\pi \mX^2}{M_c}
     \end{cases}
\end{equation}        
and  
\begin{equation}
  B ~ = ~ 
     \begin{cases}
     \displaystyle\vspace{0.2cm} \frac{(2\fhatX\mX)^{2/3}}
     {6 M_c G_\gamma^{1/3}}~~~~~
     & \displaystyle\lambda ~\gtrsim~ \frac{\pi \mX^2}{M_c} \\  
     \displaystyle \frac{(4\fhatX\mX)^{2/5}}{10 M_c G_\gamma^{1/5}}
     (1+\pi^2/y^2)^{1/5}~~~~~
     &\displaystyle \lambda ~\lesssim~ \frac{\pi \mX^2}{M_c}~,
     \end{cases}
\end{equation}
while the power-law indices $\alpha$ and $\beta$ are given by
\begin{equation}
    (\alpha,\beta) ~\approx~ 
  \begin{cases} \vspace{0.2cm}
  \displaystyle(-4/3,-2/3)~~~~~& 
     \displaystyle\lambda ~\gtrsim~ \frac{\pi \mX^2}{M_c} \\
  \displaystyle(-2/5,-4/5)~~~~~& 
     \displaystyle\lambda ~\lesssim~ \frac{\pi \mX^2}{M_c}~.
  \end{cases}
\end{equation}
Substituting these results into Eq.~(\ref{eq:wstar}), we find that
\begin{equation}
  w_\ast ~=~ \frac{\theta^2}{M_P^2 M_c\Omegatotnow}
    \left(\frac{\tMRE}{\tRH}\right)^{1/2}
    \times\begin{cases}
    \displaystyle\vspace{0.2cm}G_\gamma\mX^4 \tnow~~~~~
    &\displaystyle \lambda ~\gtrsim~ \frac{\pi \mX^2}{M_c} \\
    \displaystyle 
    \frac{3\,(2G_\gamma\mX^{8}\fhatX^{8}\tnow)^{1/5}}
    {10\,(1+\pi^2/y^2)^{6/5}}~~~~~
    &\displaystyle \lambda ~\lesssim~ \frac{\pi \mX^2}{M_c}~.
    \end{cases}
\end{equation}

Let us discuss the implications of these results.
First, it was noted in Ref.~\cite{DynamicalDM1} that the
effective equation-of-state parameter $w_{\mathrm{eff}}(t)$ for any
given dynamical dark-matter ensemble at any time $t<\tnow$ will 
always fall within the range $0 \leq w_{\mathrm{eff}}(t) \leq w_\ast$ 
as long as $\alpha + \beta < -1$.  This makes such ensembles less 
dangerous from a phenomenological point of view.  Indeed, we see 
from the results above that this criterion is satisfied for  
both the large-$\lambda$ and small-$\lambda$ regimes in the bulk-axion 
model under consideration here.  

Second, in order to convey a sense of the characteristic size of $w_\ast$ in 
the favored region of parameter space for dynamical dark matter given in 
Eq.~(\ref{eq:PreferredRegion}), we note that for the choice of $\fhatX = 10^{14}$~GeV, 
$M_c = 10^{-11}$~GeV, and $\Lambda_G = 1$~TeV, with
$g_G = \xi = \theta = 1$, we find that $w_\ast \approx 8.4\times 10^{-23}$ for
$\lambda \gtrsim \pi \mX^2/M_c$, while $w_\ast \approx 5.7\times 10^{-11}$ for 
$\lambda \lesssim \pi \mX^2/M_c$.
As these numbers are both extremely close to zero, we conclude that at present time 
our axion ensemble has an effective equation of state which can be legitimately 
interpreted as that of dark matter.  Thus our ensemble meets all three 
requirements for a self-consistent model of dynamical dark matter.   


\section{Characterizing the Ensemble:~ 
Constraints and Prospects for Detection\label{sec:Constraints}}


In the previous section, we demonstrated that an ensemble of mixed KK
excitations of a bulk axion field can collectively account for the observed
relic abundance of dark matter in our universe.  However, 
as discussed in Ref.~\cite{DynamicalDM1}, in order to be a
viable model of dynamical dark matter, the model must also comply with a 
variety of additional laboratory, astrophysical, and cosmological constraints.  
Some of these constraints are intrinsic to any theory involving large extra dimensions,
while others arise due to the physical effects of the axion field which propagates
in the bulk of those dimensions.    
A number of analyses of such constraints exist in the 
literature~\cite{ChangAxionBoundsED,DDGAxions,BambiAxion} for the specific 
case in which the bulk axion in question is identified with the QCD axion and the 
fundamental, $D$-dimensional quantum-gravity scale is taken to be roughly 
$M_D\sim\mathcal{O}(\mathrm{TeV})$.  By contrast, in the present analysis, we are 
interested in a broader class of axions which are neither required 
to couple to the fields of the SM (and in particular to hadrons) in the same manner 
as a QCD axion, nor subject to the same strict relationship between the suppression 
scale for those couplings and the axion mass.  Moreover, our primary 
motivation is not to address the hierarchy problem, but to address the
issue of what constitutes the non-baryonic dark matter in our universe.
For these reasons, we will not focus exclusively on scenarios in which $M_D$ is 
at or near the TeV scale, but also consider scenarios with much larger $M_D$.
As a consequence, exclusion limits on the parameter 
space of the more general axion scenarios
considered here can differ quite significantly from those presented in previous 
studies, and thus warrant reexamination.      

We begin our summary of the applicable constraints on our model with a brief synopsis 
of those limits which arise generically in theories with large, flat extra 
dimensions and which do not depend on the presence or
properties of the bulk axion field.  For the most part, these limits, an overview of 
which was presented in Ref.~\cite{ADDPhenoBounds}, tend to derive from 
the non-observation of physical effects related to the dynamics of KK gravitons.
These limits take many forms.    
First, there is the direct lower bound on $M_c$ quoted in Eq.~(\ref{eq:MinimumMc})
from experimental limits on modifications of Newton's law at short distances due to 
KK-graviton exchange~\cite{PDG}.  In addition, a number of constraints arise 
as a consequence of the production of these particles in the early
universe~\cite{ADDPhenoBounds,CosmoConstraintsLargeED}.  As discussed in 
Sect.~\ref{sec:Production}, these cosmological constraints can collectively 
be addressed by positing that the universe underwent a late period of cosmic 
inflation with a reheating temperature $\TRH\sim \mathcal{O}(\mathrm{MeV})$.  
Thus, by adopting an LTR cosmology with a reheating temperature of this order,
as we have done, we automatically ensure that a large number of these 
model-independent constraints are satisfied.

A number of additional constraints on theories of this sort
can be derived from observational limits on KK-graviton production in astrophysical 
sources, such as stars~\cite{HannestadRaffeltNeutronStar} and 
supernovae~\cite{KKGravitons1987A,HannestadRaffeltSupernovae}.  The most stringent
of these constraints are currently those resulting from gravitationally 
trapped KK gravitons in the halos of neutron stars either decaying to 
photons or serving as a heat source for the stars 
themselves.  In the case of $n>1$ flat extra 
dimensions with equal radii compactified on an $n$-torus, these limits supersede 
the limit on $M_c$ given in Eq.~(\ref{eq:MinimumMc}).  In particular, for $n=2$, 
the bound is $M_c \gtrsim 5.8 \times 10^{-7}$~GeV, while for $n=3$, one finds
$M_c \gtrsim 3.8 \times 10^{-10}$~GeV~\cite{HannestadRaffeltNeutronStar}.
However, if the radii of the extra dimensions differ from one another, or if 
the compactification manifold is not toroidal, these bounds can
be considerably weaker.  Furthermore, it is possible that the axion
propagates only within some number $n_a$ of the additional dimensions,
$n_a < n$.  In other words, the axion could be confined to a $(4+n_a)$-dimensional
brane within the bulk.  In this case, the effective, four-dimensional scales 
$\fhatX$ and $M_P$ are related to the fundamental, higher-dimensional scales $f_X$ and 
$M_D$ in completely different ways:
\begin{eqnarray}
  M_P^2 &=& V_n M_D^{2+n}\nonumber \\
  \fhatX^2 &=& V_{n_a} f_X^{2+n_a}~. 
\end{eqnarray} 
The upshot, then, is that na\"{i}ve limits on $M_D$ derived from KK-graviton dynamics 
under the assumption of toroidal compactification and equal radii do not necessarily 
translate in a straightforward manner into constraints on the mass scales relevant
to the physics of a bulk axion.  Fortunately, the bound in 
Eq.~(\ref{eq:MinimumMc}) is universal and is not sensitive to the 
total number of extra dimensions, unless they are each of comparable size.  
We will therefore take this bound to be the lower 
limit on $M_c$ in the $n_a = 1$ model under consideration here. 

We now turn to address those constraints which relate to the effects of the
bulk axion itself.  Indeed, a number of considerations serve to constrain 
the properties of light exotic particles with suppressed couplings to SM 
fields.  Some of these constraints derive from observational limits on the 
production of such particles in astrophysical sources such as stars and 
supernovae; others derive from limits on the decays of a cosmological population
of such fields into SM fields; and still others owe to direct experimental bounds
from microwave-cavity experiments, helioscopes, \etc~
A detailed analysis of the exclusion limits implied by these constraints on 
general bulk-axion scenarios will be presented in Ref.~\cite{DynamicalDM3}.  
Here, we merely summarize the results and discuss their implications for a mixed
KK tower of axions as a model of dynamical dark matter.

As we shall discuss further in Ref.~\cite{DynamicalDM3},
it is convenient to separate the applicable constraints into four rough classes, 
based on the origin of the constraint and on the dynamics being probed.     
The first class of constraints which apply to scenarios of this sort 
are those related to the total present-day dark-matter relic abundance $\Omegatotnow$.  
Most of these bounds have been addressed in previous sections, but it will be useful
to recapitulate them here:
\begin{itemize} 
\item The axion ensemble must yield an acceptable contribution to the 
  present-day dark-matter relic density. 
  While $\Omegatotnow < \OmegaDM$ is permitted, provided
  some additional field or fields make up the deficit, values of $\Omegatotnow$ in 
  excess of the WMAP upper bound in Eq.~(\ref{eq:OmegaWMAP}) are excluded.
\item At no time in the past may our ensemble overclose or prematurely 
  matter-dominate the universe.  
\item The present-day effective equation-of-state parameter $w_\ast$ 
  for the ensemble must not deviate significantly from zero.
\item Misalignment production must provide the dominant contribution to 
  $\Omega_\lambda$ for all $a_\lambda$, and the population of hot axions generated
  via thermal production must be negligible.  We therefore require that 
  $\Gamma_{\mathrm{prod}} \ll H$ at all times after the end of cosmic inflation,
  where $\Gamma_{\mathrm{prod}}$ is the total production rate of axions from
  interactions with SM fields in the thermal bath.    
\item We have also assumed that the population of axions generated from the decays 
  of cosmic strings associated with the breaking of the global $U(1)_X$ symmetry 
  is small compared to the population generated by misalignment production.
  We therefore impose the requirement that $f_X \gtrsim H_I$, so that 
  such strings are diluted away by inflation.  
\item Our model must respect current observational limits on isocurvature 
  fluctuations from WMAP~\cite{WMAP}. 
\end{itemize}

The last of these constraints
warrants additional discussion.  
Non-adiabatic fluctuations --- also known as isocurvature fluctuations ---
refer to fluctuations not in the total energy density (which relates 
directly to spacetime curvature) but rather in how that total 
energy density is distributed among different contributing fields
(including the collective contribution from the dark sector).
Such isocurvature fluctuations are tightly constrained by a combination of CMB observations, 
baryon-acoustic-oscillation (BAO) measurements, and supernova data~\cite{WMAP}.
Such fluctuations generically arise whenever a cosmological population of 
particles is produced in a manner such that 
its primordial density perturbations 
are uncorrelated with those of the inflaton field.  Indeed, limits on 
isocurvature fluctuations place severe constraints on the relic abundance 
of a standard QCD axion produced via vacuum 
misalignment~\cite{HertzbergAxionCosmology}, so it might reasonably
be assumed that such limits might play a significant role in constraining 
our model as well.

It turns out, however, that our model satisfies the WMAP constraints
on non-adiabatic fluctuations far more easily than do standard axion dark-matter 
models.  A detailed discussion of these constraints and how they apply to 
bulk-axion models of dynamical dark-matter will be presented in 
Ref.~\cite{DynamicalDM3}, but the gist of the argument is as follows.
Although our dynamical dark-matter ensemble comprises a large number of 
individual components $a_\lambda$, 
the fact that $\Omegatotnow \approx \OmegaDM$  
implies that 
the individual abundance $\Omega_\lambda$
associated with each of these components is actually quite small. 
Furthermore, the underlying five-dimensional nature of our KK axion tower
guarantees that the primordial density fluctuations for each $a_\lambda$ are all determined by the 
fluctuations $\delta\theta$ of the {\it same}\/ initial misalignment angle $\theta$.
For these reasons, the expected magnitude for isocurvature fluctuations 
in our model turns out to be no greater than it is in models in which 
misalignment production causes a single four-dimensional field to carry
the complete dark-matter abundance.
Moreover, if one assumes a Gaussian distribution for $\delta\theta$, it is 
straightforward to demonstrate~\cite{DynamicalDM3}
that $\langle (\delta \theta)^2\rangle \sim H_I^2/(2\pi \hat f_X)^2$.
Thus, all that is required is that 
$H_I \ll \fhatX$ within our preferred regions of parameter space.  However, as 
discussed in Eq.~(\ref{eq:PreferredRegion}), the phenomenologically preferred 
scale for $\fhatX$ in our model is roughly ${\cal O}(10^{14} - 10^{16})~{\rm GeV}$.
For $\fhatX$ at or around this scale, it turns out that current constraints on 
isocurvature fluctuations can be satisfied, provided that 
$H_I \lsim {\cal O}(10^9 - 10^{10})$~GeV.  Such a scale for $H_I$ 
is easy to realize in 
traditional cosmological scenarios, and is even more natural in 
LTR cosmologies wherein the reheating temperature is $\mathcal{O}(\mathrm{MeV})$.  
Thus, in our model, it is not difficult to satisfy current isocurvature 
bounds while simultaneously obtaining a total relic abundance 
$\Omegatotnow \approx \OmegaDM$.   

The underlying reason why our model easily evades these non-adiabatic constraints is that 
within the preferred region of parameter space in Eq.~(\ref{eq:PreferredRegion}),
the five-dimensional axion in our model is not the standard QCD axion.  In particular, we see
that the scale $\Lambda_G$ 
is significantly larger than $\Lambda_{\rm QCD}$.
Our model is
thus freed from the implicit parametric dependence on $\Lambda_{\rm QCD}$ which
afflicts more traditional models of axion dark matter, and allows the
corresponding non-adiabatic fluctuations to have a much smaller scale.


A second class of constraints comprises those observational limits on
processes in which axions are produced via their 
interactions with the fields of the SM and then subsequently detected via 
those same interactions.  These include:
\begin{itemize}
\item Limits from helioscope experiments, such as CAST~\cite{CAST}, which
  search for axions produced by interactions with SM particles in the sun via
  their ``conversion'' to photons in the presence of a magnetic field.
\item Limits from light-shining-through-walls (LSW) experiments
  (see Ref.~\cite{JaeckelReview} for a thorough review), including
  those by the BEV and GammaeV collaborations.  
\end{itemize}
The most stringent of these bounds is currently that from CAST; we shall
therefore take the CAST bound as representative of this class.

The physical processes to which this second class of limits applies  
are all subject to a particular effect which arises universally
in models with both brane and bulk mass terms.
This is the phenomenon of decoherence discussed in   
Refs.~\cite{DDGAxions,DynamicalDM1}.  This decoherence phenomenon 
can substantially suppress the cross-sections for such processes in
our model, and thereby significantly weaken the bounds on $\fhatX$,
$M_c$, and $\Lambda_G$.  To summarize,   
the cross-section for any process in which
axions are produced at some time $t_0$ and then subsequently detected at a later
time $t$ is given by 
\begin{equation}
  \sigma(t) ~\propto~ \frac{N^2}{\fhatX^4} P(t)~,
  \label{eq:CouplerToCouplerXSecBasic}
\end{equation} 
where $N\sim f_X/M_c$ is the number of modes contributing in the sum and
where $P(t)$ is the detection probability at time $t$.  This latter quantity 
is given in the relativistic limit by~\cite{DDGAxions}
\begin{equation}
  P(t) ~=~ \frac{1}{N^2} \left[\sum_\lambda \wtl^8 A_\lambda^4 e^{-\Gamma_\lambda t} +
     \sum_\lambda \sum_{\lambda'\neq\lambda}\wtl^4\wtl'^4A_\lambda^2A_{\lambda'}^2
     e^{-(\Gamma_\lambda + \Gamma_{\lambda'})t/2}
     \cos \left(\frac{(\lambda^2-\lambda'^2)(t-t_0)}{2p}\right) \right]~,
  \label{eq:Poft}
\end{equation}
where $p$ is the initial momentum of the axion.  For any reasonable choice of 
model parameters, the sum in the second term decoheres on time scale so rapid
as to be effectively instantaneous~\cite{DDGAxions}.  As a result, $\sigma(t)$ 
is suppressed, relative to the na\"{i}ve expectation, by an additional factor of
$N$.  This effect considerably weakens the constraints in this class. 

A third class of constraints can be derived from processes in which 
axions are produced via their interactions with SM fields but not 
subsequently detected.  Instead, the presence of the axions is made manifest 
by their ability to carry away momentum and energy from a given system.  These
constraints include:
\begin{itemize}
\item Observational limits on the energy loss in supernovae, and, in 
  particular, on the fraction of the energy released by SN1987A in the 
  form of light exotic fields~\cite{RaffeltSN1987ABound}.  
\item Limits related to the effects of energy dissipation by axions on stellar
  lifetimes.  The most stringent such limits currently come from observations
  of globular-cluster stars~\cite{PDG}, but similar limits have also been 
  derived from constraints on the lifetimes or energy-loss rates 
  of other astrophysical bodies (\eg, the sun~\cite{SolarLifetimeBounds} 
  and white dwarfs~\cite{RaffeltWhiteDwarf}).  
\item Constraints from the absence of observed signals in channels such as 
  $j + \displaystyle{\not}E_T$ and $\gamma + \displaystyle{\not}E_T$ at 
  particle colliders.  In general, the constraints on axion production in 
  these channels are analogous to the well-known constraints on KK-graviton 
  production~\cite{KKGravitonsAtColliders}.   
\item Limits on the branching fractions in particular exotic decay channels for
  certain hadrons~\cite{MassoALPsLongPaper}.
\end{itemize}
The degree to which many of these limits constrain the 
parameter space of bulk-axion scenarios depends quite crucially on how the
axion in question couples to the fields of the SM.  Moreover, many of the constraints in
this class are considerably relaxed in regions of parameter space in which $y\lesssim 1$,
due to the coupling-suppression phenomenon discussed in Ref.~\cite{DynamicalDM1}.  This
effect will be discussed in greater detail in Ref.~\cite{DynamicalDM3}.
 
A fourth and final class of constraints is related to the interactions and decays of a cosmological 
population of axions.  Depending on the cosmological epoch during which such 
decays occur, they can result in a number of potential signals, none of which have
been observed to date.  For example, these include:
\begin{itemize}
\item Decays of cosmic axions which occur after the beginning of the BBN epoch (at around
  $t\sim 1$~s), but before last scattering (at around $t\sim 10^{13}$~s).  These could 
  disrupt nucleosynthesis and affect the abundances of light elements~\cite{BBNLimits}. 
\item  Photoproduction (either primary or secondary) from any axion decays that 
  occur between the epoch of electron-positron annihilation (at around $t\sim 10^{3}$~s) 
  and last scattering.  These can lead to observable distortions of the 
  CMB~\cite{HuAndSilk}. 
\item Photoproduction (either primary or secondary) from any axion decays that 
  occur after last scattering.  These can lead to peaks and other indicative features 
  in the diffuse X-ray and gamma-ray spectra~\cite{YanagidaPhotonLimits}, but such 
  features have not been observed by FERMI~\cite{FERMIdiffGRB}, EGRET~\cite{EGRETdiffGRB}, 
  COMPTEL~\cite{COMPTELdiffXRB,SreekumarDiffuseBG}, or any other X-ray or gamma-ray   
  telescope~\cite{HEAOdiffXRB,ChandraDeepFieldData,ChandraPowerLawFitHickox}.
\item Entropy production from late axion decays.  This can have observational effects on
  cosmological parameters, such as the rate of cosmic expansion. 
\item Limits from microwave-cavity-detector experiments such as 
  CARRACK~\cite{CARRACK} and ADMX~\cite{ADMX}, 
  which search for cosmic axions via their ``conversion'' to 
  photons in the presence of strong magnetic fields.
\end{itemize} 

It should be reiterated that the vast majority of the constraints enumerated above
are highly model-dependent.  The standard energy-dissipation limit from 
SN1987A~\cite{RaffeltSN1987ABound}, for example, provides one of the most stringent 
limits on the parameter space of a QCD axion.  However, these limits are predicated 
on the assumption that the axion couples to hadrons with significant strength, 
and that processes such as $NN\rightarrow N Na$ consequently dominate the 
axion-production rate.  A purely photonic axion, on the other hand, lacks such 
couplings, and hence can only be generated via interactions such as the Primakoff process
$e^-\gamma\rightarrow e^-a$, for which the rate is much smaller.  As a result, 
the bounds on $\fhatX$, $M_c$, and $\Lambda_G$ for such an axion are considerably 
weaker than those for a hadronic axion (see, for example, 
Refs.~\cite{MassoALPsShortPaper,MassoALPsLongPaper} for an analysis of this 
constraint for a four-dimensional photonic axion).  A variety of
other constraints, including bounds from monojet searches at hadron colliders and
from the requirement that misalignment production of axions dominates over thermal
production, also differ markedly depending on whether or not the axion in question
couples to hadrons.  Still other bounds, such as that from energy
loss in white dwarfs~\cite{RaffeltWhiteDwarf}, depend sensitively 
on whether or not a given axion couples to leptons.  

\begin{figure}[t!]
\begin{center}
  \epsfxsize 2.25 truein \epsfbox {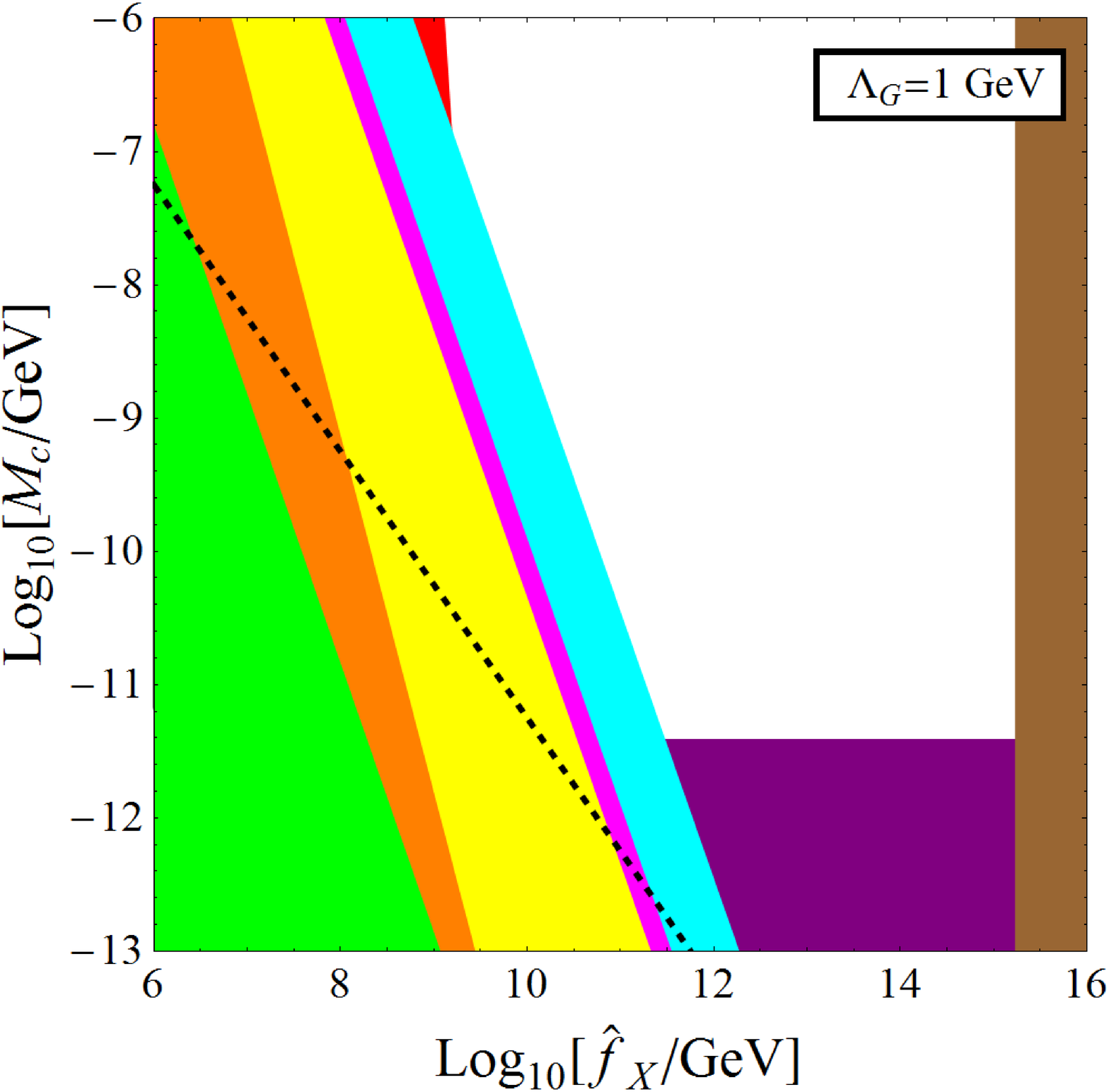}
  \epsfxsize 2.25 truein \epsfbox {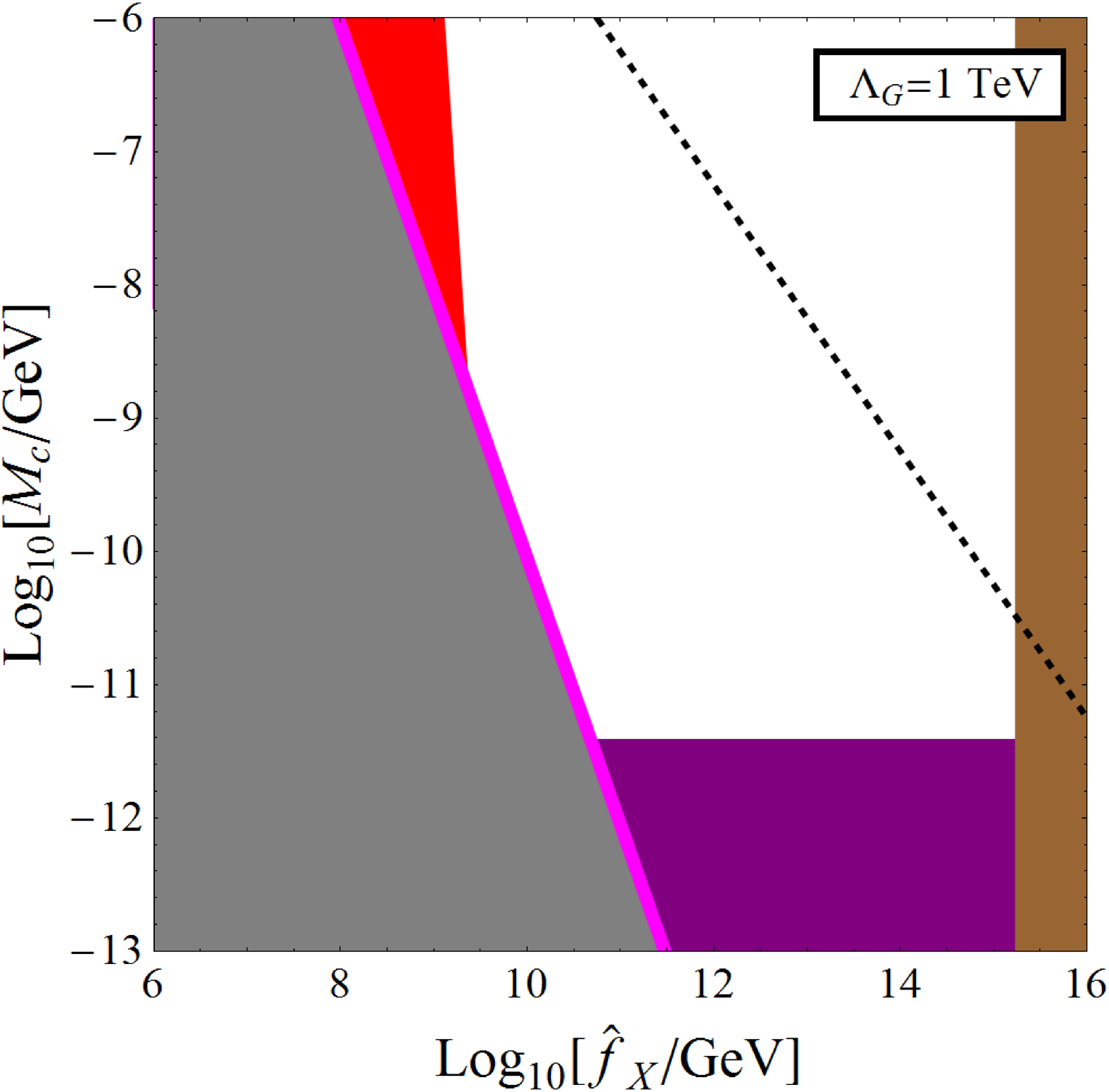}
  \epsfxsize 2.25 truein \epsfbox {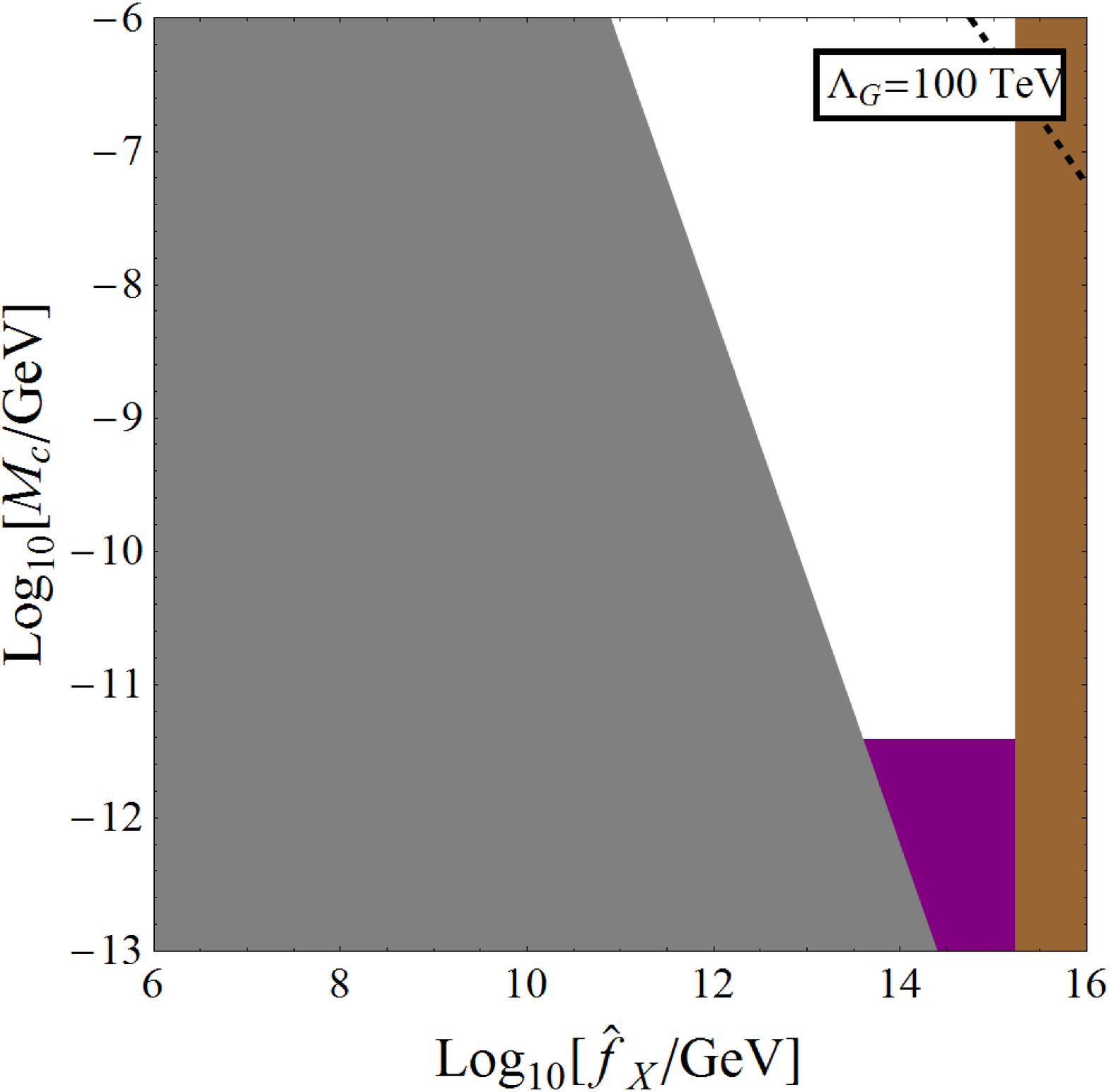}
\end{center}
\caption{Exclusion contours associated with all applicable phenomenological constraints 
for our bulk-axion model with $\Lambda_G = 1$~GeV (left panel), 
$\Lambda_G = 1$~TeV (middle panel), and $\Lambda_G = 100$~TeV (right panel).
In each case, we have taken $\xi=g_G=1$, with $\TRH = 5$~MeV and $H_I = 10^{-3}$~GeV,
and we have assumed that the axion only couples to the photon field.
The shaded regions are respectively excluded by
data from helioscope measurements with CAST (red), collider considerations (magenta), 
tests of Newton's-law modifications at E\"{o}tv\"{o}s-type experiments (purple), 
measurements of the diffuse extragalactic X-ray and gamma-ray spectra (orange), 
observations of the lifetimes of globular-cluster stars (yellow), energy-loss
limits from supernova SN1987A (cyan), the model-consistency requirement that 
$\Lambda_G < f_X$ (gray), and the upper bound on the dark-matter
relic abundance from WMAP (brown).  The black, dashed line corresponds 
to the condition $y=\pi$. 
\label{fig:MasterConstraintPlotPhotonic}}    
\begin{center}
  \epsfxsize 2.25 truein \epsfbox {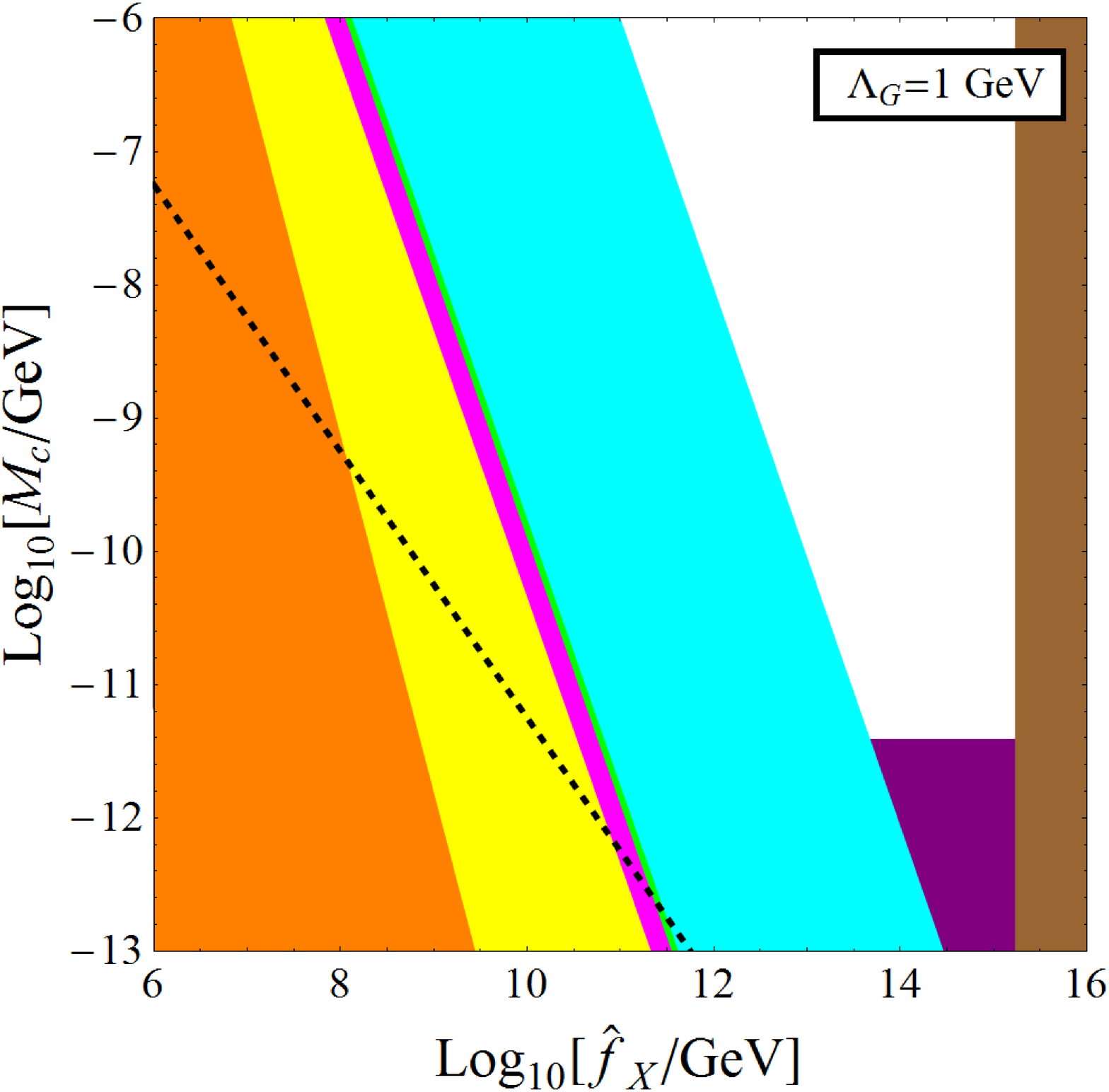}
  \epsfxsize 2.25 truein \epsfbox {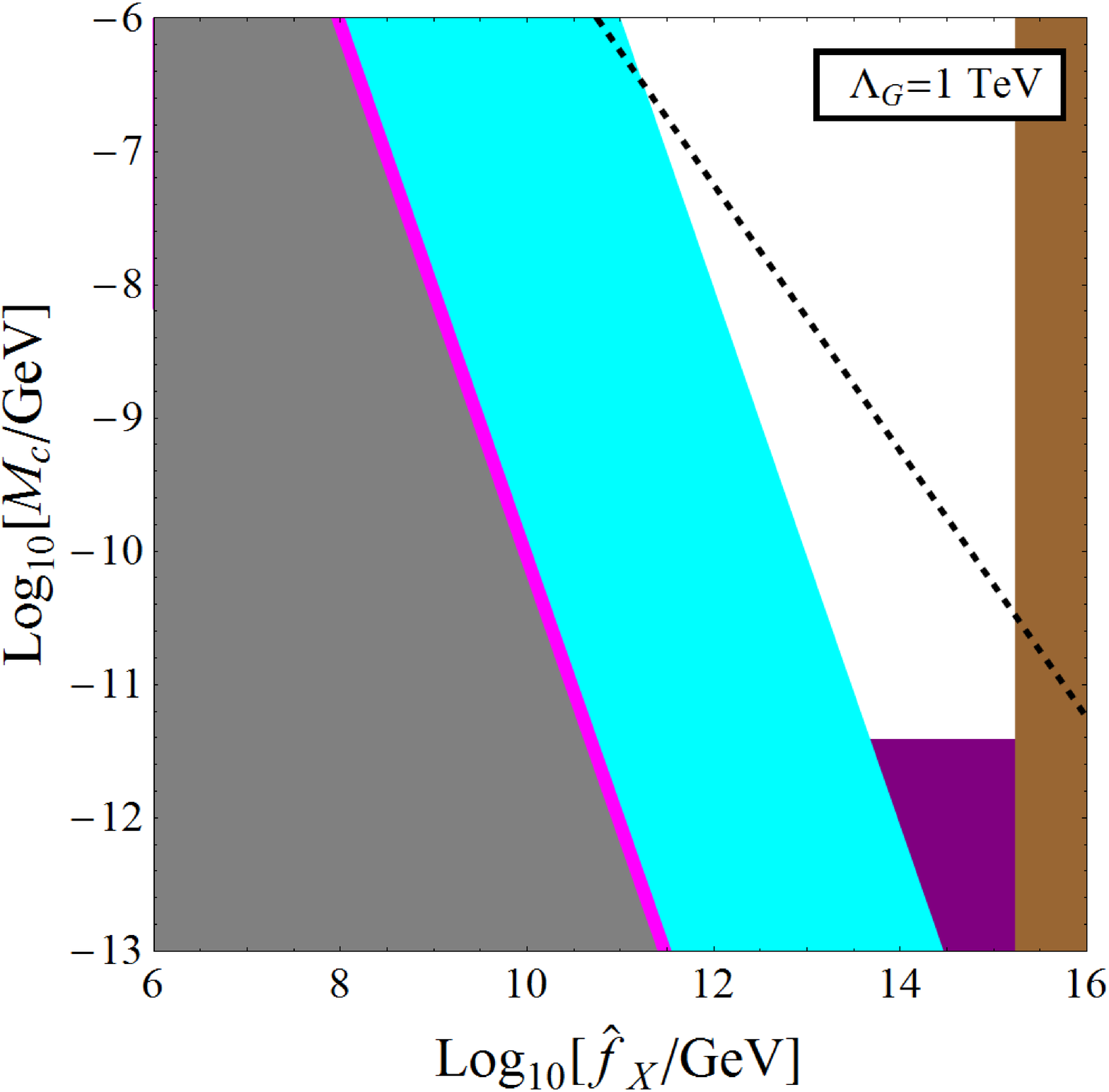}
  \epsfxsize 2.25 truein \epsfbox {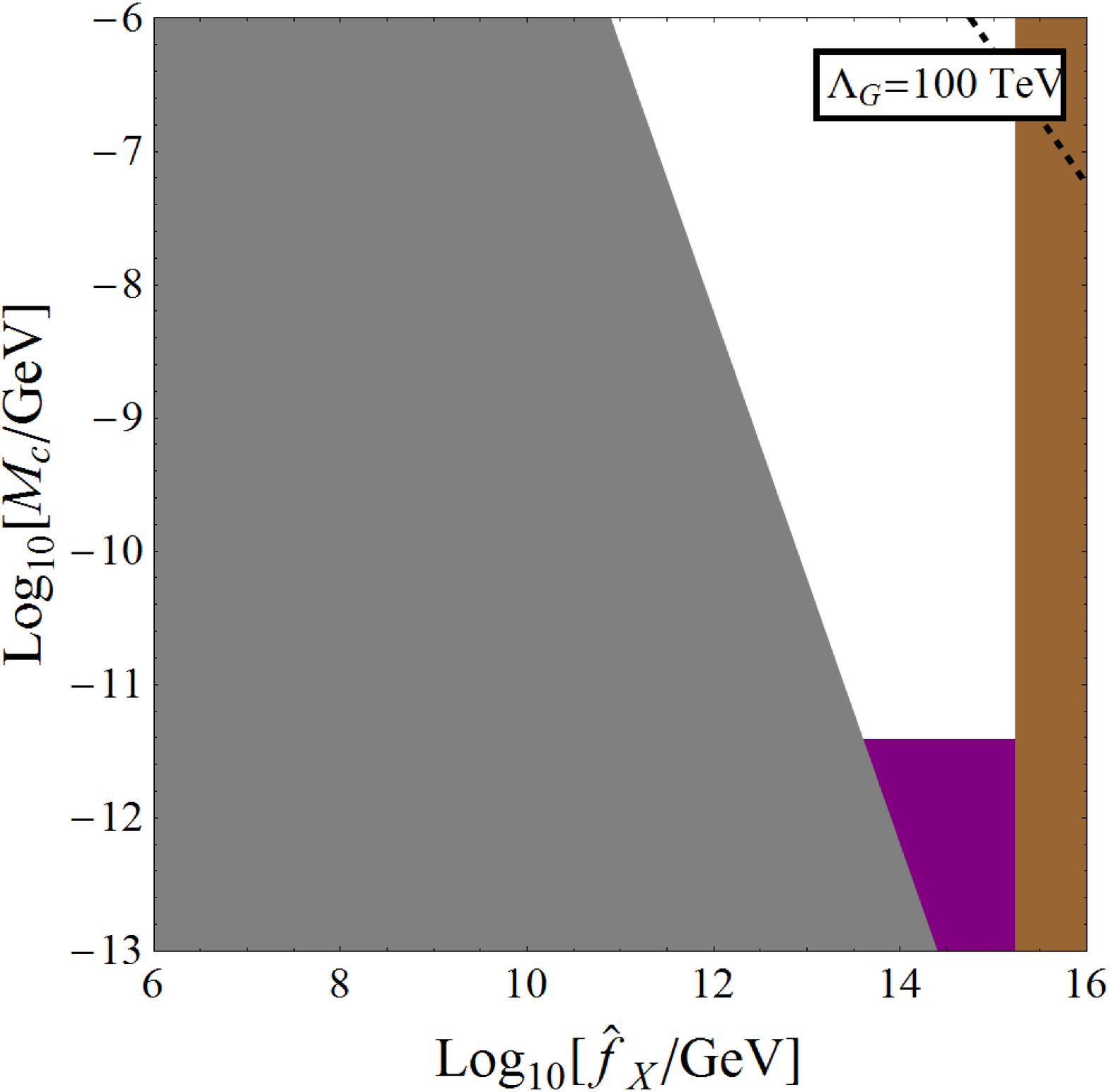}
\end{center}
\caption{Same as in Fig.~\protect\ref{fig:MasterConstraintPlotPhotonic}, but for a 
``hadronic'' axion --- \ie, an axion coupled both to the photon and to the gluon 
field (and hence to pions, nucleons, \etc), but not directly to SM quarks or leptons.
\label{fig:MasterConstraintPlotHadronic}}    
\end{figure} 

In Figs.~\ref{fig:MasterConstraintPlotPhotonic} and~\ref{fig:MasterConstraintPlotHadronic},
we display a series of exclusion plots in $(\fhatX,M_c)$ space, 
taken from Ref.~\cite{DynamicalDM3}, which 
indicate the regions of parameter space excluded by the considerations enumerated above.
The three panels in Fig.~\ref{fig:MasterConstraintPlotPhotonic} correspond to
$\Lambda_G = \{1\mathrm{~GeV}, 1\mathrm{~TeV},100\mathrm{~TeV}\}$ for the case of
a photonic axion with $c_\gamma = 1$, while the three panels in Fig.~\ref{fig:MasterConstraintPlotHadronic} correspond to the same choices of 
$\Lambda_G$, but for a hadronic axion with $c_\gamma = c_g = 1$.  In each case, 
we have taken $\xi=g_G=\theta=1$, with $\TRH = 5$~MeV and $H_I = 10^{-3}$~GeV;
for the hadronic case, we have also assumed that $C_{a\pi}$, $C_{a\pi N}$, \etc, take 
the values given in Eqs.~(\ref{eq:CapAndCan}) and~(\ref{eq:CaNAndCapi}).
The shaded regions in each panel are excluded by the battery of 
constraints discussed above.  The red region is excluded by CAST data, the 
magenta region by limits on collider processes in which axions appear as missing 
energy, the purple region by limits on modifications of Newton's law from 
E\"{o}tv\"{o}s-type experiments, the orange region by limits on distinguishable
features in the diffuse extragalactic X-ray and gamma-ray background spectra,
the yellow region by observations of the lifetimes of globular-cluster stars,
the cyan region by energy-loss limits from supernova SN1987A, the gray region
by the model-consistency requirement that $\Lambda_G < f_X$, and the brown region
by the upper bound on the dark-matter relic abundance from WMAP.  
A black, dashed line corresponding to the condition $y=\pi$ has also been
included in each panel for reference.  Note that each of the exclusion regions
shown, with the exception of that from WMAP,
differs from the corresponding exclusion region for a four-dimensional axion.  The
exclusion regions shown are those appropriate for the five-dimensional axion on which
our model is based, and are derived in Ref.~\cite{DynamicalDM3}.    

The constraints enumerated above for which no exclusion contour has 
been included in these figures are generally subleading.   
For example, the applicable constraints from exotic hadron 
decays~\cite{MassoALPsLongPaper} are generally far weaker than the constraints
from SN1987A, thermal production, \etc, for any given choice of parameters.   
The constraints arising from observational limits on distortions of the CMB
are not particularly stringent either, and turn out not to constrain any portion of the 
model parameter space shown in any of the panels appearing in  
Figs.~\ref{fig:MasterConstraintPlotPhotonic} and~\ref{fig:MasterConstraintPlotHadronic}.
This is because the regions of parameter space in which the $\Gamma_\lambda$ are
sizeable are those in which $\fhatX$ is quite small, meaning that the $\Omega_\lambda$ 
are also quite small, as is evident from Fig.~\ref{fig:OmegaTotPanelsLTR}. 
Constraints related to the effects of late-decaying $a_\lambda$ on BBN were not
explicitly calculated in Ref.~\cite{DynamicalDM3}.  However,    
exclusion contours derived from BBN constraints on late-decaying particles are 
expected to be roughly similar to those derived from CMB constraints, and consequently 
such constraints are not expected to rule out any additional region of model parameter    
space not already excluded by other considerations.  Limits on the effective
equation-of-state parameter $w_\mathrm{eff}(t)$ 
are not particularly constraining either.  This should
come as no surprise, given that we showed in Sect.~\ref{sec:Abundances} that the
effect of decays on $\Omegatotnow$ was negligible within the region of parameter   
space relevant for dynamical dark matter.
A number of additional constraints not listed above also serve to constrain very light 
axions and axion-like particles~\cite{StringAxiverse}; however the particles for which 
these constraints apply typically involve values of $\mX$ far smaller than those of
interest here.   

It is evident from these figures that the most stringent constraints on both 
photonic and hadronic axions are those from SN1987A (cyan) and from 
collider limits on missing-energy processes (magenta).
Nevertheless, it is also evident that a hadronic axion is significantly more 
constrained than a purely photonic axion.  As discussed above, the primary 
reason for this is that the rate of axion production in a thermal setting via
interactions with nuclei, pions, \etc, is far larger than the corresponding 
rate of production via the electron Primakoff process and other interactions
which involve the coupling of an axion to photons alone.    

Having assessed the phenomenological constraints on a bulk axion, we
are now able to definitively address the question as to whether or not 
our model is a viable model of dynamical dark-matter.
In order for this to be so, we require that at least some part of the preferred 
region in Eq.~(\ref{eq:PreferredRegion}) be consistent with the constraints 
discussed above.  Inspecting Figs.~\ref{fig:MasterConstraintPlotPhotonic} 
and~\ref{fig:MasterConstraintPlotHadronic}, we see that indeed our preferred 
region is compatible with all of these constraints 
in both the photonic and hadronic axion cases for $\Lambda_G \gtrsim 100$~GeV,
with $M_c$ above the lower bound from Newton's-law modification, but small enough
so that $y \lesssim \pi$.  Furthermore, we also see from 
Figs.~\ref{fig:MasterConstraintPlotPhotonic} 
and~\ref{fig:MasterConstraintPlotHadronic} that the phenomenological constraints
even permit us  to reach deeply into the $y\ll 1$ region.  Note that this
represents a radical departure from the QCD-axion results presented in 
Ref.~\cite{DDGAxions} --- a departure which is enabled 
because $\Lambda_G$ is a free parameter in our model.
{\it Thus, we conclude that within this region of parameter 
space, our bulk-axion model constitutes a viable, explicit model of dynamical 
dark matter.}

It should be stated that in addition to the limits discussed above, 
certain additional astrophysical bounds may also serve to constrain 
the parameter space of bulk-axion scenarios. 
For example, it has recently been shown~\cite{GiannottiDuffy} 
that limits on gamma-ray signals from decaying axions with masses of
$\mathcal{O}(10-100\mathrm{~MeV})$ produced in supernovae 
can yield an even more stringent limit than that arising from energy-dissipation
considerations alone.  While these bounds are once again
model-dependent (and directly applicable only to cases in which the axion in question
couples directly to hadrons with significant strength, and not to a photonic 
axion), they could provide an 
important additional constraint on the parameter space of dynamical dark-matter 
scenarios involving bulk axions.  Furthermore, it is also possible that
comparable bounds could be obtained from an analysis of photo-emission limits 
and cooling-rate constraints from neutron stars, similar to that performed for
KK gravitons in Ref.~\cite{HannestadRaffeltNeutronStar}.  

It is important to note that while additional bounds related to axion production in 
supernovae may serve to further constrain the parameter space of bulk-axion scenarios,
these constraints cannot rule out axion models of dynamical dark matter
entirely.  This is due to the fact that for any given choice of model parameters 
$\fhatX$, $M_c$, and $\LambdaG$, the couplings of 
any mode for which $\lambda \lesssim \pi \mX^2/M_c$ to the SM fields will 
be suppressed by mixing effects, as discussed in Ref.~\cite{DynamicalDM1}.  
Indeed, because we can reach deeply into the $y\ll 1$ region, the magnitude of 
this coupling suppression can be quite significant. 
For example, for $y\sim \mathcal{O}(10^{-3})$, we find that the first twenty 
axion mass eigenmodes have coupling suppressions $\wtl^2A_\lambda\sim 10^{-6}$.
If the coupling suppressions are significant for those $a_\lambda$ with masses in the 
``dangerous'' range $10\mathrm{~MeV} \lesssim \lambda\lesssim 100$~MeV discussed
above, such $a_\lambda$ will be produced in supernovae at a negligible rate, and 
thus all supernova bounds on axion production can be evaded.  This can be 
arranged by demanding that $\pi \mX^2/M_c \gtrsim 1$~GeV, so that all modes with
masses $\lambda \ll 1$~GeV are effectively in the small-$\lambda$ regime.  Therefore, 
since $\Omegatotnow$ is essentially independent of $\Lambda_G$ within our preferred
region of parameter space, satisfying this condition is simply a matter of 
choosing a sufficiently large value for $\Lambda_G$. 
Indeed, setting $\fhatX = 10^{14}$~GeV in accord with Eq.~(\ref{eq:PreferredRegion}), 
we find that all axion-production constraints from supernovae can be avoided for 
\begin{equation}
  \LambdaG ~\gtrsim~ (56\mathrm{~TeV}) \times 
   \left(\frac{M_c}{10^{-11}\mathrm{~GeV}}\right)^{1/4}~.
\end{equation}
We emphasize that this rough bound is not a necessary condition for consistency 
with supernova data, but a sufficient one.  Furthermore, since the neutron-star 
cooling and photo-emission bounds on KK gravitons rest on the assumption that a 
population of gravitationally-bound particles of this sort was generated by the 
supernova whose core-collapse produced a given neutron star, any similar bound on 
axions would also cease to apply in this regime.  We also note that
since the fundamental scale $f_X$ is still roughly an order of magnitude larger 
than the value of $\Lambda_G$ required to satisfy this condition, given the input 
values of $\fhatX$ and $M_c$, no theoretical inconsistency results from positing a
confinement scale of this order.

The fact that this coupling-suppression phenomenon is capable of 
rendering our model consistent with supernova bounds despite the 
large multiplicity of light modes attests to the importance of this effect
in brane/bulk theories.  A more detailed overview of
this phenomenon and its physical implications will be provided in 
Ref.~\cite{DynamicalDM3}.


\section{Discussion and Conclusions\label{sec:Conclusions}}


The aim of this paper has been to present an explicit realization of the 
dynamical dark-matter framework presented in Ref.~\cite{DynamicalDM1}.   
To that end, we have shown that an ensemble consisting of the KK excitations of a light, 
axion-like field can indeed provide such a realization.  Indeed, we have shown
that despite the fact that the masses, decay widths, and relic abundances of 
all of these particles are controlled by
only three dimensionful parameters, the ensemble to which they give rise is 
simultaneously able to reproduce the observed value of $\OmegaDM$ and satisfy 
all applicable constraints from laboratory experiments, astrophysics, and cosmology.  
As such, this model provides a ``proof of concept'' for dynamical dark matter as
a viable alternative framework for dark-matter physics.  In addition, it also provides 
a method of addressing the dark-matter question which does not require the introduction 
of any additional stabilizing symmetry.    

Many qualifications, extensions, and possible generalizations of our dynamical 
dark-matter framework were discussed at the end of Ref.~\cite{DynamicalDM1}; here, 
we shall restrict our attention to five points which are specific to the bulk-axion 
model presented in this paper.

\begin{itemize}    

\item
First, in this work, we have made use of the rapid-turn-on approximation in
Eq.~(\ref{eq:Heaviside})
in calculating the relic abundances of the $a_\lambda$.  As discussed in 
Sect.~\ref{sec:Production}, this approximation is well motivated, since the
instanton-generated mass term $\mX(T)$ falls rapidly with temperature when
$T\gtrsim \Lambda_G$.  Furthermore, the primary results of this paper are 
essentially insensitive to this approximation.  This is because  
the fields which contribute significantly to $\OmegaDM$ in regions of 
parameter space which yield a realistic dark-matter relic abundance begin 
oscillating only well after $\mX(T)$ has already settled into its constant, late-time 
value.  However, the relic abundance of any field which begins oscillating 
before $\mX(T)$ takes this late-time value will, in general, depend on the 
details of how this mass evolves in time.  The quantitative effect on the 
abundance of a single field has long been appreciated~\cite{Turner}, but 
in our model, the effects are more complicated and more subtle
because we have a coupled system of mixed scalars with different masses and 
therefore different oscillation times.  It would be interesting to examine how
a more rigorous treatment of the turn-on of $\mX(T)$ would affect $\Omegatot$
and $\eta$ in situations in which these quantities are sensitive to the
time-dependence of this brane-mass term.  Such a study would have important
implications for more general scenarios involving other 
kinds of light bulk scalars.  Indeed, the relationship between the 
size of the brane-mass term for such scalars and the time at
which that brane mass is dynamically generated may differ significantly 
from the relationship which holds for axions.              

\item
Second, as alluded to in Sect.~\ref{sec:Constraints}, it may be possible 
to further test or constrain the parameter space of bulk-axion models of dynamical
dark matter in a number of ways.  We have already mentioned one potential
constraint which derives from limits on high-energy photons resulting from
the decays of axions produced in supernovae~\cite{GiannottiDuffy}.
Other considerations
also merit investigation.  For example, a detailed analysis of the limits 
imposed by BBN on scenarios involving multiple decaying fields with different 
lifetimes and abundances could provide important constraints on 
dynamical dark-matter models in general.  In addition, other considerations, 
such as limits on mass loss and decreases in the dark-matter density in the halos 
of dwarf galaxies~\cite{DwarfGalaxyDecayingDM}, could also be used to constrain 
dynamical dark-matter models.  Indeed, while a number of standard 
constraints on individual unstable relic particles in the early universe have
been revisited in a dynamical dark-matter context~\cite{DynamicalDM3}, it would
be interesting to see how other constraints would apply in this context as well.

\item 
Third, we note that we have not specified a particular model of
inflation as part of the cosmological context for our model.  Indeed,
other than requiring a low reheating temperature $\TRH \sim 
\mathcal{O}(\mathrm{MeV})$, we have remained largely agnostic about the
details of the inflationary model, the form of the inflaton potential, or
even the scale $H_I$.  For the most part, our model does not depend on
these particulars.  However, certain consistency conditions do place
meaningful restrictions on the set of inflationary scenarios with which
our model is compatible.  One such condition can be derived from the fact
that vacuum fluctuations during inflation generically give rise to a
background value $\langle \phi^2 \rangle \approx H_I^3 t_I/4\pi^2$ for any
scalar $\phi$ with a mass $m_\phi \ll H_I$, where $t_I$ is the duration
of inflation.  This implies that the relationship between the mass $\lambda$
and initial energy
density $\rho_\lambda$ in Eq.~(\ref{eq:RhoInitCondits}) in our model is
truly valid only for the lighter $a_\lambda$ in a given tower --- \ie, those for
which $\theta^2 A_\lambda^2 \fhatX^2 \gtrsim H_I^3 t_I/4\pi^2$.  By contrast,
any heavier $a_\lambda$ which still satisfy $\lambda \ll H_I$ receive
the leading contributions to their background values from vacuum fluctuations
during inflation, and thus effectively acquire an initial abundance
$\rho_\lambda \sim \lambda^2 H_I^3 t_I$.  In typical scenarios, we expect
$H_I t_I \approx N_e \sim \mathcal{O}(60)$, where $N_e$ is the number of
$e$-foldings of inflation.  The results for $\Omegatot$ derived in
Sect.~\ref{sec:Abundances} therefore remain consistent, provided that
$\fhatX^2 \gg H_I^2$.  Indeed, since $\fhatX \sim 10^{14} -
10^{15}$~GeV within the preferred region of parameter space specified in
Eq.~(\ref{eq:PreferredRegion}), we see that $\fhatX \gg H_I$ is certainly not
inconsistent with our model and is in fact even expected.  However, this condition
on $H_I$ has non-trivial implications for inflationary models.
While a low scale for $H_I$ is certainly not excluded 
(see, \eg, Refs.~\cite{LTRAxionsKamionkowski,lowscaleinflation}), extremely small
values of $H_I$ tend to be rather non-generic~\cite{TensorToScalarNonGeneric}
among typical classes of inflationary potentials, and thus
require either substantial tuning or careful construction.  Indeed, any
consistent inflationary model of this sort must give rise to density
fluctuations on a scale consistent with constraints from CMB
data~\cite{WMAP}, such as those on the spectral index $n_s$, and must also
satisfy other observational constraints.  The development of explicit
inflationary scenarios of this sort is therefore an interesting topic for
future investigation.

\item
Fourth, we note that while we have chosen in this paper to focus on the case
in which the ensemble of fields reproducing $\OmegaDM$ are the KK excitations 
of a bulk axion field, such a field is by no means unique in possessing the 
characteristics necessary to give rise to such an ensemble.  Indeed, as 
discussed in Ref.~\cite{DynamicalDM1},     
much of the analysis presented here pertains to any light bulk scalar for which
a mass term is dynamically generated via its interactions with brane-localized 
fields.  Furthermore, for a generic bulk scalar, the relationship between the time 
at which this mass term is dynamically generated and the 
magnitude of this mass term itself may differ from that which relates $t_G$ 
and $\mX$ for a bulk axion.  As a result, much more freedom may exist for 
constructing viable models within the dynamical dark-matter framework.  For 
example, light moduli could also, in principle, provide a viable model of 
dynamical dark matter.  

\item
Finally, we emphasize that the presence of additional axion-like 
fields is fairly generic, and perhaps even expected, in many 
theoretically motivated scenarios for physics beyond the Standard 
Model (see, \eg, Ref.~\cite{WittenStringAxion}).  Moreover, it has even been argued
that many of these axion-like fields are likely to be light~\cite{StringAxiverse}.  
Thus, the discovery of a vast ensemble of axion-like particles 
could provide important insight into what physics looks like at high scales.  
Indeed, if many of these axions have relatively small masses, we find ourselves 
in the intriguing situation in which most of the matter in the universe is 
simultaneously both light and dark. 

\end{itemize} 

Our goal in this work has been to provide an existence proof for dynamical dark
matter --- \ie, to provide a model in which lifetimes are balanced against 
abundances in such a 
way that the ensemble of dark-matter particles successfully reproduces $\OmegaDM$ 
while at the same time satisfying all phenomenological constraints.  As we have seen 
in this paper, our bulk-axion model indeed passes this test.  In one sense, our 
model does so in the most interesting way possible: with $y\ll 1$ (signifying that
our tower of axion KK modes is highly mixed) and with a tower fraction $\eta$ which 
is significantly different from zero.  In another sense, however,
this model is fairly conservative: those modes which contribute most to 
$\Omegatotnow$ turn out to be rather long-lived, and likewise our numerical result for $w_\ast$ 
within the preferred region of parameter space turns out to be rather close to 
zero.  Indeed, at first 
glance, one might suspect that these latter properties are in fact generic for 
dynamical dark-matter models, or even that such models are therefore really no different 
from traditional dark-matter models in terms of their abundance and stability 
requirements.
  
This is not the case, however, for the balancing of lifetimes against 
abundances --- which is the hallmark of the dynamical dark-matter framework ---
is precisely why this framework does {\it not} require such a degree of 
stability, much less the existence of a stabilizing symmetry.
While certain accidental features of
our bulk-axion model result in a preferred region of parameter space
which is somewhat conservative, we emphasize that these features are not 
generic even to theories with bulk scalars, much less 
realistic dynamical dark-matter models as a whole.  Note, for example,   
that a particular relationship exists in bulk-axion models between the mass 
$\lambda$ of a given KK mass eigenstate $a_\lambda$, the strength of its effective 
coupling to SM fields, and the overall magnitude of its relic abundance 
$\Omega_\lambda$ through the dependence of these quantities on 
$\fhatX$.  Even for other bulk scalars (\eg, moduli), these 
relationships do not necessarily hold.  There is therefore no reason to 
expect dynamical dark-matter models based around such fields to be as 
conservative as the axion model we have presented here.  

In this connection, there is an even more important point that deserves
emphasis.  In dynamical dark-matter scenarios, we have no single
characteristic decay width $\Gamma$ nor abundance $\Omega$, but rather an 
entire {\it spectrum}\/ of widths $\Gamma_\lambda$ and abundances 
$\Omega_\lambda$.  This therefore begs the fundamental question:  if our 
``proof of concept'' model presented here is to be viewed as somewhat
conservative, how far from the conservative limit can we go?

At first glance, one might try to answer this question by attempting to 
determine, for each time $t$ during the evolution of the universe, 
the maximum abundance $\Omega_{\mathrm{max}}(t)$ that a given
component in a dark-matter ensemble may have if it has a 
lifetime $\tau \sim t$.  In other words, given the entirety of the 
cosmological constraints from BBN, CMB distortions, \etc, there exists a 
{\it function}\/ $\Omega_{\mathrm{max}}(\Gamma)$ which describes the 
maximum abundance any dark-matter constituent may have as a function of 
its decay width.  It might therefore seem that knowledge of this function 
would uniquely determine the full range of possibilities inherent in 
our dynamical dark-matter framework.

Such an approach to answering our fundamental question is, in a sense, 
already a departure from the usual manner of approaching dark-matter physics.
However, even the notion of such a function $\Omega_{\rm max}(\Gamma)$
relies too strongly on a single-particle perspective.
One of the critical features of our dynamical dark-matter framework
is that it involves a vast {\it ensemble}\/ of dark-matter components.
Some of these components might decay earlier in cosmological evolution,
while others might decay later. 
As a result, the maximum abundance that a given component may have if 
it decays on a characteristic time scale $\tau$ will itself
be directly affected not only by the abundances of all of the
other components with earlier characteristic decay times $\tau' < \tau$,
but even the components with $\tau' > \tau$.  Moreover,
as we have seen, most phenomenological constraints on dark-matter decays
are sensitive not merely to what happens at a specific moment in time, but
to the integrated effects of such decays over a broad range of time scales.
In other words, our dynamical dark-matter framework 
teaches us that astrophysical and cosmological constraints do not lead 
to a single function $\Omega_{\rm max}(\Gamma)$,
but rather a more subtle set of intertwined constraints on lifetimes and 
abundances across our entire dark-matter ensemble as a whole.

Clearly, this issue has not been studied in any detail in the literature.
However, it is readily apparent that this is indeed the only proper way
in which one should express constraints on particle decays from
a generic dark sector.  Viewed from this perspective, then, the
existence of even one viable dynamical dark-matter model --- no matter
how ``conservative'' it might be --- gives us strong motivation to re-examine 
cosmological and astrophysical constraints within this framework.
Indeed, it is only in this way that we will be able to fully explore
our dynamical dark-matter framework, and understand its full range
of phenomenological possibilities.


\begin{acknowledgments}


We would like to thank K.~Abazajian, Z.~Chacko, D.~Chung, M.~Drees, J.~Feng, 
J.~Kost, J.~Kumar, R.~Mohapatra, S.~Su, T.~Tait, X.~Tata, and N.~Weiner for discussions. 
This work was supported in part by the Department of Energy under Grants 
DE-FG02-04ER41291 and DE-FG02-04ER-41298. 
The opinions and conclusions expressed here are those of 
the authors, and do not represent either the Department of Energy or
the National Science Foundation.

\end{acknowledgments}

\bigskip

\appendix


\section{~Evolution of a Decaying Axion Field\label{app:ExactSolsaAppendix}}


For completeness, in this Appendix we provide exact solutions to 
Eq.~(\ref{eq:TheDoubleDotEqnWithGammaLambda}) for a real-valued 
function $a_\lambda(t)$.  These take the form
\begin{equation}
  a_\lambda(t) ~ = ~ 
    c_\lambda^{(M)}\widetilde{M}_\kappa(\lambda,t) + c_\lambda^{(U)}
      \widetilde{U}_\kappa(\lambda,t)~,
    \label{eq:MasterSolsArbitrary}
\end{equation}
where $c_\lambda^{(M)}$ and $c_\lambda^{(U)}$ are undetermined 
constants, and
\begin{eqnarray}
  \widetilde{M}_\kappa(\lambda,t) &\equiv& e^{-(k_\lambda+\Gamma_\lambda)t/2}\bigg[
      M\left(\frac{\kappa(k_\lambda+\Gamma_\lambda)}{2k_\lambda},\kappa,k_\lambda t\right) +
      e^{k_\lambda t}
      M\left(\frac{\kappa(k_\lambda-\Gamma_\lambda)}{2k_\lambda},\kappa,-k_\lambda t\right)
      \bigg]\nonumber\\
  \widetilde{U}_\kappa(\lambda,t) &\equiv& e^{-(k_\lambda+\Gamma_\lambda)t/2}\bigg[
      U\left(\frac{\kappa(k_\lambda+\Gamma_\lambda)}{2k_\lambda},\kappa,k_\lambda t\right) +
      e^{k_\lambda t}U\left(\frac{\kappa(k_\lambda-\Gamma_\lambda)}{2k},\kappa,-k_\lambda
       t\right)\bigg]~.
  \label{eq:DefOfFU}
\end{eqnarray}
In these expressions, $k_\lambda\equiv\sqrt{\Gamma_\lambda^2-4\lambda^2}$, $\kappa$
was defined in Eq.~(\ref{eq:DefOfnForH}), 
$M(a,b,x)$ denotes Kummer's confluent hypergeometric function
\begin{equation}
  M(a,b,x) ~=~ \sum_{m=0}^{\infty} \frac{(a)_m x^m}{(b)_m m!}~,
\end{equation}
where $(x)_n = (x+n-1)!/(x-1)!$ is the Pochhammer function, and 
$U(a,b,x)$ denotes the Tricomi confluent hypergeometric function 
\begin{equation}
  U(a,b,x) ~=~ \frac{\Gamma(1-b)}{\Gamma(a-b+1)}M(a,b,x) + 
     \frac{\Gamma(b-1)}{\Gamma(a)}x^{1-b}M(a-b+1,2-b,x)~.
\end{equation}
It can be verified upon setting $\Gamma_\lambda = 0$ that 
Eq.~(\ref{eq:MasterSolsArbitrary}) reduces to the exact form 
obtained for a tower of stable KK axions in Ref.~\cite{DDGAxions}.

The values of $c_\lambda^{(M)}$ and $c_\lambda^{(U)}$ in
Eq.~(\ref{eq:MasterSolsArbitrary}) are determined by the initial
conditions chosen for $a_\lambda(t)$ and $\dot{a}_\lambda(t)$ at $t=t_0$, where
$t_0$ is some initial time.  Expressed in terms of these initial values, this
equation takes the general form   
\begin{equation}
  a_\lambda(t) ~ = ~ \frac{
    \Big[\dot{a}_\lambda(t_0)  {\widetilde{U}}_\kappa(\lambda,t_0) - 
      a_\lambda(t_0)\dot{\widetilde{U}}_\kappa(\lambda,t_0)\Big]\widetilde{M}_\kappa(\lambda,t) - 
    \Big[\dot{a}_\lambda(t_0)\widetilde{M}_\kappa(\lambda,t_0) - 
      a_\lambda(t_0)\dot{\widetilde{M}}_\kappa(\lambda,t_0)\Big]\widetilde{U}_\kappa(\lambda,t)}{
    \dot{\widetilde{M}}_\kappa(\lambda,t_0)\widetilde{U}_\kappa(\lambda,t_0) - 
      \dot{\widetilde{U}}_\kappa(\lambda,t_0)\widetilde{M}_\kappa(\lambda,t_0)}~,
    \label{eq:MasterSolsGeneralInit}
\end{equation}
where the time derivatives of $\widetilde{M}_\kappa(\lambda,t)$ 
and $\widetilde{U}_\kappa(\lambda,t)$
have the explicit forms
\begin{eqnarray}
  \dot{\widetilde{M}}_\kappa(\lambda,t)&=&
     e^{-(k_\lambda+\Gamma_\lambda)t/2}(k_\lambda-\Gamma_\lambda)\Bigg[
      M\left(\frac{\kappa(k_\lambda+\Gamma_\lambda)}{2k_\lambda},\kappa,k_\lambda t\right) - 
     e^{k_\lambda t} 
     M\left(\frac{\kappa(k_\lambda-\Gamma_\lambda)}{2k_\lambda}+1,\kappa+1,-k_\lambda
      t\right)\Bigg]\nonumber\\
  \dot{\widetilde{U}}_\kappa(\lambda,t)&=& -\frac{1}{2}
      e^{-(k_\lambda+\Gamma_\lambda)t/2}\Bigg\lbrace
      (k_\lambda+\Gamma_\lambda)\left[
      U\left(\frac{\kappa(k_\lambda+\Gamma_\lambda)}
      {2k_\lambda},\kappa,k_\lambda t\right)+
      \kappa\, U\left(\frac{\kappa(k_\lambda+\Gamma_\lambda)}
      {2k_\lambda}+1,\kappa+1,k_\lambda t\right) \right] 
      \nonumber \\ & & ~~~~~~~~~\, - 
      e^{k_\lambda t}(k_\lambda-\Gamma_\lambda)\left[
      U\left(\frac{\kappa(k_\lambda-\Gamma_\lambda)}
      {2k_\lambda},\kappa,-k_\lambda t\right) +
      \kappa\, U\left(\frac{\kappa (k_\lambda-\Gamma_\lambda)}
      {2k_\lambda}+1,\kappa+1,-k_\lambda t\right)\right]\Bigg\rbrace~.
\end{eqnarray}
Once again, if we set $\Gamma_\lambda =0$ in this
expression (which also implies that $k_\lambda=2i\lambda$), we recover the result 
\begin{equation}
  a_\lambda(t)~\stackrel{\Gamma_\lambda\rightarrow 0}{\longrightarrow}~ -\frac{\pi}{\sqrt{2}}
      a_\lambda(t_0)\lambda t_0^{5/4}t^{-1/4}\left[J_{-5/4}(\lambda t_0)J_{1/4}(\lambda t)+
      J_{5/4}(\lambda t_0)J_{-1/4}(\lambda t)\right]~,
   \label{eq:MasterEqnSolDDGLimit}
\end{equation}
which agrees with the result obtained in Ref.~\cite{DDGAxions}.

In the rapid-turn-on approximation, in which $\mX(t)$ takes the Heaviside
form specified in Eq.~(\ref{eq:Heaviside}), the initial conditions for 
$a_\lambda$ and $\dot{a}_\lambda$ at $t_0 = t_\lambda$ take the form given in 
Eq.~(\ref{eq:alambdaInitCondits}). 
Upon substituting these initial conditions into Eq.~(\ref{eq:MasterSolsGeneralInit}), 
we find that during the cosmological epoch in which coherent oscillations of a given
$a_\lambda$ begin, we have
\begin{equation}
  a_\lambda(t) ~ = ~ \theta\fhatX A_\lambda\,\frac{
    \dot{\widetilde{M}}_\kappa(\lambda,t_\lambda)\widetilde{U}_\kappa(\lambda,t)
    - \dot{\widetilde{U}}_\kappa(\lambda,t_\lambda)\widetilde{M}_\kappa(\lambda,t)}{
    \dot{\widetilde{M}}_\kappa(\lambda,t_\lambda)\widetilde{U}_\kappa(\lambda,t_\lambda) - 
      \dot{\widetilde{U}}_\kappa(\lambda,t_\lambda)\widetilde{M}_\kappa(\lambda,t_\lambda)}~.
    \label{eq:MasterSolsInitSub}
\end{equation}
The value of any $a_\lambda$ during subsequent epochs can then be obtained
iteratively from this relation.


\end{document}